\RequirePackage{rotating} %
\documentclass[structabstract]{aa}
\usepackage[table]{xcolor}
\usepackage{siunitx}
\usepackage{graphicx}
\usepackage{tabularx}
\usepackage{array}
\usepackage{nccmath}
\usepackage{natbib}
\usepackage{longtable}
\usepackage{caption}
\usepackage{subcaption}
\DeclareUnicodeCharacter{2212}{\textendash}
\usepackage[colorlinks=true, allcolors=blue]{hyperref}
\usepackage{txfonts}

\newcommand{\kms}{km~s$^{-1}$}
\newcommand{\M}{\textbf{Maser}}
\newcommand{\Th}{th.}
\newcommand{\Bl}{bl.}
\newcommand{\BO}{\textbf{Maser}$+$th.}
\newcommand{\ND}{n.d.}
\newcommand{\NO}{---}
\begin{document}

   \title{Vibrationally excited HCN transitions in circumstellar envelopes of carbon-rich AGB stars}

   %\subtitle{I. Overviewing the $\kappa$-mechanism}

   \author{Manali Jeste
          \inst{1}\fnmsep\thanks{Member of the International Max Planck Research School (IMPRS) for  Astronomy  and  Astrophysics  at  the  Universities  of  Bonn  and Cologne.}, 
          %\and
          Yan Gong\inst{1},
          Ka Tat Wong\inst{2},
          Karl M. Menten\inst{1},
          Tomasz Kami\'nski\inst{3}, and
          Friedrich Wyrowski\inst{1}
          }

   \institute{Max-Planck-Institut f\"ur Radioastronomie, Auf dem H\"ugel 69, D-53121 Bonn, Germany\\
              \email{mjeste@mpifr.de}
        \and
             Institut de Radioastronomie Millimétrique, 300 rue de la Piscine, 38406 Saint Martin d’Hères, France
        \and
             Nicolaus Copernicus Astronomical Center, Polish Academy of Sciences, Rabia{\'n}nska 8, 87-100 Toru{\'n}, Poland
             }

   \date{Received ---; accepted ---}

% \abstract{}{}{}{}{} 
% 5 {} token are mandatory
 
  \abstract
  % context heading (optional)
  % {} leave it empty if necessary  
   {HCN is the most abundant molecule after H$_{2}$ and CO in the circumstellar envelopes (CSEs) of carbon-rich asymptotic giant branch (AGB) stars. Its rotational lines within vibrationally excited states are exceptional tracers of the innermost region of carbon-rich CSEs.}
  % aims heading (mandatory)
   {We aim to constrain the physical conditions of CSEs of carbon-rich stars using thermal lines of the HCN molecule. Additionally, we also search for new HCN masers and probe the temporal variations for HCN masers, which should shed light on their pumping mechanisms.}%expanding the knowledge of the lines in the carbon-rich stars.}%expand the knowledge of HCN masers in the entire carbon-rich class. 
  % methods heading (mandatory)
   {We observed 16 carbon-rich AGB stars in various HCN rotational transitions within the ground and 12 vibrationally excited states with the Atacama Pathfinder EXperiment (APEX) 12 meter submillimeter telescope.}
  % results heading (mandatory)
  {We detect 68 vibrationally excited HCN lines from 13 carbon-rich stars, including 39 thermal transitions and 29 maser lines, which suggests that vibrationally excited HCN lines are ubiquitous in carbon-rich stars. Population diagrams constructed, for two objects from the sample, for thermal transitions from different vibrationally excited states give excitation temperature around 800$−-$900 K, confirming that they arise from the hot innermost regions of CSEs (i.e., $r<$20$R_{*}$). Among the detected masers, 23 are newly detected, and the results expand the total number of known HCN masers lines toward carbon-rich stars by 47\%. In particular, the $J=2$--$1$ (0, 3$^{1e}$, 0), $J=3$--$2$ (0, 2, 0), $J=4$--$3$ (0, 1$^{1f}$, 0) masers are detected in an astronomical source for the first time. Our observations confirm temporal variations of the 2--1 (0, 1$^{1e}$, 0) maser on a timescale of a few years. Our analysis of the data suggests that all detected HCN masers are unsaturated. A gas kinetic temperature of ${\gtrsim}700$~K and an H$_{2}$ number density of $>$10$^{8}$~cm$^{-3}$ are required to excite the HCN masers. In some ways, HCN masers in carbon-rich stars might be regarded as an analogy of SiO masers in oxygen-rich stars.}
  % conclusions heading (optional), leave it empty if necessary 
   {}


\titlerunning{Vibrationally excited HCN transitions in circumstellar envelopes of carbon-rich AGB stars}

\authorrunning{Jeste et al.}
   
   \maketitle

\section{Introduction}

Low- and intermediate-mass stars will reach the Asymptotic Giant Branch (AGB) phase in a late stage of their evolution, in which they play a crucial role in the cosmic matter cycle \citep[e.g.,][]{2012sse..book.....K,2018A..hoef..olof&ARv..26....1H}. Because of the mass-loss processes in this phase, circumstellar envelopes (CSEs) are created around these AGB stars. Depending on the element ratio [C]/[O] in the stellar photosphere, AGB stars can belong to three types: oxygen-rich stars (also known as M-type stars, [C]/[O]$<$1), S-type stars ([C]/[O]$\approx$1),  carbon-rich star (also known as C-type stars, [C]/[O]$>$1) \citep{1996A&ARv...7...97H,2018A..hoef..olof&ARv..26....1H}. In light of different chemical compositions for different AGB types \citep{1964AnTok...9.....T,2006A&A...456.1001C,2013A&A...550A..78S}, different tracers should be carefully selected to study the physical conditions of CSEs toward different types of AGB stars.

%HCN is a linear molecule and its lowest rotational emission line was first detected by \citet{1971ApJ..snyder..163L..47S}. HCN has been widely detected in various environments including star-forming regions \citep[e.g.,][]{2014ApJ...787...74G}, CSEs of AGB stars \citep[e.g.,][]{1971ApJ...170L.109M,cerni..2011A&A...529L...3C}, planetary nebula \citep[e.g.,][]{2001ApJ...562..824H}, and supernova remnants \citep[e.g.,][]{1992ApJ...399..114T}. 
HCN is of great interest for studying the CSEs of carbon-rich stars, because it is one of the parent molecules formed in the stellar atmosphere, and is also the most abundant molecule after H$_{2}$ and CO in such stars, as suggested by chemical models \citep[e.g.,][]{1964AnTok...9.....T,2006A&A...456.1001C}. Observations have also confirmed that the HCN fractional abundances relative to H$_{2}$ are as high as $\sim$3$\times 10^{-5}$ for carbon-rich stars \citep[e.g.,][]{fonfria2008detailed,2013A&A...550A..78S}. HCN is thus widely used to study the CSEs of carbon-rich stars.

\begin{table*}[!htbp]
\setlength{\tabcolsep}{3pt}
\centering
\caption{Relevant parameters of the 16 C-rich stars selected for this study.}\label{tab:sample}
%\small
\hspace*{-0.8cm}\begin{tabular}{lcccccccccc}
\hline
\hline
 Source name & R.A. & Dec.  & Distance & $V_\text{LSR}$  &  $\dot{M}$ & 2$V^{*}_\text{inf}$& Observation Time & Period  & $L_{\star}$\\
&  J2000 & J2000 & (pc) & (\kms) & (M$_\odot$ $\text{yr}^{-1} $) & (\kms) & (hours) & (d) & (L$_\odot$)
\\
\hline
W Ori$^e$     & 05:05:23.72 & +01:10:39.5    & 490  & $-$1.0  & $3.1 \times {10^{-7}}$   & 22  & 19.33 & 212& 11600\\
S Aur$^e$     &  05:27:07.45 & +34:08:58.6    & 1099  & $-$17.0    & $5.9 \times {10^{-6}}$   & 51 & 7.08 &  590.1 & 24000 \\
V636 Mon$^c$  & 06:25:01.43 & $-$09:07:15.9  & 880  & +10.0  & $5.8 \times {10^{-6}}$   & 40 & 11.06 & 543 &8472\\ 
 R Vol$^b$     & 07:05:36.20 & $-$73:00:52.02 & $880_{-176}^{+149}$ & $-$11.0 & $(2.9 \pm{0.68}) \times {10^{-6}}$& 36 & 16.02 & 453 &8252 \\
IRC +10216$^a$ & 09:47:57.4  & +13:16:44      & 140  & $-$26.5 & $2-4\times {10^{-5}}$ & 36 & 3.56 & 630 & 8750$^c$ \\
RAFGL 4211$^a$ & 15:11:41.9  & $-$48:20:01    & 950  & $-$3.0   & $0.93 \times {10^{-5}}$ & 42 & 14.11 & 632&\\
II Lup$^a$    & 15:23:05.7  & $-$51:25:59    & 640  & $-$15.0 & $1 \times {10^{-5}}$ &  46  & 13.21 & 576 & 8900$^d$\\
IRC +20370$^c$ & 18:41:54.39 & +17:41:08.5    & 600  & $-$0.8  &  $3.0 \times {10^{-6}}$   & 28 & 7.07 &524 &7900  \\
V Aql$^c$     & 19:04:24.15 & $-$05:41:05.4  & 330  & +53.5  & $1.4 \times {10^{-7}}$   & 16 & 4.11 & 407 & 6500\\
IRC +30374$^c$ & 19:34:09.87 & +28:04:06.3    & 1200 & $-$12.5  & $1.0 \times {10^{-5}}$   & 50 & 3.37  & --&9800\\
AQ Sgr$^b$    & 19:34:18.99 & $-$16:22:27.04 & $330_{-60}^{+95}$   & +20.0  &$2.5 \times {10^{-6}}$ & 25 & 7.59 & 199.6&2490 \\
CRL 2477$^c$  & 19:56:48.43 & +30:43:59.9    & 3380 & +5.0     & $1.1 \times {10^{-4}}$   & 40 & 2.40 & -- & 13200\\
CRL 2513$^c$  & 20:09:14.25 & +31:25:44.9    & 1760 & +17.5  & $2.0 \times {10^{-5}}$   & 51 & 2.57 & 706 & 8300\\
RV Aqr$^c$    & 21:05:51.74 & $-$00:12:42.0  & 670  & +0.5   & $2.3 \times {10^{-6}}$   & 30 & 10.7 & 453 & 6800\\
Y Pav$^b$     & 21:24:16.74 & $-$69:44:01.96 & $400_{-77}^{+125}$  &  0.0   &$(2.8 \pm{0.96}) \times {10^{-6}}$& 22 & 12.08 & 449&5076  \\
CRL 3068$^c$  & 23:19:12.24 & +17:11:33.4    & 1300 & $-$31.5  & $2.5 \times {10^{-5}}$   & 29 & 16.11 & 696&10900  \\
 %W Ori$^c$     & 05:05:23.72 & +01:10:39.5    & 220  & $-$1.0  & $7.0 \times {10^{-8}}$   & 22  & 15.26 & 212& 3500\\

 %S Aur$^c$     &  05:27:07.45 & +34:08:58.6    & 300  & $-$17.0    & $4.0 \times {10^{-7}}$   & 49 & 4.53 &  590.1 & 8900 \\
\hline
\end{tabular}
\bigskip
\tablefoot{This table gives information about the parameters of the sample of the carbon-rich stars. References: $^a$\citet{2018AA...613A..49M} -- distance considered here is D$^\text{PLR}$, where the distance is derived from the Mira period-luminosity relation; $^b$\citet{2017A&A..rau..600A..92R} (Mass loss rate from \citealt{Loup..1993A&AS...99..291L}); $^c$\citet{2018A&A..massalkhi..611A..29M}. $^d$\citet{2014Ramstedt}; $^{e}$ \citet{2001A&A...368..969S}; The distances for W Ori (220 pc) and S Aur (820 pc) were derived from the Hipparcos measurements and the period-luminosity relation in \citet{2001A&A...368..969S}, respectively. The distances are updated by the more accurate GAIA parallax measurements \citep{2018A&A...616A...1G}. Their mass-loss rates and luminosities are also updated by scaling the new distances. All periods are from the AAVSO (\url{https://www.aavso.org/}) with the exception of CRL 2513 and IRC+10216 (\citealt{2006A&A...460..539K}; \citealt{menten..2012A&A...543A..73M}; \citealt{Pardo2018}).}
\normalsize
\end{table*}

The HCN molecule has three vibrational modes, corresponding to the CH stretching mode, $\nu_{1}$ , the doubly degenerate bending mode, $\nu_{2}$, and the CN stretching mode, $\nu_{3}$, together denoted as ($\nu_{1}$, $\nu_{2}$, $\nu_{3}$). The wavelengths of the fundamental transitions of these modes are 3.0, 14.0, and 4.8 $\mu$m, which correspond to excitation temperatures of 4764, 1029, and 3017 K, respectively \citep[e.g.][]{1934PhRv...45..277A,2014MNRAS.437.1828B,2018AA...613A..49M}. These infrared ro-vibrational lines are difficult to observe with ground-based telescopes due to the Earth's atmosphere. In contrast, numerous HCN rotational transitions from many  vibrational states are easily accessible with ground-based millimeter and submillimeter telescopes. Many such HCN transitions have  been widely detected in CSEs of carbon-rich stars \citep{cerni..2011A&A...529L...3C,Bieging_2001}, the preplanetary nebula CRL 618 \citep{2003ApJ...586..338T}, star-forming regions \citep{2011A&A...529A..76R}, and even external galaxies \citep{2010ApJ...725L.228S,2021A&A...656A..46M}. Because of the high extinction and density, the innermost regions of CSEs are difficult to study in lower $J$ rotational transitions from the vibrational ground state, which become optically thick; where $J$ is the angular momentum quantum number.  The HCN rotational transitions in vibrationally excited  states usually have much lower opacities. These properties, together with their high energy levels and critical densities, make them exceptional tracers to investigate the innermost regions of CSEs around carbon-rich stars, which are still poorly constrained. Similar to the observations of vibrationally excited SiO and H$_{2}$O lines toward oxygen-rich stars \citep{2016A&A...590A.127W}, observations of vibrationally excited HCN lines may pave the way toward understanding the acceleration zone and the atmosphere around the photosphere of carbon-rich stars. Furthermore, detections of multiple transitions of HCN will place constraints on the excitation conditions of the maser lines arising in these regions.

Studies of vibrationally excited HCN transitions have been still limited. A large number of HCN $J=3$--$2$ rotational transitions
from 28 vibrational states have been investigated toward IRC+10216 \citep{cerni..2011A&A...529L...3C}. Using \emph{Herschel}/PACS and SPIRE data, \citet{2018AA..Nicolaes...618A.143N} performed the population diagram analysis of HCN vibrationally excited lines for 8 carbon-rich stars, but they could not resolve the line profiles due to the coarse spectral resolution of their data. Both studies infer high excitation temperatures, supporting the idea that these transitions arise from the innermost region of CSEs. High spectral resolution observations toward more carbon-rich stars can provide more statistical constraints on the role of vibrationally excited HCN transitions in studying the entire class of carbon-rich AGB stars. 

Masers are important tools to study CSEs \citep[e.g.,][]{1996A&ARv...7...97H,2014ARA&A..52..339R}. OH, SiO, and H$_{2}$O masers are common in CSEs of oxygen-rich stars, but these masers have not been detected toward carbon-rich stars so far \citep{maser..book..2012msa..book.....G}. Instead, HCN and SiS transitions are commonly found to show maser actions toward carbon-rich stars \citep[e.g.,][]{lucas1988new,Bieging_2001,1983ApJ...267..184H,2006ApJ...646L.127F,schilke..mehringer..menten2000ApJ...528L..37S,sis..Gong_2017,2018ApJ...860..162F,2018AA...613A..49M}. Compared with SiS masers, HCN masers are found to be more widespread \citep{2018AA...613A..49M}. However, previous maser studies have been limited to certain HCN transitions and a low number of targets (see Table~\ref{tab:catalogue}). Extending the search for HCN masers to other vibrationally excited HCN transitions and toward more carbon-rich stars may shed light on the pumping mechanism.

With the motivations mentioned above, studies of HCN rotational lines in vibrationally excited states with high spectral resolutions are indispensable to investigate the innermost region of carbon-rich CSEs. We therefore undertake new observations of multiple vibrationally excited transitions with different rotational quantum numbers $J$ toward a selected sample of carbon-rich stars. In this paper, we start with an overview of our sample of carbon stars in Section~\ref{tab:sample}, followed by the observations and data reduction of the stellar sample in Sections \ref{sec:sample} and \ref{sec:obs}. We present our results in Section \ref{sec:results} and discuss in detail all the sources (some individually) in Section \ref{sec:discussion}, followed by a brief discussion on the aspect of summary and outlook of the project in Section \ref{sec:summary}.

\section{Sample of carbon stars}
\label{sec:sample}

\begin{table}[!htbp]
%\centering
%\caption{\textbf{Summary of observations}}
\caption{List of observed HCN lines in ground and vibrationally excited states with the APEX telescope.}\label{tab:lines}
\begin{tabular}{cccccc}
\hline
$ \Delta J$ & $\nu$  & ($\nu_{1}$, $\nu_{2}$, $\nu_{3}$) & $E_{\rm u}$ & $S_{\rm ul} \mu^2$& $\log_{10}(A_{\rm ul})$ \\
     & (MHz) &     & (K) & (D$^2$) &     \\
 (1) & (2)   & (3) & (4) & (5)     & (6) \\            
\hline
2--1 & 176011.260 & (1, 0, 0) & 4777.1 & 54.6 & $-3.64$ \\
     & 176052.380 & (0, 0, 1) & 3029.5 & 53.3 & $-3.65$ \\
     & 177238.656 & (0, $1^{1e}$, 0) & 1037.1 & 38.9 & $-3.77$ \\
     & 177261.111 & (0, 0, 0) & 12.8 & 53.6 & $-3.63$ \\
     & 177698.780 & (0, $3^{1e}$, 0) & 3053.6 & 36.7 & $-3.80$ \\
     & 178136.478 & (0, $1^{1f}$, 0) & 1037.2 & 38.9 & $-3.77$ \\
     & 178170.380 & (0, $2^{0}$, 0) & 2043.5 & 50.4 & $-3.66$ \\
\hline
%33 & 248573.053 & (0, 1, 0) & 3417.0 & 1.6 & $-3.56$ \\
3--2 & 264011.530 & (1, 0, 0) & 4789.8 & 81.9 & $-3.08$ \\
     & 264073.300 & (0, 0, 1) & 3042.2 & 79.9 & $-3.09$ \\
     & 265852.709 & (0, $1^{1e}$, 0) & 1049.9 & 69.2 & $-3.14$ \\
     & 265886.434 & (0, 0, 0) & 25.5 & 80.5 & $-3.08$ \\
     & 266540.000 & (0, $3^{1e}$, 0) & 3066.4 & 65.2 & $-3.16$ \\
     & 267109.370 & (0, $2^{2f}$, 0) & 2078.1 & 42.0 & $-3.35$ \\
     & 267120.020 & (0, $2^{2e}$, 0) & 2078.1 & 42.0 & $-3.35$ \\
     & 267199.283 & (0, $1^{1f}$, 0) & 1050.0 & 69.2 & $-3.14$ \\
     & 267243.150 & (0, $2^{0}$, 0) & 2056.3 & 75.6 & $-3.10$ \\
     & 269312.890 & (0, $3^{1f}$, 0) & 3066.6 & 65.2 & $-3.15$ \\
\hline
4--3 & 354460.435 & (0, $1^{1e}$, 0) & 1066.9 & 97.4 & $-2.73$ \\
     & 354505.478 & (0, 0, 0) & 42.5 & 107.3 & $-2.69$ \\
     & 355371.690 & (0, $3^{1e}$, 0) & 3083.4 & 91.8 & $-2.75$ \\
     & 356135.460 & (0, $2^{2f}$, 0) & 2095.2 & 75.6 & $-2.83$ \\
     & 356162.770 & (0, $2^{2e}$, 0) & 2095.2 & 75.6 & $-2.83$ \\
     & 356255.568 & (0, $1^{1f}$, 0) & 1067.1 & 97.4 & $-2.72$ \\
     & 356301.178 & (0, $2^{0}$, 0) & 2073.4 & 100.8 & $-2.71$ \\
     & 356839.554 & (0, $3^{3e}$, 0) & 3127.2 & 42.8 & $-3.08$ \\
     & 356839.650 & (0, $3^{3f}$, 0) & 3127.2 & 42.8 & $-3.08$ \\
\hline
\end{tabular}
\tablefoot{(1) Rotational transition. (2) Rest frequency from the CDMS database \citep[CDMS;][]{2003JMoSp.220..223Z,2005JMoSt.742..215M}. (3) Vibrational state. (4) Upper-level energy. (5) Line Strength. (6) Einstein $A$ coefficient.
%The instrument used is indicated below the rotational transition: SEPIA180 for $J=2-1$; PI230 for $J=3-2$ and the direct $l$-type transition in $J=33$; FLASH345 for $J=4-3$. See Table~\ref{setup_table} for the receiver setups.
}
\label{summary_obs}
\end{table}

Table~\ref{tab:sample} lists the information of the 16 carbon-rich stars in our survey. These sources are selected by applying the following two criteria: (1) the targets should be accessible with the Atacama Pathfinder EXperiment\footnote{This publication is based on data acquired with the Atacama Pathfinder EXperiment (APEX). APEX is a collaboration between the Max-Planck-Institut f\"ur Radioastronomie (MPIfR), the European Southern Observatory, and the Onsala Space Observatory.} telescope (APEX); (2) the carbon-rich nature of these stars and the detection of HCN have already been confirmed by previous studies. We ended up with a 16 object sample, three of them from \citet{2018AA...613A..49M}, where the authors explored 177 GHz HCN maser emission toward a sample of 13 carbon-rich AGB stars, three other targets from \citet{2017A&A..rau..600A..92R}, and the remaining ten sources from \citet{2018A&A..massalkhi..611A..29M}. These stars have mass-loss rates on the order of 10$^{-7}$ to 10$^{-4}$~$M_{\odot}$~yr$^{-1}$, spanning about three orders of magnitude. Additionally, we estimated the pulsation phase of the observed stars for different observational epochs (see Table \ref{tab:phase}), which is discussed further in Appendix \ref{app:phasetext}.

\section{Observations and Data Reduction}
\label{sec:obs}

We carried out spectral line observations toward the selected 16 carbon-rich stars with the APEX 12-m telescope \citep{2006A&A...454L..13G} from 2018 June 8 to 2018 December 20 under the projects M-0101.F-9519A-2018 and M-0102.F-9519B-2018. The total observing time is 185.06 hours. We observed a total of 26 rotational transitions of HCN, $J=2$--$1$, $J=3$--$2$, and $J=4$--$3$, in various vibrationally excited states. Table~\ref{tab:lines} shows the list of the observed HCN lines. We made use of SEPIA \citep[a Swedish-ESO PI receiver;][]{2012ITTST...2..208B,2016Msngr.165...13I},  PI230, and FLASH$^{+}$ \citep[both MPIfR receivers;][]{2014ITTST...4..588K}, as the frontend instruments to observe HCN transitions in $J=2$--$1$, 3$-$2, and 4$-$3, respectively. Table \ref{tab:setup} gives a brief summary of the instrumental setups of our observations. All HCN lines in the same frequency band were observed and calibrated simultaneously in a single setup. The Fast Fourier Transform Spectrometers (FFTSs) were connected as backend to analyze the spectral data \citep{2006A&A...454L..29K}. These FFTS modules provide us spectral resolutions of 76~kHz (or 0.13~\kms), 61~kHz (or 0.07~\kms), 38~kHz (or 0.03~\kms) for lines at 177~GHz, 267~GHz, and 355~GHz, respectively. %better than 0.1~\kms: $-$0.13~\kms (for $J=2-1$), 0.07~\kms (for $J=3-2$), and $-$ 0.03~\kms (for $J=4-3$).

Additional observations of IRC +10216, II Lup, and RAFGL 4211, i.e. the three targets from \citet{2018AA...613A..49M}, were observed again with the SEPIA receiver ($\eta_{\rm mb}$ = 0.85) in October 2021 under the project M-0108.F-9515B-2021 to study the variability of the vibrationally excited HCN lines in $J=2$--$1$ (Sect.~\ref{sec:variability}).

\begin{table}[!htb]
\centering
\caption{APEX receiver setups of the HCN observations.}
\label{tab:setup}
\begin{tabular}{ccccccc}
\hline
 Receiver & Tuning $\nu$ (SB) &  $\Delta J$ & IF range &  $\eta_{\rm mb}$& Jy/K\\
          & (GHz)    &            & (GHz)    &      \\
    (1)   & (2)      &  (3)       &  (4)     &         (5) & (6)\\
\hline
SEPIA180\tablefootmark{a} & 177 (USB) & 2--1 &  4$-$8  &0.77& 33\\
PI230\tablefootmark{b}    & 266 (USB) & 3--2 & 4$-$12 &0.62& 46\\
FLASH345\tablefootmark{c} & 355 (USB) & 4--3 & 4$-$8  &0.66& 53\\
\hline
\end{tabular}
\tablefoot{The instrumental information is obtained from the APEX website, \url{https://www.apex-telescope.org/instruments/}. (1) Receiver name. (2) Tuning frequency (Sideband). The tuning frequency corresponds to an IF frequency of 6 GHz for all of the receivers. (3) Rotational transitions of the covered HCN lines. (4) Intermediate frequency range. (5) Main-beam efficiencies. (6) Jansky to Kelvin conversion factor. \\
\tablefoottext{a}{\citet{2018A&A...611A..98B,2018A&A...612A..23B}.} \tablefoottext{b}{\url{https://www.eso.org/public/teles-instr/apex/pi230/}; also Sect. 2.1 of \citet{2020A&A...636A..39B}.} \tablefoottext{c}{\citet{2014ITTST...4..588K}.}}
\end{table}

The spectra were obtained with a secondary mirror, wobbling at a rate of 1.5 Hz and a beam throw of $\pm$60\arcsec\,in azimuth. Total integration times for different targets are summarized in Table~\ref{tab:sample}. The chopper-wheel method was used to calibrate the antenna temperature, $T_{\rm A}^{*}$ \citep{1976ApJS...30..247U}, and the calibration was done every 8--10 minutes. We converted the antenna temperatures to main-beam temperatures, $T_{\rm mb}$, with the relationship, $T_{\rm mb}= T_{\rm A}^{*}\eta_{\rm f}/\eta_{\rm mb}$, where $\eta_{\rm f}$ and $\eta_{\rm mb}$ are the telescope's forward efficiency and main-beam efficiency, respectively. We obtained the values of $\eta_{\rm f}$, $\eta_{\rm mb}$ , and the Jansky to Kelvin (Jy/K) conversion factor for HCN $J=3$--$2$ from the telescope's report\footnote{\url{https://www.apex-telescope.org/telescope/efficiency/}}. To get precise $\eta_{\rm mb}$ for the $J=2$--$1$ and $J=4$--$3$ transitions of HCN that was not included in the telescope's report, we calculated the antenna temperature from the cross scans of Mars taken during the observations and compared it with the modeled brightness temperature of Mars at that time\footnote{\url{https://lesia.obspm.fr/perso/emmanuel-lellouch/mars/}}. Table~\ref{tab:setup} lists the main-beam efficiency and Jy/K factor for each HCN line. The telescope's focus was checked after sunrise and sunset, while the pointing, based on nearby pointing sources, was accurate to within $\sim$2\arcsec. The full-width at half-maximum beam sizes, $\theta_{\rm b}$, are 36\arcsec, 24\arcsec, and 18\arcsec\,at 177 GHz, 267 GHz, and 356 GHz, respectively. Velocities are given with respect to the local standard of rest (LSR). A brief summary of observed HCN rotational lines in various vibrational states is given in Table~\ref{tab:lines}.

Data reduction was performed with the GILDAS\footnote{\url{https://www.iram.fr/IRAMFR/GILDAS/}} software package \citep{2005sf2a.conf..721P}. First order polynomial baseline was subtracted for all observed spectra. The spectra were smoothed at the expense of velocity resolution to improve the signal-to-noise (S/N) ratios.

\section{Results}   
\label{sec:results}

\begin{figure*}[!htbp]%[!htb]
\centering
\includegraphics[width=0.33\textwidth]{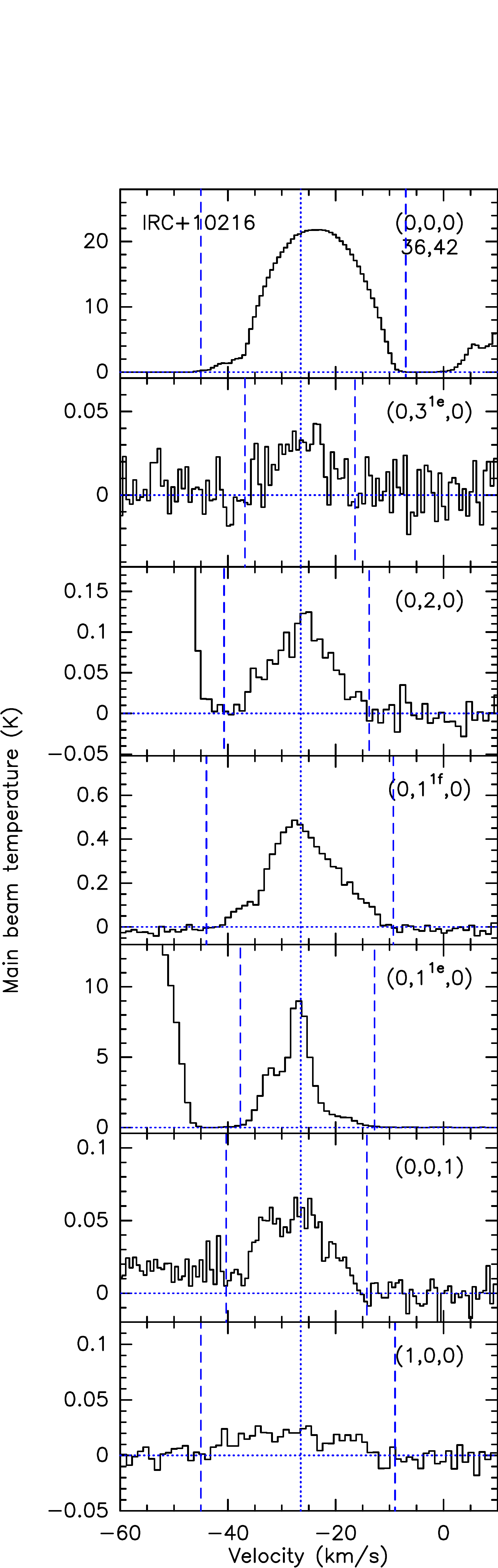}\quad
\includegraphics[width=0.62\textwidth]{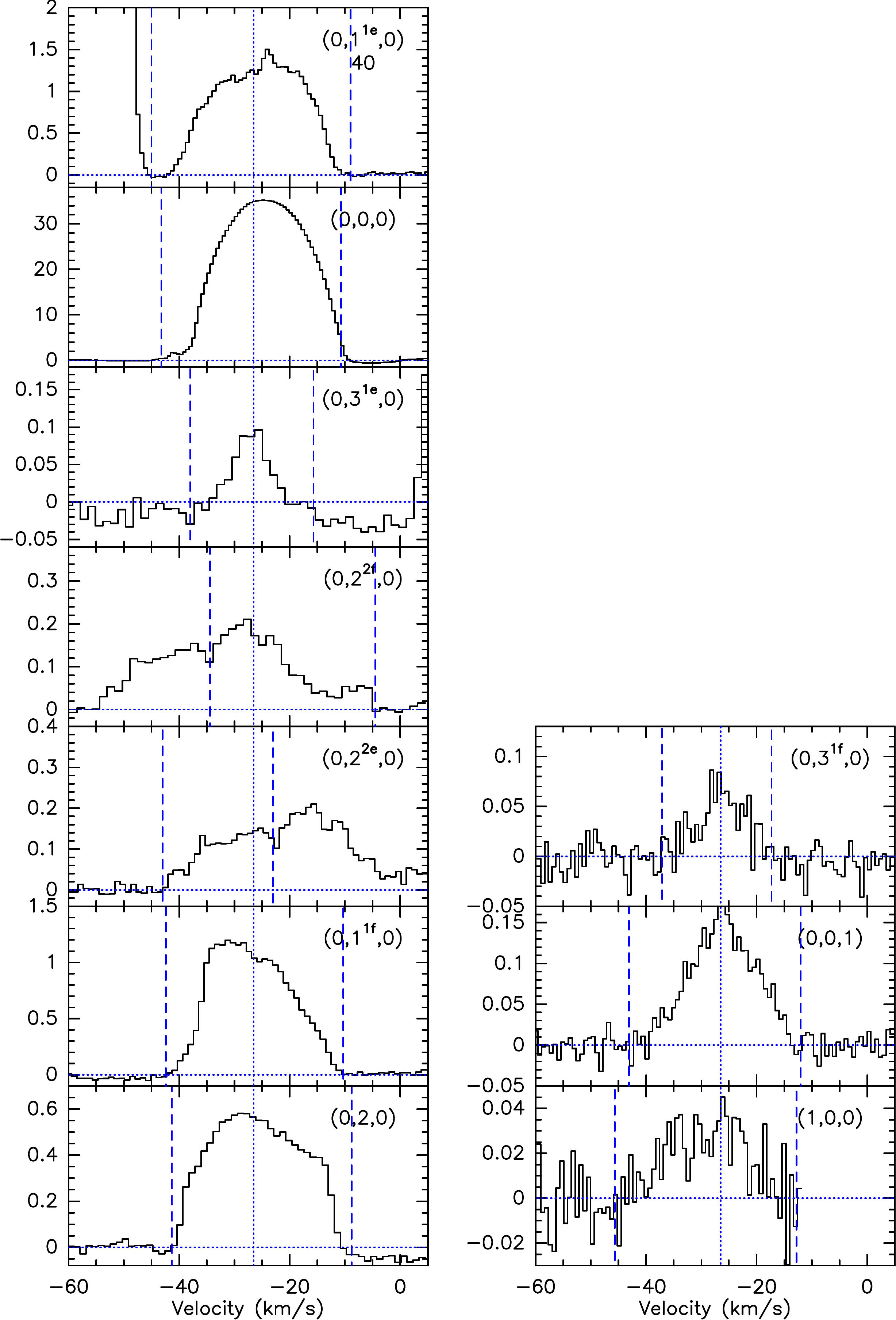}
\caption{\textit{From left to right:} Spectra for $J=2$--$1$ (first) and 3--2 (second and third) transitions of HCN toward IRC +10216. The vibrational state of each transition is indicated in the top-right corner of the respective spectrum. The vertical dashed blue lines denote the range of the emission in V$_\text{LSR}$, the vertical dotted line shows the systemic V$_\text{LSR}$ of the star, and the horizontal dotted line is the baseline. The number below the vibrational quantum numbers in the panel is a code assigned to the date of observation of the particular transition. The corresponding dates of the codes are listed in the Notes of Table \ref{tab:ground_hcn_results_shell}.}
\label{fig:irc_all_spec}
\end{figure*}

\begin{figure*}[t]%[!htbp]%[!htb]
\centering
\includegraphics[width=0.64\textwidth]{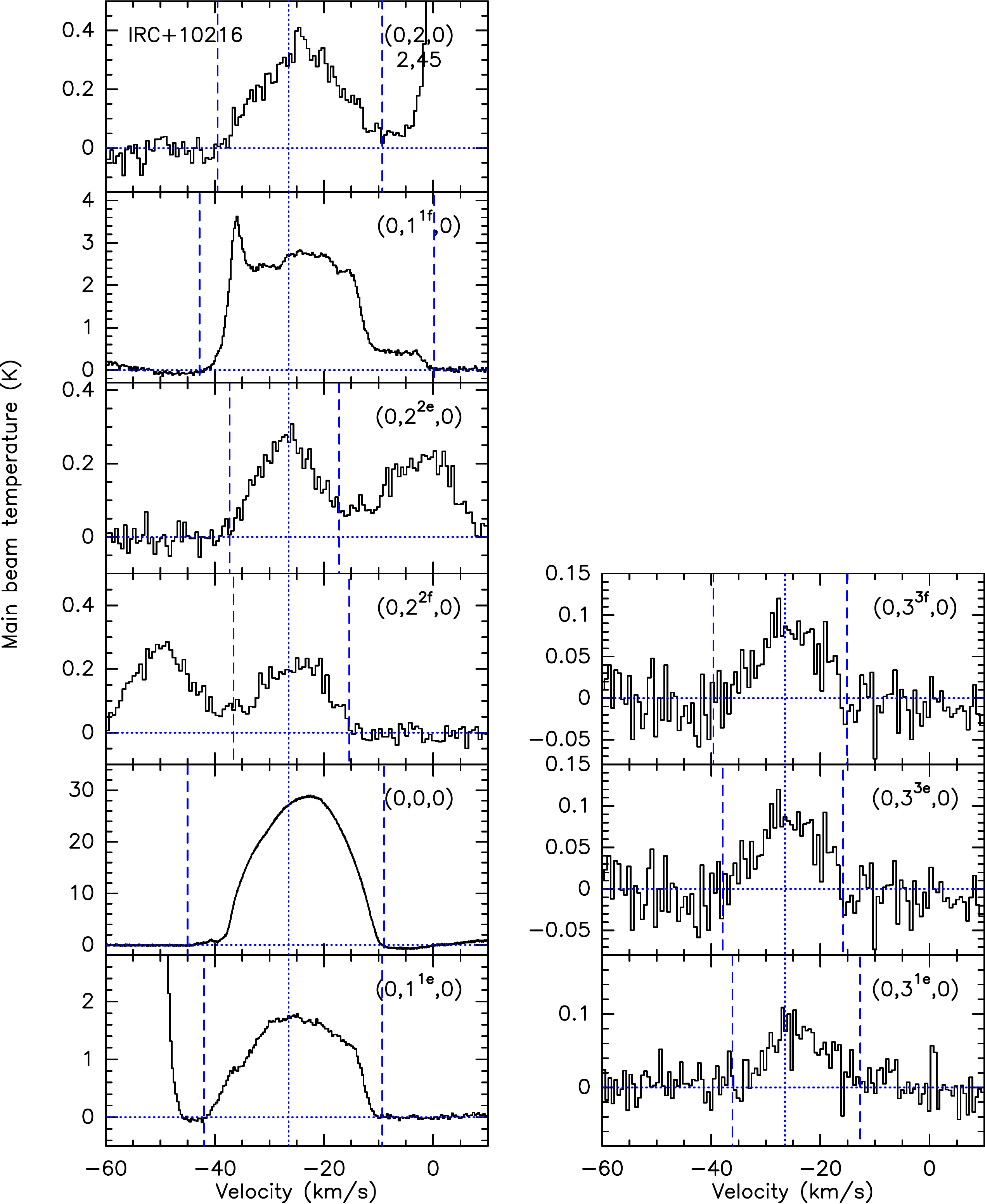}
\caption{Spectra for $J=4$--$3$ transitions of HCN towards IRC +10216. Description of the figure is the same as in the Fig. \ref{fig:irc_all_spec} caption.}
\label{fig:irc_second_spec}
\end{figure*}

\begin{figure*}[t]%[!htbp]
\centering
\includegraphics[width=0.32\textwidth]{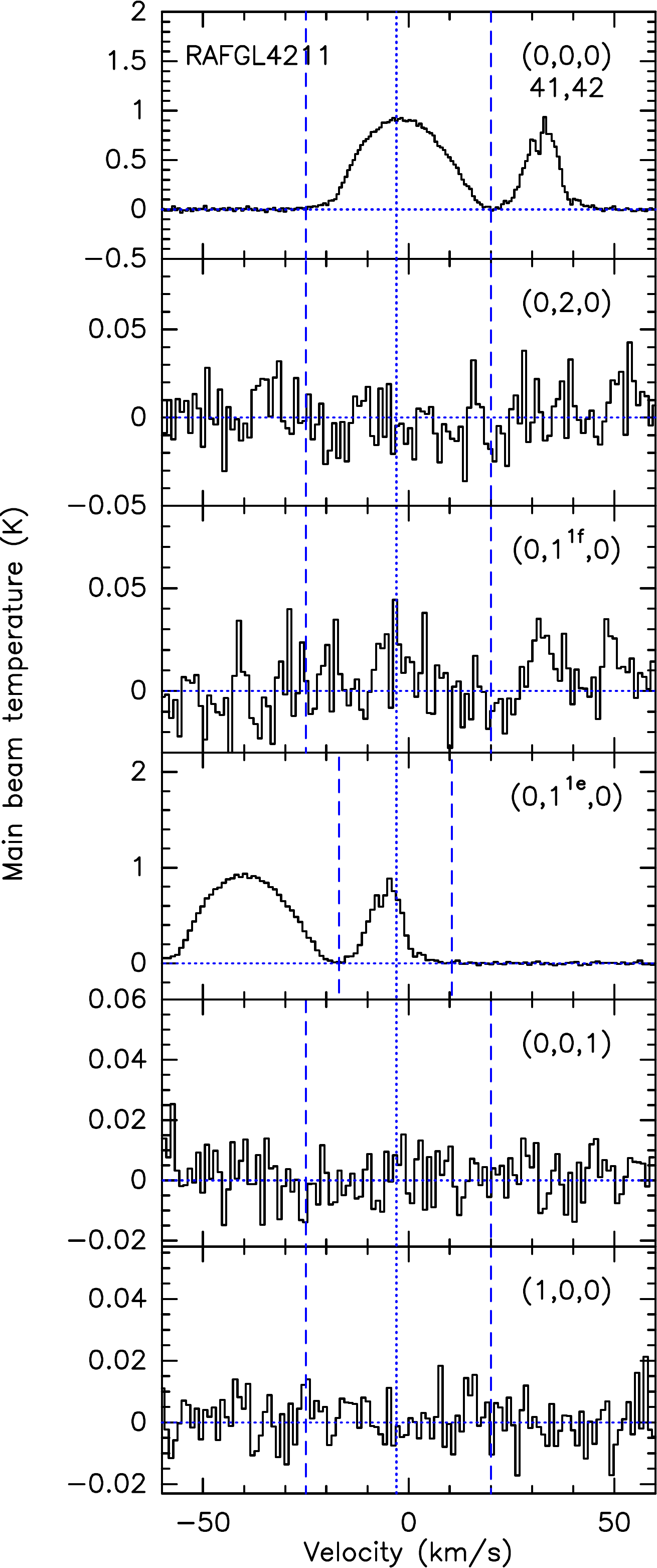}\quad
\includegraphics[width=0.32\textwidth]{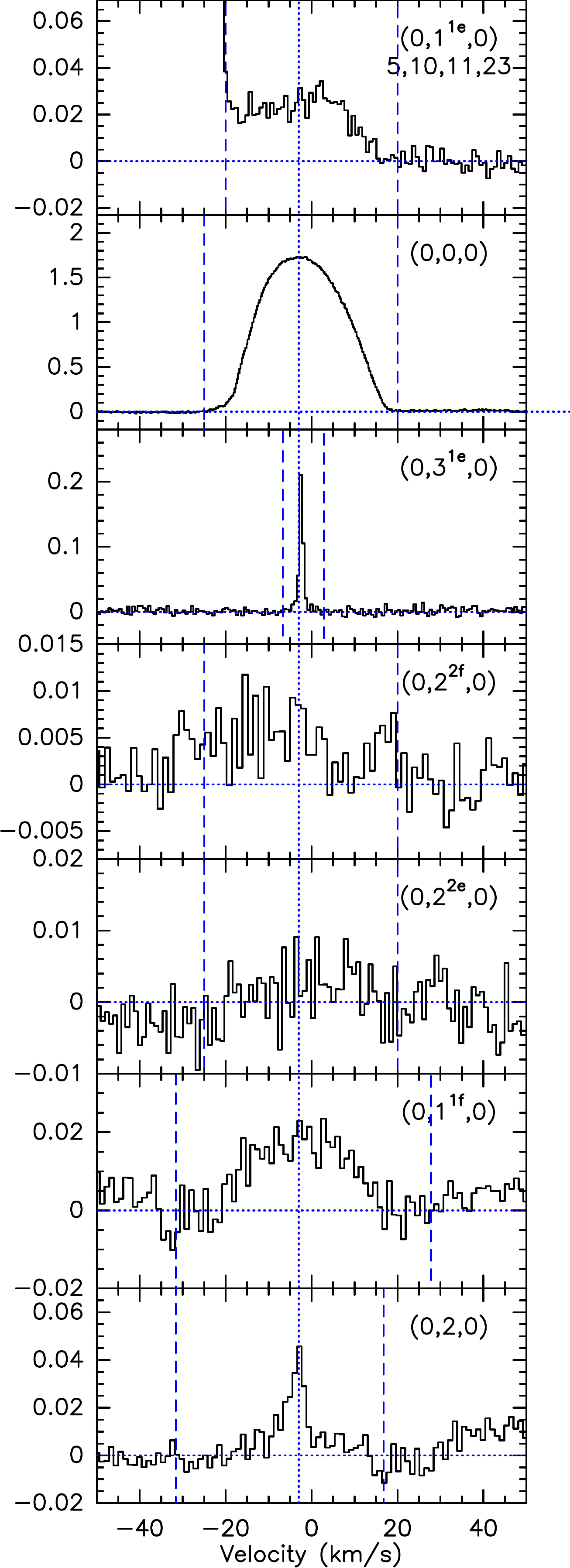}\quad
\includegraphics[width=0.29\textwidth]{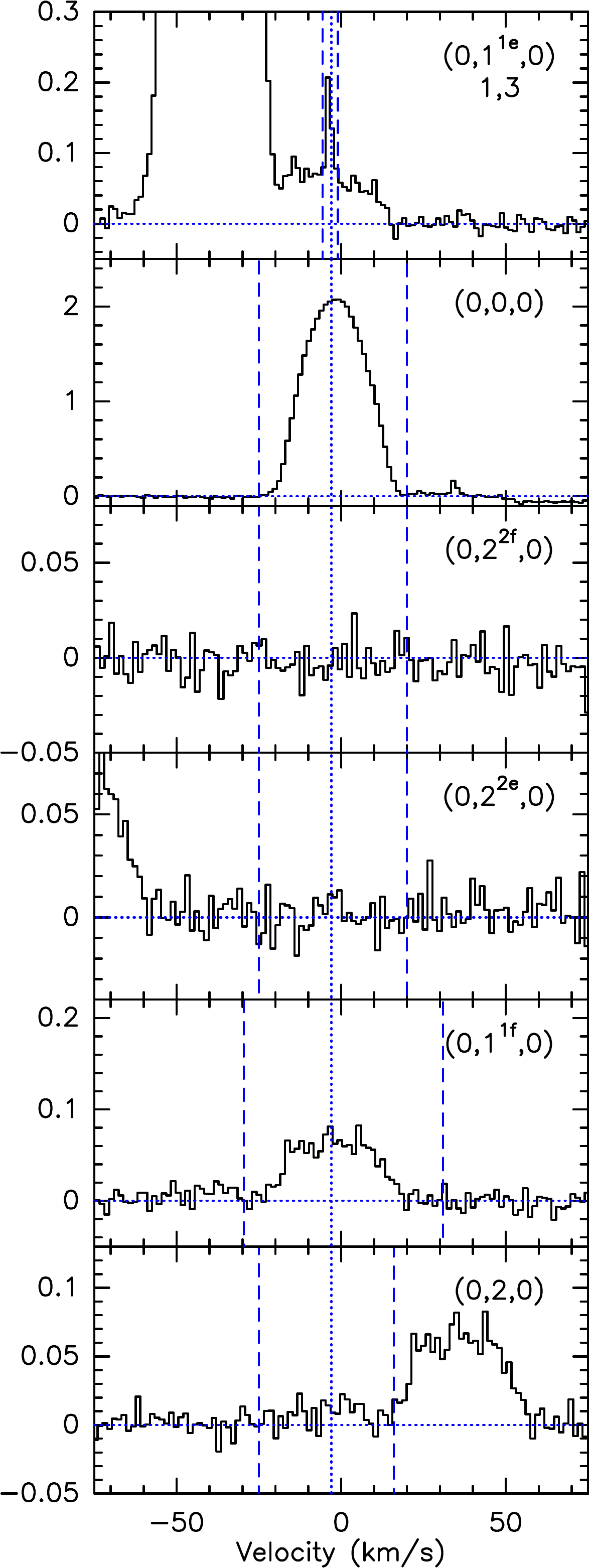}
\caption{Spectra for $J=2$--$1$ (left), 3--2 (center) and 4--3 (right) transitions of HCN towards RAFGL 4211. Description of the figure is the same as in the Fig. \ref{fig:irc_all_spec} caption.}
\label{fig:rafgl-all}
\end{figure*}

\begin{figure*}[t]%[!htbp]
\centering
\includegraphics[width=0.32\textwidth]{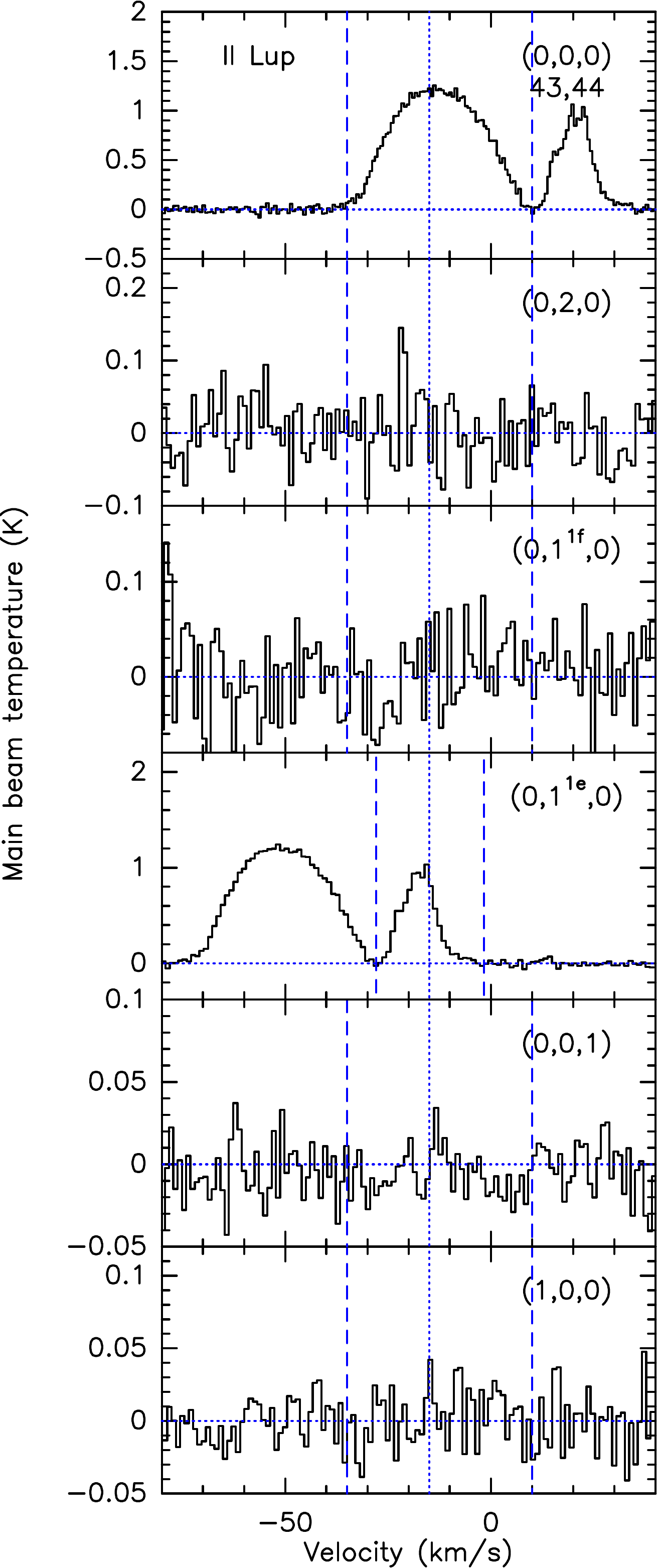}\quad
\includegraphics[width=0.30\textwidth]{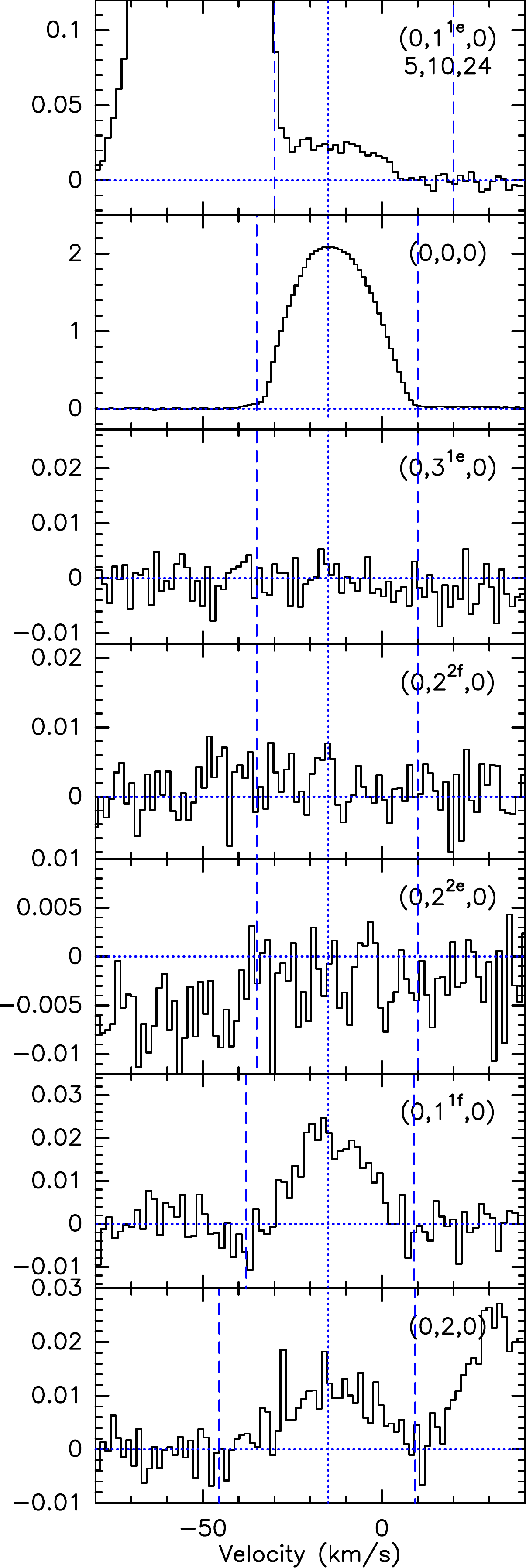}\quad
\includegraphics[width=0.29\textwidth]{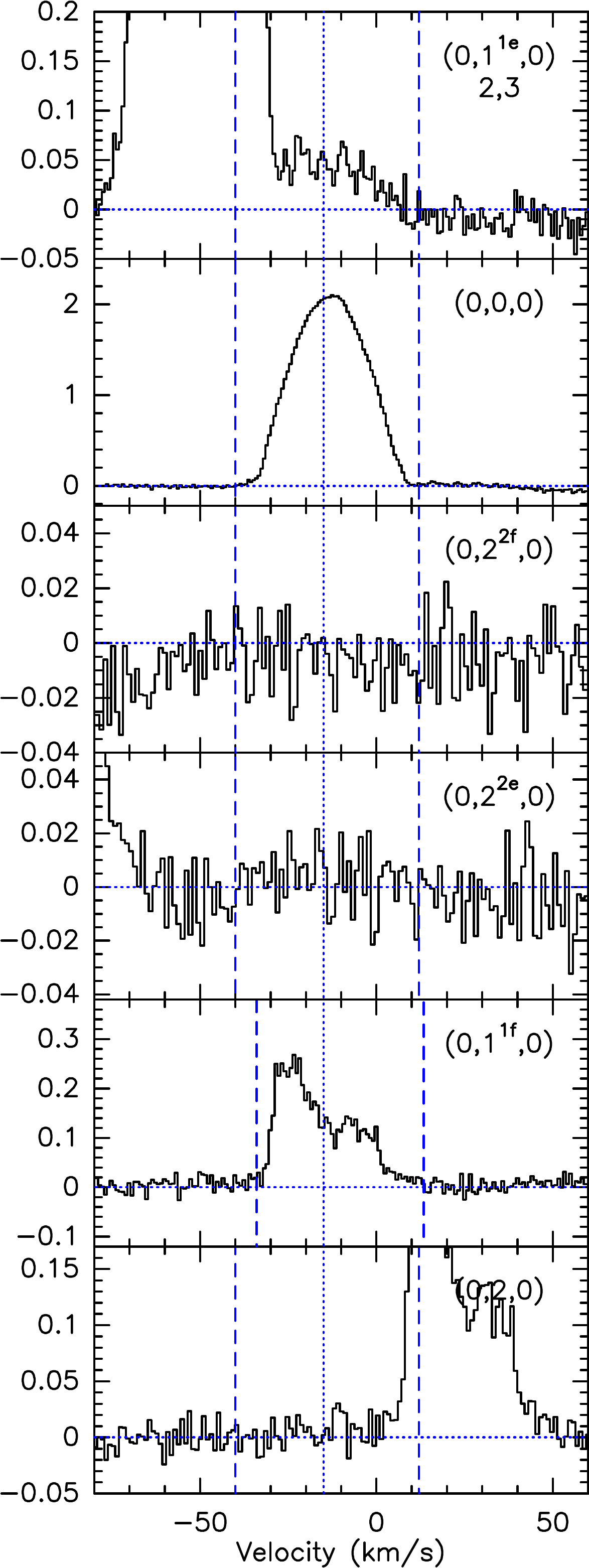}
\caption{Spectra for $J=2$--$1$ (left), 3--2 (center) and 4--3 (right) transitions of HCN towards II Lup. Description of the figure is the same as in the Fig. \ref{fig:irc_all_spec} caption.}\label{fig:ii-lup_all_spec}
\end{figure*}

Figures \ref{fig:irc_all_spec}--\ref{fig:irc_second_spec}, \ref{fig:rafgl-all}, and \ref{fig:ii-lup_all_spec} present the observed spectra towards the three exemplary sources, IRC +10216, RAFGL 4211, and II Lup, respectively, while the spectra of the other 13 sources are presented in Appendix \ref{app:spectra}. In our study, the detection criteria is set to be an S/N ratio of at least three. 

\subsection{Ground-state HCN transitions}
\label{sec:ground-results}

\begin{figure}[t]%[!htbp]%[!htb]
\centering %\resizebox{\hsize}{!}
\includegraphics[width = 0.45 \textwidth]{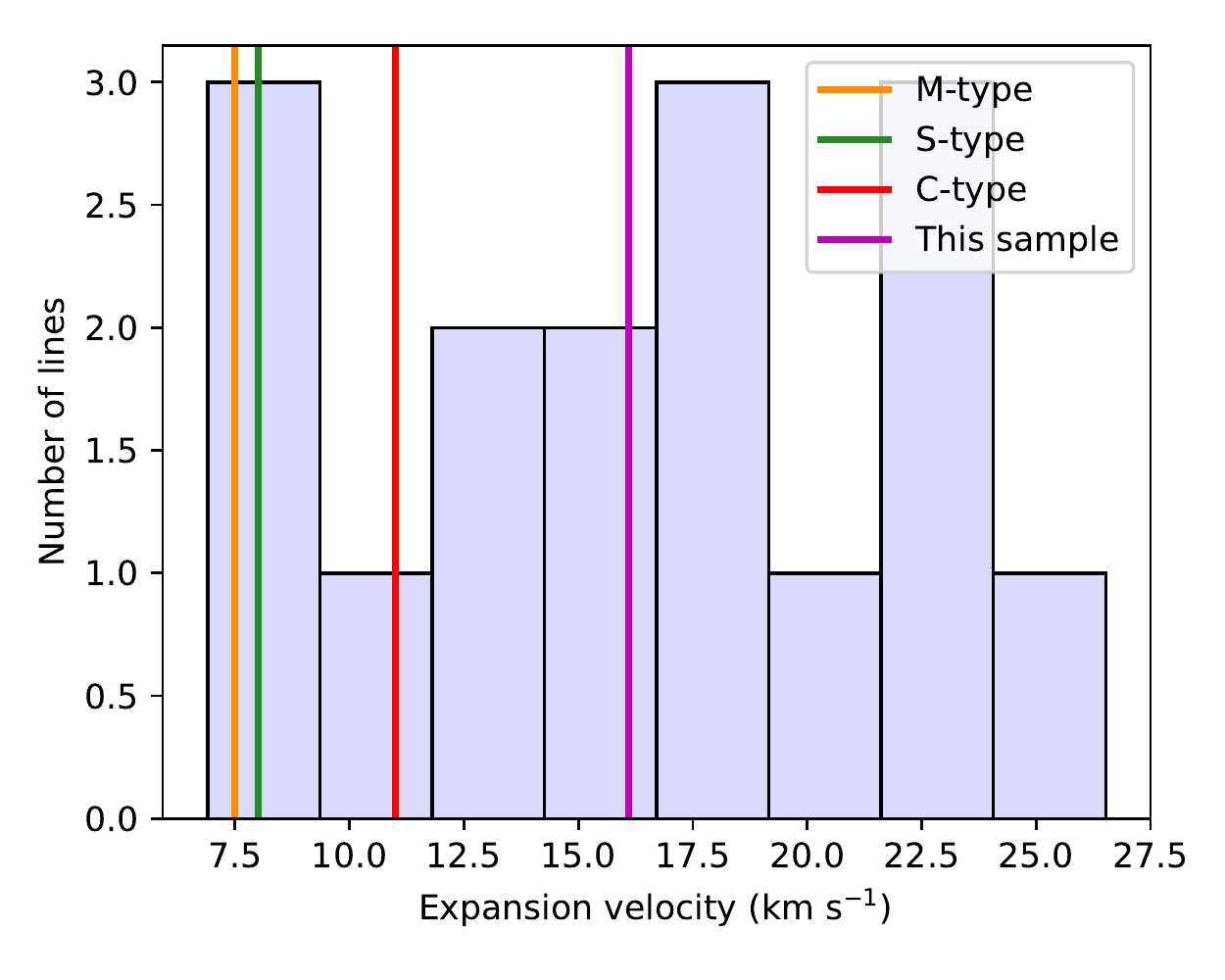}
\caption{Histogram distribution of expansion velocities derived from $J=3$--$2$ ground state HCN lines. Magenta line marks the median for HCN $J=3$--$2$ from this work. Median calculated by \citet{ramstedt2009} towards M-type stars are marked in orange, S-type in green and C-type in red.}
\label{fig:vexp}
\end{figure}

HCN $J=2$--$1$, 3$-$2, and 4$-$3 transitions in the ground state are detected toward all 16 sources. These spectral line profiles are parabolic, which is the characteristic of a saturated line from an expanding CSE  \citep[e.g.,][]{1975ApJ...197..603M,1982A&A...107..128O}. We fit the line profiles in the GILDAS/CLASS software using the \textsc{shell} method to obtain line parameters including velocity-integrated intensity ($\int T_{\rm mb}{\rm d}V$), systemic LSR velocity ($V_{\rm LSR}$), and expansion velocity ($V_{\rm exp}$). The results are given in Table~\ref{tab:ground_hcn_results_shell}. Compared with the integrated HCN $J=2$--$1$ (0, 0, 0) intensities of IRC +10216, II Lup, and RAFGL 4211 in \citet{2018AA...613A..49M}, our values are lower by a factor of $\sim$50\% on the $T_{\rm mb}$ scale. Regular measurements of calibration sources with APEX in the HCN $J=2$--$1$ line find a scatter of intensities of about 15\%. Given the large spatial extent of the HCN emission \citep{dayal+1995,2018AA...613A..49M}, variability of the vibrational ground state HCN lines is unlikely. We therefore attribute this difference to calibration uncertainties\footnote{During 2017--2018, there had been continuous upgrades of the APEX telescope including the replacements of panels, subreflector, and wobbler. The calibration, optics, and position of the SEPIA receiver has also changed.}. Comparing with the published results (see Table~\ref{tab:sample}), we find that our derived $V_{\rm LSR}$ are generally higher than the systemic LSR velocities in the literature by 1$-$3~\kms\,except for V636 Mon. As already explained by previous studies \citep{1982A&A...107..128O,2018AA...613A..49M}, this velocity difference is attributed to the presence of blueshifted self-absorption caused by low-excitation gas in the outer parts of CSEs. Our observations demonstrate that such HCN blueshifted self-absorption is ubiquitous in carbon-rich stars. Figure~\ref{fig:vexp} presents the distribution of expansion velocities derived from the HCN $J=3$--$2$ ground-state line toward our sample. The median value of the expansion velocity is 16.1~\kms, higher than previously reported medians in carbon-rich stars in \citet{ramstedt2009} because our sample includes objects with higher mass-loss rates than the Ramstedt et al. study. 

\begin{table*}[t]
\centering
\caption{Observed parameters derived from the \textsc{shell} method for the HCN rotational transitions in the ground vibrational state.}\label{tab:ground_hcn_results_shell}
\begin{tabular}{cccccc}
\hline
 Source & $\Delta J$   & $\int T_{\rm mb} {\rm d}V$   & $V_{\rm exp}$ & $V_{\rm LSR}$  & Date obs. \\
  & & (K km\,s$^{-1}$) & (km\,s$^{-1}$) & (km\,s$^{-1}$) & \\
\hline
IRC+10216 & 2--1  & 437.43 (5.21) & 14.8 (1.5) & $-$23.6 (1.0) &36, 42 \\ 
          & 3--2 & 706.15 (1.7) & 14.7 (0.3) & $-$24.3 (0.2)&40 \\
          & 4--3 & 553.62 (0.4) & 14.4 (0.08) & $-$24.06 (0.0)&2, 45  \\ 
RAFGL4211 & 2--1  & 21.89 (0.64)  &17.86 (1.6) &$-$1.36 (3.14)&41, 42 \\ 
          & 3--2 & 41.30 (0.05)&17.82 (0.19)  &$-$1.93 (0.07) &5, 10, 11, 23\\
          & 4--3 & 47.83  (0.22) &17.45 (1.19) &$-$1.74 (1.02)&1, 3  \\ 
II Lup & 2--1 &  33.22 (0.8)  & 20.36 (1) &$-$12.76 (2.7)& 43, 44 \\ 
       & 3--2 & 57.00 (0.08) &20.42 (0.2)& $-$13.45 (0.0)&5, 10, 24\\
       & 4--3 &  54.58 (0.14) &20.02 (0.7) &$-$13.21 (0.32)&2, 3   \\ 
R Vol & 2--1 & 5.41 (0.1) & 17.59 (1.2) & $-$9.06 (2.0)&36, 38, 42\\ 
      & 3--2 & 9.44 (0.06) & 17.5 (0.9) & $-$9.97 (0.66)&31 \\
      & 4--3 & 10.27 (0.04) & 16.81 (0.07) & $-$9.99 (0.06)&12, 14, 27, 30, 45\\ 
Y Pav & 2--1 &0.45 (0.04)  & 9.1 (1.7) & $-$2.84 (3.5) &30 \\ 
      & 3--2 & 0.39 (0.2)  & 7.1 (1.9) & $-$2.85 (4.6)&5, 7, 9, 24, 29\\
      & 4--3 & 0.39 (0.04)& 5.7 (1.6) &$-$2.49 (0.69) &26$-$28  \\ 
AQ Sgr & 3--2 &  0.88 (0.01)  & 8.3 (1.1) &23.1 (1.1)&5, 6, 17, 21, 24  \\
       & 4--3 & 2.9 (0.4)  & 6.7 (0.2) & 24.5 (0.38) &14, 16\\ 
CRL 3068 & 2--1 & 18.42 (0.39) & 13.35 (0.8) & $-$28.8 (1.6) &30, 36$-$38 \\ 
         & 3--2 & 27.98 (0.13) &12.87 (0.5) & $-$29.4 (0.03)&4, 8, 17, 24 \\
         & 4--3 & 33.48 (0.06) &12.57 (0.2) & $-$29.2 (0.05) &14$-$16, 34  \\  
IRC+30374 & 2--1 &  12.04 (0.21)  &24.2 (3.2)  &$-$10.5 (2.7) &40  \\ 
          & 3--2 &  24.32 (0.13) & 23.5 (1)& $-$10.9 (0.6)&5  \\
          & 4--3 &  9.99 (0.12)  & 21.7 (0.8) & $-$10.7 (2.5)&25  \\ 
V636 Mon & 2--1 &  4.88  (0.06)  & 22.7 (1.6) &998 (1.9) &33, 37  \\ 
         & 3--2 & 6.95  (0.03)  &22.8 (0.8) &10.06 (0.7)& 28\\
         & 4--3 & 7.92 (0.03)&21.5 (0.06)  &10.56 (0.04) &12, 14, 20, 30 \\
IRC+20370 & 2--1 & 8.94 (0.14) &13.7 (1.3)  &1.36 (2.0)&31 \\ 
          & 3--2 &  14.09 (0.07)  & 13.4 (0.5) & 0.78 (0.45)&5, 10, 23, 24  \\
          & 4--3 &   17.79 (0.16) & 13.2 (1.7) & 0.82 (0.73)&34\\ 
W Ori & 2--1 & 2.21 (0.14) & 11.42 (6.1) & 1.10 (4.4)&32, 37  \\
      & 3--2 & 4.67 (0.2) & 10.60 (1.0) & 0.3 (2.7) & 4, 8$-$10, 21, 31\\ 
      & 4--3 &  5.55 (0.02) &10.53 (0.27) &0.7 (0.17) &12, 14, 19, 27, 35  \\ 
S Aur & 2--1 &  3.85 (0.29) & 24.01(8.2) & $-$13.98 (3.9)&37  \\ 
      & 3--2 & 5.92 (0.13)  &21.96 (1.3)  & $-$14.76 (0.82)&24  \\
      & 4--3 & 6.56 (0.06) & 20.61 (0.3) &$-$14.64 (1.03)&13, 35\\ 
V Aql & 3--2 & 4.44  (0.04) & 6.9 (0.6) &  55.04 (0.5)&5, 8, 22  \\
      & 4--3 & 4.40 (0.03)  &7.54 (0.54)  & 55.06 (0.3)&13, 14 \\
CRL 2513 & 2--1 & 4.44 (0.06)  & 24.6 (2) &19.17 (0.2)& 31, 37  \\ 
         & 3--2 &  8.67 (0.11) & 26.52 (2.9) &18.77 (2.4)& 5, 8\\
CRL 2477 & 2--1 & 3.39 (0.13)  &12.5 (1.4)  & 5.42 (1.35)&41  \\ 
         & 3--2 & 6.88 (0.07)  &18.8 (2.5)  &6.92 (1.1)& 17\\
RV Aqr & 2--1 & 6.61 (0.1)  & 14.8 (2.3) &2.94 (1.5)&31, 32  \\ 
       & 3--2 &  8.6 (0.18)  & 14.6 (0.4) &2.13 (1.8)&8, 9, 18 \\
       & 4--3 &  10.8 (0.06) &13.58 (0.02)  &2.3 (0.03) &15, 25, 27, 28  \\ 
\hline
\end{tabular}
\tablefoot{The last column refers to the identifier assigned to the date of the observation in 2018 of the corresponding transition toward the source. The dates in 2018 are in the format ``DD Month'': 1 = 8 Jun, 2 = 2 Jul, 3 = 3 Jul, 4 = 4 Jul, 5 = 8 Jul, 6 = 11 Jul, 7 = 12 Jul, 8 = 13 Jul, 9 = 14 Jul, 10=15 Jul, 11 = 16 Jul, 12 = 3 Sep, 13 = 4 Sep, 14 = 5 Sep, 15 = 6 Sep, 16 = 7 Sep, 17 = 8 Sep, 18 = 9 Sep, 19 = 11 Sep, 20 = 12 Sep, 21 = 13 Sep, 22 = 15 Sep, 23 = 5 Oct, 24=6 Oct, 25 = 23 Nov, 26 = 24 Nov, 27 = 25 Nov, 28 = 26 Nov, 29 = 27 Nov, 30 = 28 Nov, 31 = 29 Nov, 32 = 30 Nov, 33 = 2 Dec, 34 = 6 Dec, 35 = 7 Dec, 36 = 9 Dec, 37 = 10 Dec, 38 = 11 Dec, 39 = 13 Dec, 40 = 14 Dec, 41 = 15 Dec, 42 = 16 Dec, 43 = 17 Dec, 44 = 18 Dec, 45 = 20 Dec.}
\normalsize
\end{table*}

\subsection{Vibrationally excited HCN transitions}
\label{sec:vibstate-results}

HCN rotational transitions in vibrational states with upper energy levels of up to 4789 K (see Table~\ref{tab:lines}) were targeted in our study. We detected HCN rotational lines in the following 11 excited vibrational states: (1, 0, 0), (0, 0, 1), (0, 1$^{1e}$, 0), (0, 1$^{1f}$, 0), (0, 2$^{2e}$, 0), (0, 2$^{2f}$, 0), (0, 2$^{0}$, 0), (0, 3$^{1e}$, 0), (0, 3$^{1f}$, 0), (0, 3$^{3e}$, 0), and (0, 3$^{3f}$, 0). Table~\ref{tab:summ_det} summarizes the detection statuses in each source. Vibrationally excited HCN transitions are found to show more complex line profiles, because both thermal and maser emission may contribute to the observed line profiles. The properties of the thermal and maser emission are discussed in Sect.~\ref{sec:thermal} and Sect.~\ref{sec:maser}, respectively.
%
%\begin{table}[!htbp]
%\centering
%\caption{\textbf {Line detections } }
%\caption{Summary of the lines detected in number of sources from our sample.}\label{tab:det_rate}
%\begin{tabular}{cccc}
%\hline
%\hline
% ($\nu_{1}$, $\nu_{2}$, $\nu_{3}$) & 2--1 & 3--2 & 4--3 \\  
% \hline
%(0, 0, 0) & 16 & 16 & 16 \\
%(1, 0, 0) & 0 & 1 & 0  \\
%(0, 0, 1) &1& 2 & 0\\
%(0, 1, 0)$^{\star}$ &0 &0 &0 \\
%(0, 1$^{1e}$, 0) &11 &9 &5 \\
%(0, 1$^{1f}$, 0) &4& 5 & 5\\
%(0, 2, 0) & 1&4 & 1\\
%(0, 2$^{2e}$, 0) &0&  1&2\\
%(0, 2$^{2f}$, 0) &0& 1&2\\
%(0, 3$^{1e}$, 0) & 1 & 2& 1\\
%(0, 3$^{1f}$, 0) &0 &1 & 0\\
%(0, 3$^{3e}$, 0) &0 &0 & 1\\
%(0, 3$^{3f}$, 0) & 0&0 & 1\\
%\hline
%\end{tabular}
%\tablefoot{Summary of the lines detected in number of sources from our sample.}
%\end{table}

% vvv - if landscape page needed - vvv %
%\onecolumn
%\begin{landscape}
%\noindent
%\begin{table}[!htb]
%\begin{minipage}{\linewidth}
% ^^^ - if landscape page needed - ^^^ %
%\newpage
%\onecolumn
\begin{table*}%[htb]
\caption{Summary of our HCN line observations and the nature of detection}
  \begin{subtable}{\textwidth}%[h]
%\scriptsize
    \centering
    %\caption{\textbf{Observed parameters}}\label{resulttable}
    \caption{$J$ = 2$-$1}
    \begin{tabular}{c|ccccccc}
    \hline
    \hline
     ($\nu_1 \nu_2 \nu_3$) & $000$ & $01^{1e}0$ & $01^{1f}0$ & $020$ & $001$ & $03^{1e}0$ & $100$ \\
     \hline
      $E_\text{u}$ (K) & 13 & 1037 & 1037 & 2044 & 3030 & 3054 & 4777 \\
      \hline
    IRC +10216 & \Th & \M & \M & \Th & \Th & \Th & \ND \\
    RAFGL 4211 & \Th & \M & \ND & \ND & \ND & \ND & \ND \\
    II Lup & \Th & \M & \ND & \ND & \ND & \ND & \ND \\
    R Vol & \Th & \M & \M & \ND & \ND & \ND & \ND  \\
    Y Pav & \Th & \ND & \ND & \ND & \ND & \ND & \ND \\
    AQ Sgr & \NO & \NO & \NO & \NO & \NO & \NO & \NO \\
    CRL 3068 & \Th & \M & \Th & \ND & \ND & \ND & \ND \\
    IRC +30374 & \Th & \Bl & \ND & \ND & \ND & \ND & \ND \\
    V636 Mon & \Th & \M & \ND & \ND & \ND & \ND & \ND \\
    IRC +20370 & \Th & \M & \ND & \ND & \ND & \ND & \ND \\
    W Ori & \Th & \M & \ND & \ND & \ND & \ND & \ND \\
    S Aur & \Th & \M & \ND & \ND & \ND & \ND & \ND \\
    V Aql & \NO & \NO & \NO & \NO & \NO & \NO & \NO \\
    CRL 2513 & \Th & \ND & \ND & \ND & \ND & \ND & \ND \\
    CRL 2477 & \Th & \ND & \ND & \ND & \ND & \ND & \ND \\
    RV Aqr & \Th & \M & \M & \ND & \ND & \ND & \ND \\
    \hline
    {\#}sources det. & 14 & 11 & 4 & 1 & 1 & 1 & 0 \\
      \hline
    \end{tabular}
    
    %\label{tab:week1}
\end{subtable}
\\
\par\bigskip
%\hfill
\begin{subtable}{\textwidth}%[h]
    \centering
    \caption{$J$ = 3$-$2}
    %\caption{\textbf{Observed parameters}}\label{resulttable}
    \begin{tabular}{c|cccccccccc}
    \hline
    \hline
    ($\nu_1 \nu_2 \nu_3$) & $000$ & $01^{1e}0$ & $01^{1f}0$ & $020$ & $02^{2e}0$ & $02^{2f}0$ & $001$ & $03^{1e}0$ & $03^{1f}0$ & $100$ \\
    \hline
    $E_\text{u}$ (K) & 26 & 1050 & 1050 & 2056 & 2078 & 2078 & 3042 & 3066 & 3067 & 4790 \\
    \hline
     IRC +10216 & \Th & \BO & \BO & \Bl & \Bl & \Bl & \Th & \Th & \Th & \Bl \\
     RAFGL 4211 & \Th & \Bl & \Th & \M & \ND & \ND & \ND & \M & \ND & \ND \\
     II Lup & \Th & \Bl & \Th & \Bl & \ND & \ND & \ND & \ND & \ND & \ND \\
     R Vol & \Th & \M & \ND & \ND & \ND & \ND & \ND & \ND & \ND & \ND \\
     Y Pav & \Th & \ND & \ND & \ND & \ND & \ND & \ND & \ND & \ND & \ND \\
     AQ Sgr &\Th & \M & \ND & \ND & \ND & \ND & \ND & \ND & \ND & \ND \\
     CRL 3068 & \Th & \Th & \Th & \Th & \ND & \ND & \Th & \ND & \ND & \ND \\
     IRC +30374 &\Th & \ND & \ND & \ND & \ND & \ND & \ND & \ND & \ND & \ND \\
     V636 Mon & \Th & \ND & \ND & \ND & \ND & \ND & \ND & \ND & \ND & \ND \\
     IRC +20370 & \Th & \M & \ND & \ND & \ND & \ND & \ND & \ND & \ND & \ND \\
     W Ori & \Th & \M & \Th & \ND & \ND & \ND & \ND & \ND & \ND & \ND \\
     S Aur & \Th & \Th & \ND & \ND & \ND & \ND & \Th & \ND & \ND & \ND \\
     V Aql & \Th & \M & \ND & \ND & \ND & \ND & \ND & \ND & \ND & \ND \\
     CRL 2513 & \Th & \ND & \ND & \ND & \ND & \ND & \ND & \ND & \ND & \ND \\
     CRL 2477 & \Th & \ND & \ND & \ND & \ND & \ND & \ND & \ND & \ND & \ND \\
     RV Aqr & \Th & \M & \ND & \ND & \ND & \ND & \ND & \ND & \ND & \ND \\
     \hline
     {\#}sources det. & 16 & 11 & 5 & 4 & 1 & 1 & 3 & 2 & 1 & 1  \\
    \hline
    \end{tabular}
    
   % \label{tab:week2}
 \end{subtable}
 \\
\par\bigskip
\begin{subtable}{\textwidth}%[h]
    \centering
    \caption{$J$ = 4$-$3}
    %\caption{\textbf{Observed parameters}}\label{resulttable}
    \begin{tabular}{c|ccccccccc}
    \hline
    \hline
    ($\nu_1 \nu_2 \nu_3$)& $000$ & $01^{1e}0$ & $01^{1f}0$ & $020$ & $02^{2e}0$ & $02^{2f}0$ & $03^{1e}0$ & $03^{3e}0$ & $03^{3f}0$ \\ 
    \hline
    $E_\text{u}$ (K) & 43 & 1067 & 1067 & 2074 & 2095 & 2095 & 3083 & 3127 & 3127 \\
    \hline
     IRC +10216 & \Th & \Th & \BO & \Th & \Bl & \Bl & \Th & \Bl & \Bl \\
     RAFGL 4211 & \Th & \BO & \Bl & \ND & \ND & \ND & \ND & \ND & \ND \\
     II Lup & \Th & \Bl & \BO & \ND & \ND & \ND & \ND & \ND & \ND \\
     R Vol & \Th & \M & \ND & \ND & \ND & \ND & \ND & \ND & \ND \\
     Y Pav & \Th & \ND & \ND & \ND & \ND & \ND & \ND & \ND & \ND \\
     AQ Sgr & \Th & \ND & \ND & \ND & \ND & \ND & \ND & \ND & \ND \\
     CRL 3068 & \Th & \Th & \BO & \Th & \Bl & \Bl & \ND & \ND & \ND \\
     IRC +30374 & \Th & \ND & \ND & \ND & \ND & \ND & \ND & \ND & \ND \\
     V636 Mon & \Th & \ND & \ND & \ND & \ND & \ND & \ND & \ND & \ND \\
     IRC +20370 & \Th & \ND & \ND & \ND & \ND & \ND & \ND & \ND & \ND \\
     W Ori & \Th & \M & \Th & \ND & \ND & \ND & \ND & \ND & \ND \\
     S Aur & \Th & \ND & \ND & \ND & \ND & \ND & \ND & \ND & \ND \\
     V Aql & \Th & \ND & \ND & \ND & \ND & \ND & \ND & \ND & \ND \\
     CRL 2513 & \NO & \NO & \NO & \NO & \NO & \NO & \NO & \NO & \NO \\
     CRL 2477 & \NO & \NO & \NO & \NO & \NO & \NO & \NO & \NO & \NO \\
     RV Aqr & \Th & \Th & \ND & \ND & \ND & \ND & \ND & \ND & \ND \\
     \hline
     {\#}sources det. & 14 & 7  & 5 & 2 & 2 & 2 & 1 & 1 & 1 \\
    \hline
    \end{tabular}
    %\label{tab:week2}
   
 \end{subtable}
  \tablefoot{{``\M''}: maser emission, {``\Th''}: thermal emission, {``\Bl''}: blended line profile, {``\ND''}: non-detection, {``\NO''}: transition not observed.}
 \label{tab:summ_det}
 \end{table*}

\normalsize % test

\begin{table*}[!htb]
\centering
\caption{Observed properties of detected HCN transitions in vibrationally excited states.}\label{tab:vib_hcn_results_updated}
\scriptsize
\begin{tabular}{cccccccc}
\hline
\hline
 Source name & $J$ & $(\nu_{1}, \nu_{2}, \nu_{3})$ & $\int T_{\rm mb} {\rm d}V$ & $\Delta V$ & $V_\text{LSR}$ & $V_\text{LSR}$ range & Comments \\  
 &  &  & (K km\,s$^{-1}$) & (km\,s$^{-1}$) & (km\,s$^{-1}$) & (km\,s$^{-1}$) & \\
\hline
IRC+10216 & 2--1 & (0, 0, 1) & 0.85 (0.03)  & 18.4 (2.5) & $-$28.7 (0.8) & [$-$37.3, $-$14.2]& \\
          &      & (0, 1$^{1e}$, 0) & 66.54 (0.06) & 8.0 (0.0) & $-$27.6 (0.0) & [$-$37.7, $-$12.8] & maser \\
          &      & (0, 3$^{1e}$, 0) & 0.39 (0.05) & 10.8 (1.6) & $-$26.3 (0.7) & [$-$36.8,$-$16.4] & \\
          &      & (0, 1$^{1f}$, 0) & 7.36 (0.06) & 15.0 (0.1) & $-$26.2 (0.0) & [$-$44.0, $-$9.3] & maser\\
          &      & (0, 2, 0) & 1.42 (0.05) & 12.6 (0.5) & $-$26.3 (0.2) & [$-$40.7, $-$13.8] & \\
%          & 33 & (0, 1$^{1f}$, 0)--(0, 1$^{1e}$, 0) & 2.09 (0.05) & --&--&[$-$45.0, $-$9.0]& \\
          & 3--2 & (1, 0, 0) & 0.51 (0.06) & -- &-- & [$-$45.7, $-$12.8] & blended\tablefootmark{a} \\
          &      & (0, 0, 1) & 2.6 (0.03) & 15.7 (0.5) & $-$26.4 (0.2) & [$-$40.7, $-$9.3] & \\
          &      & (0, 1$^{1e}$, 0) & 23.84 (0.3) & 17.8 (0.6) & $-$25.04 (0.3) & [$-$42.7, $-$7.3] & maser + c\\
          &      & (0, 3$^{1e}$, 0) & 1.16 (0.05)& 9.4 (0.6) & $-$26.9 (0.2) & [$-$38.0, $-$15.7] & \\
          &      & (0, 2$^{2f}$, 0) & 2.95 (0.05)  & 17.91 (1.2) & $-$28.3 (0.4) & [$-$34.4, $-$4.5] & blended\tablefootmark{b} \\
          &      & (0, 2$^{2e}$, 0) & 2.05 (0.05) & 18.43 (0.8) & $-$27.7 (0.4) & [$-$43.0, $-$23.0] & blended\tablefootmark{b} \\
          &      & (0, 1$^{1f}$, 0) & 20.24 (0.1) & 16.52 (0.2) & $-$27.53 (0.08) &[$-$41.5,$-$8.5] & maser + c \\
          &      & (0, 2, 0) & 13.5 (0.07) & -- & -- & [$-$41.3, $-$8.8] & blended\tablefootmark{c} \\
          &      & (0, 3$^{1f}$, 0) & 0.65 (0.06) & 9.3 (1.0) & $-$26.2 (0.5) & [$-$37.1, $-$17.3] & \\
          & 4--3 & (0, 1$^{1e}$, 0) & 36.45 (0.08) & 19.0 (0.05) & $-$24.8 (0.02) & [$-$42.0, $-$9.3] & \\
          &      & (0, 3$^{1e}$, 0) & 1.06 (0.07) & 13.6 (1.4) & $-$24.12 (0.5) & [$-$36.1, $-$12.7] & \\
          &      & (0, 2$^{2f}$, 0) & 3.23 (0.09) & 14.35 (0.6) & $-$25.6 (0.2) & [$-$36.6, $-$15.4] & blended\tablefootmark{d} \\
          &      & (0, 2$^{2e}$, 0) &3.5 (0.08) & 13.5 (0.5)& $-$26.5 (0.2) & [$-$37.3, $-$17.2] & blended\tablefootmark{d} \\
          &      & (0, 1$^{1f}$, 0) & 70.41 (0.12) & 22.5 (0.04) & $-$25.4 (0.0) &  [$-$42.8, 0.2] & maser + c\\
          &      & (0, 2, 0) & 5.8 (0.09)  & 16.5 (0.4) & $-$23.9 (0.1) & [$-$39.5, $-$9.3] & \\
          &      & (0, 3$^{3e}$, 0) & 1.10 (0.09) & 11.6 (1.1) & $-$25.8 (0.5) & [$-$37.9, $-$15.8] & blended\tablefootmark{d} \\
          &      & (0, 3$^{3f}$, 0) & 1.09 (0.1) & 11.6 (1.1) & $-$25.8 (0.5) & [$-$39.6, $-$15.1] & blended\tablefootmark{d} \\
\hline
 RAFGL4211 & 2--1 & (0, 1$^{1e}$, 0) & 7.66 (0.09) &8.97 (0.07) & $-$5.42 (0.03) & [$-$16.9, 10.5] & maser \\
           & 3--2 & (0, 1$^{1e}$, 0) & 0.62 (0.02) & -- & -- & [$-$20.0, 20.0] & blended\tablefootmark{e} \\
           &      & (0, 3$^{1e}$, 0) & 0.28 (0.01)  &0.89 (0.02)& $-$2.37 (0.01)& [$-$6.7, 2.9] & maser \\
           &      & (0, 1$^{1f}$, 0) & 0.51 (0.03) & 24.30 (1.4) & $-$2 (0.7) & [$-$21.6, 27.8] & \\
           &      & (0, 2, 0) & 0.5 (0.02) & 14.9 (1.7) & $-$3.4 (0.4) & [$-$31.6, 16.8] & maser \\
           & 4--3 & (0, 1$^{1e}$, 0) & 0.62 (0.02) & 24.41 (0.9) & $-$5.83 (0.4) & [$-$5.7, $-$1.0] & maser + c \\
           &      & (0, 1$^{1f}$, 0) & 1.82 (0.08) & 26.7 (1.3) & $-$2.5 (0.6)& [$-$29.6, 30.9] & blended\tablefootmark{f} \\
 \hline
II Lup & 2--1 & (0, 1$^{1e}$, 0) &9.71 (0.3)  & 9.3 (0.13) & $-$17.51 (0.06) & [$-$27.9, $-$1.7] & maser \\
       & 3--2 & (0, 1$^{1e}$, 0) & 0.82 (0.03) & -- & -- &[$-$30.0, 20.0]  & blended\tablefootmark{e} \\
       &      & (0, 1$^{1f}$, 0) & 0.43 (0.03) & 20.06 (1.2) & $-$14.0 (0.6)& [$-$37.9, 9.0] & \\
       &      & (0, 2, 0) & 0.35 (0.03)  & -- & -- & [$-$45.4, 9.3] & blended\tablefootmark{c} \\
       & 4--3 & (0, 1$^{1e}$, 0) & 1.48 (0.07) & -- & -- & [$-$30.0, 12.0]& blended\tablefootmark{e} \\
       &      & (0, 1$^{1f}$, 0) & 5.27 (0.08) & 25.6 (0.59) & $-$19.03 (0.25) & [$-$34.0, 13.4] & maser + c \\
\hline
R Vol & 2--1 & (0, 1$^{1e}$, 0) & 0.86 (0.03) & 3.85 (0.11)& $-$12.99 (0.05) & [$-$22.7, $-$1.4] & maser \\
      &      & (0, 1$^{1f}$, 0) & 0.19 (0.06)  & 5.06 (1.6)& $-$12.87 (0.49) & [$-$32.7, 0.0] & maser \\
      & 3--2 & (0, 1$^{1e}$, 0)  & 0.25 (0.03) &4.8 (0.5)& $-$14.3 (0.23) & [$-$21.1, $-$8.7] & maser \\
      & 4--3 & (0, 1$^{1e}$, 0) & 0.25 (0.03)  & 4.2 (0.9) & $-$11.9 (0.3) & [$-$18.2, $-$6.2] & maser \\
\hline
AQ Sgr & 3--2 & (0, 1$^{1e}$, 0) & 0.09 (0.01)  & 7.7 (1.3) & 21.5 (0.5) & [13.4, 30.1] & maser \\
\hline
CRL 3068 & 2--1 & (0, 1$^{1e}$, 0) & 5.84 (0.02) & 10.1 (0.04) & $-$34.1 (0.02) & [$-$46.6, $-$16.2] & maser \\
         &      & (0, 1$^{1f}$, 0) & 0.45 (0.03) & 12.1 (0.9) & $-$29.4 (0.4) & [$-$40.3, $-$19.8] & \\
         & 3--2 & (0, 0, 1)        & 0.19 (0.01) & 16.1 (1.7)   & $-$30.18 (0.8) & [$-$48.2, $-$14.8] & \\
         &      & (0, 1$^{1e}$, 0) & 1.31 (0.02) & 15.5 (0.7) & $-$29.6 (0.4) & [$-$47.0, $-$16.0] & \\
         &      & (0, 1$^{1f}$, 0) & 1.25 (0.01) & 14.36 (0.4) & $-$31.4 (0.03) & [$-$45.9, $-$17.1] & \\
         &      & (0, 2, 0) & 0.59 (0.02) & 17.3 (0.8) & $-$30.7 (0.3) & [$-$45.5, $-$15.8] & \\
         & 4--3 & (0, 1$^{1e}$, 0) & 3.0 (0.04)  & 15.9 (0.3) & $-$30.6 (0.1) & [$-$46.3, $-$15.5] & \\
         &      & (0, 2$^{2f}$, 0) & 0.44 (0.04) & 13.8 (1.6) & $-$30.2 (0.7) & [$-$40.1, $-$20.6]& blended\tablefootmark{d} \\
         &      & (0, 2$^{2e}$, 0) & 0.56 (0.04)  & 14.6 (1.3) & $-$31.0 (0.6) & [$-$40.1, $-$21.1] & blended\tablefootmark{d} \\
         &      & (0, 1$^{1f}$, 0) & 10.23 (0.03) & 19.8 (0.1) & $-$34.8 (0.04) &[$-$45.9, $-$16.9] & maser + c\\
         &      & (0, 2, 0) & 0.63 (0.05) & 15.3 (1.2) & $-$31.1 (0.5) & [$-$50.0, $-$10.0] & \\
\hline
IRC+30374 & 2--1 & (0, 1$^{1e}$, 0) & 1.01 (0.1)  & 14.4 (2.4) & $-$15.8 (1.1) & [$-$24.4, $-$7.5] & blended\tablefootmark{e} \\
\hline
V636 Mon & 2--1 & (0, 1$^{1e}$, 0) & 0.2 (0.02)  & 5.1 (1.5) & 6.2 (0.4) & [2.7, 12.7] & maser \\
\hline
IRC+20370 & 2--1 & (0, 1$^{1e}$, 0) & 1.14 (0.04) & 5.7 (0.2) & $-$1.9 (0.08) & [$-$7.1, 5.08] & maser \\
          & 3--2 & (0, 1$^{1e}$, 0) & 0.86 (0.03)  & 8.3 (0.4) & $-$2.5 (0.15) & [$-$9.9, 6.1] & maser \\
\hline
W Ori & 2--1 & (0, 1$^{1e}$, 0)  & 1.84 (0.04)  &4.25 (0.08)  & $-$0.08 (0.03) &[$-$10.3, 8.3] & maser \\
      & 3--2 & (0, 1$^{1e}$, 0) & 2.03 (0.01) & 6.04 (0.04)& $-$1.15 (0.01) & [$-$12.0, 9.6] & maser \\
      &      & (0, 1$^{1f}$, 0) & 0.13 (0.01) & 5.71 (0.7) & $-$1.51 (0.23) & [$-$9.3, 7.6] &   \\
      & 4--3 & (0, 1$^{1e}$, 0) & 0.53 (0.02) & 6.19 (0.28) & $-$1.18 (0.12) & [$-$7.9, 8.3] & maser \\
      &      & (0, 1$^{1f}$, 0) & 0.25 (0.03) & 8.7 (1.0) & $-$1.25 (0.43) & [$-$10.3, 10.7] & \\
\hline
S Aur & 2--1 & (0, 1$^{1e}$, 0) & 1.42 (0.07) & 5.21 (0.2) & $-$16.61 (0.09) & [$-$25.0, $-$3.9] & maser \\
      & 3--2 & (0, 1$^{1e}$, 0) & 0.55 (0.07) & 7.7 (2.3) & $-$16.88 (0.62) & [$-$25.9, $-$8.4] & \\
      &      & (0, 0, 1)        & 0.024 (0.003) & 0.19 (0.04) & $-$17.15 (0.02) & [$-$19.5, $-$15.1] & \\
\hline
V Aql & 3--2 & (0, 1$^{1e}$, 0) & 0.4 (0.03)  & 5.0 (0.51) & 55.31 (0.19) & [41.8, 62.0] & maser\\
\hline
RV Aqr & 2--1 & (0, 1$^{1e}$, 0) & 1.11 (0.01) & 4.02 (0.05) & $-$1.0 (0.02) & [$-$5.2, 7.6] & maser \\
       &      & (0, 1$^{1f}$, 0) & 0.17 (0.03) & 3.1 (0.4) & $-$1.1 (0.2) & [$-$6.2, 5.5] & maser \\
       & 3--2 & (0, 1$^{1e}$, 0) & 1.09 (0.02) & 3.63 (0.09) & $-$1.0 (0.03) & [$-$7.4, 4.8] & maser\\
       & 4--3 & (0, 1$^{1e}$, 0) & 0.53 (0.1) & 7.3 (1.0)  & $-$0.16 (0.4) & [$-$5.7, 4.0] &   \\
\hline
\end{tabular}
\tablefoot{The integrated intensities ($\int T_{\rm mb} {\rm d}V$) are calculated by integrating over the intensities within the LSR velocity range of the detection, whereas the line widths ($\Delta V$) and velocity centroids ($V_\text{LSR}$) are obtained from Gaussian fit. The last column gives our interpretation on the properties of the emission: ``maser" indicates that the line profile is dominated by maser emission; ``maser + c" indicates that the maser is contaminated with a thermal component; ``blended'' indicates that the line may be blended with another line. \\
\tablefoottext{a}{$J=3$--$2$ (1,0,0): potentially blended with another unspecified line \citep[IRC +10216;][]{cerni..2011A&A...529L...3C}.}
\tablefoottext{b}{$J=3$--$2$ (0,2$^{2e}$,0) and (0,2$^{2f}$,0): two $l$-type doublet components blended with each other and multiple lines of C$_4$H, C$_3$N, and $^{13}$CCCN \citep[IRC +10216;][]{2017ApJ...845...38H}.}
\tablefoottext{c}{$J=3$--$2$ (0,2,0): potentially blended with $^{29}$SiS (15--14) at 267242.2178 MHz \citep[IRC +10216 and II Lup;][]{cerni..2011A&A...529L...3C,2017ApJ...845...38H}.}
\tablefoottext{d}{Two $l$-type doublet components blended with each other.}
\tablefoottext{e}{$v_2=1^{1e}$ line blended with the corresponding ground-state emission.}
\tablefoottext{f}{$J=4$--$3$ (0, 1$^{1f}$, 0): potentially blended with $^{29}$SiS (20--19) at 356242.4290 MHz (RAFGL 4211 only).}
 }
 % $J=3-2$ transitions in the (0, 2$^{2e}$, 0) and (0, 2$^{2f}$, 0) states are blended with each other + many lines in IRC +10216, 
 % $J=4-3$ transitions in the (0, 2$^{2e}$, 0) and (0, 2$^{2f}$, 0) states are blended with each other in IRC +10216 and CRL 3068. 
 % $J=4-3$ transitions in the (0, 3$^{3e}$, 0) and (0, 3$^{3f}$, 0) states are blended with each other in IRC +10216.
 % $J=2-1$ (0, 1$^{1e}$, 0) in IRC +30374. 
 % $J=3-2$ (0, 1$^{1e}$, 0) in II Lup and RAFGL 4211, 
 % $J=4-3$ (0, 1$^{1e}$, 0) in II Lup.
\end{table*}
\normalsize

\subsubsection{Thermal emission}\label{sec:thermal}
We detect 39 vibrationally excited HCN thermal lines toward IRC +10216, II Lup, RAFGL 4211, S Aur, W Ori, RV Aqr, IRC +30374, and CRL 3068, while thermal lines are not detected toward the other sources. Most of these sources have high mass-loss rates of $\gtrsim$1$\times 10^{-5}$~M$_{\odot}$~yr$^{-1}$, except for RV Aqr and W Ori, which have lower mass-loss rates of $<$3$\times 10^{-6}$~M$_{\odot}$~yr$^{-1}$. These thermal lines display Gaussian-like profiles. In order to derive the observed parameters of the thermal emissions, we assume a single Gaussian component to fit the observed line profile. The results are shown in Table \ref{tab:vib_hcn_results_updated}. The peak intensities are within a range of 0.001--0.5 K for these lines. The fitted velocities are generally in a good agreement with the systemic velocities in the literature within $\sim$3$\sigma$ (see also Table~\ref{tab:sample}). 

\begin{figure*}[!htbp]
\centering
\includegraphics[width=0.45\textwidth]{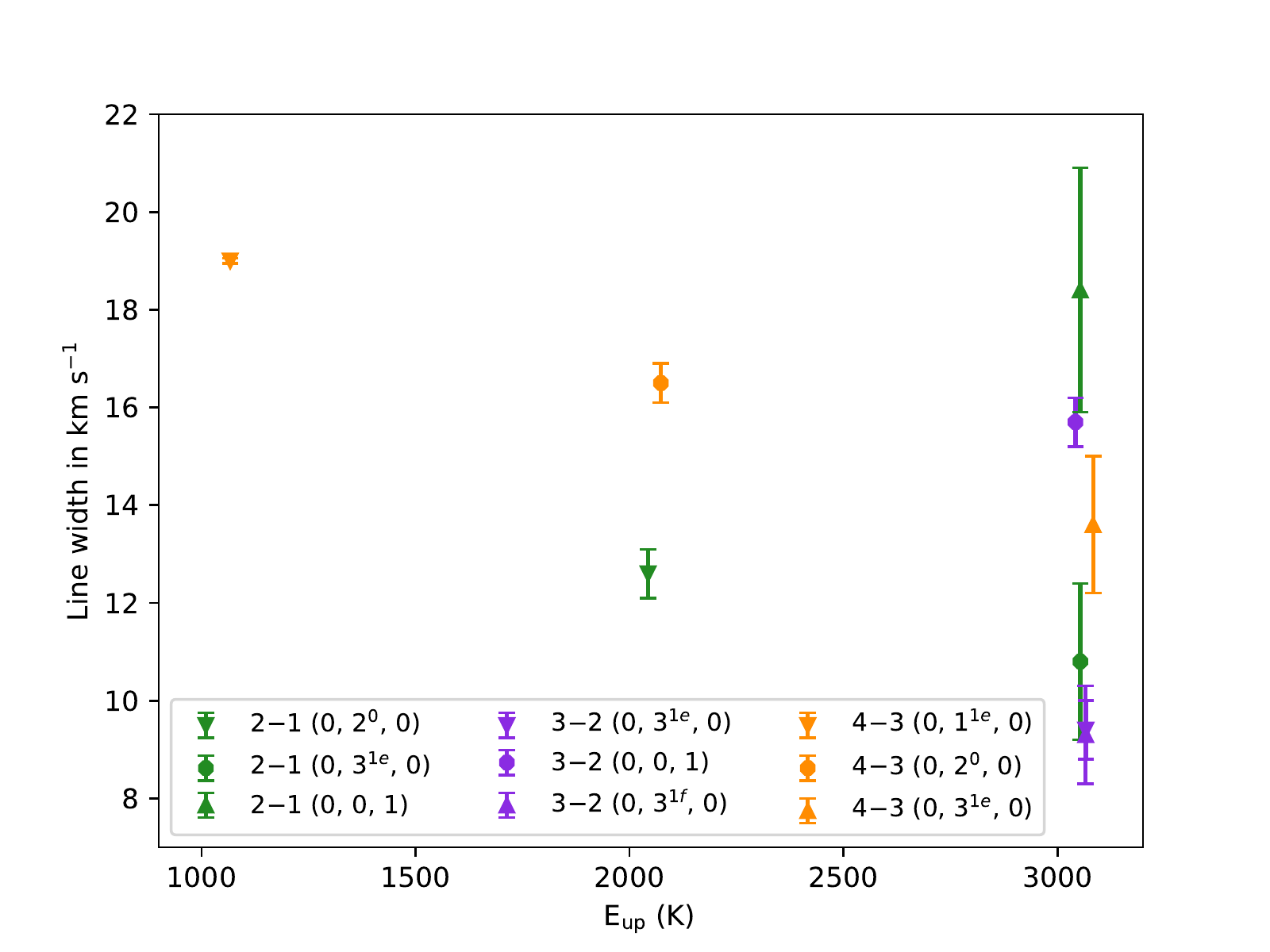}
\includegraphics[width=0.45\textwidth]{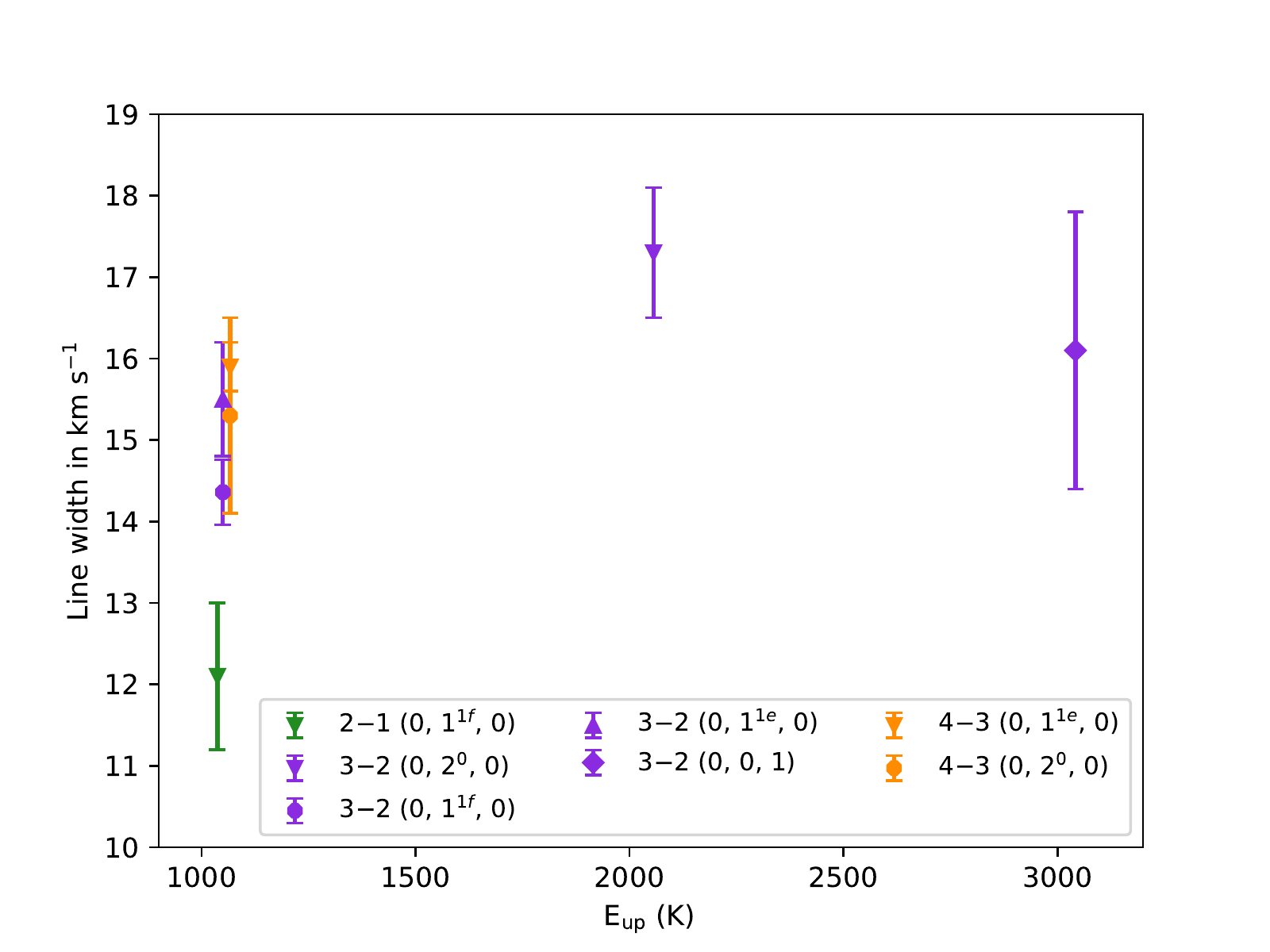}
\caption{FWHM line widths as a function of upper energy levels for IRC +10216 (left) and CRL 3068 (right). The colors represent the rotational transitions and the symbols represent the various thermal vibrationally excited lines as indicated in the legends.}\label{fig:linewid}
\end{figure*}

All of these lines have FWHM line widths of $>$8~\kms. We also note that the velocity spread of thermal vibrationally excited HCN lines can be as broad as that of the ground-state line (see Figs.~\ref{fig:irc_all_spec} and \ref{fig:crl-3068_all_spec} for example). Because we only detected enough thermal vibrationally excited HCN lines in IRC +10216 and CRL 3068, we investigate the fitted line widths as a function of upper energy levels for these two sources. The results are shown in Fig.~\ref{fig:linewid}. The line widths do not have an apparent dependence on either their upper energy levels or their Einstein $A$ coefficients (see Table \ref{tab:lines}). Even though a couple of transitions have similar upper energy levels, their line widths can vary by a factor of 1.5. Particularly, the $\nu_{2}=$3 lines are found to be the narrowest with expansions of $\sim$5~\kms in IRC +10216 due to their high energy levels and critical densities. The line widths are comparable to those of other high energy transitions from parent molecules \citep[e.g., HCN, CS, SiS;][]{2000ApJ...544..881H,2008ApJS..177..275H,nimesh..2009ApJ...692.1205P}. Comparing with the models of the velocity structure in IRC +10216 \citep[e.g., Fig.~5 in][]{2015A&A...574A...5D}, we find that the $\nu_{2}=$ 3 lines might arise from the wind acceleration zone, which is about 8 to 10~$R_{*}$ from the star.

The innermost regions of CSEs have high extinctions, volume densities, and temperatures. Such regions are poorly studied because they are not easily probed with the standard dense-gas tracers in the ground-state lines due to effects like photon trapping and self-absorption. Our observations extend the spectrally resolved detection of the vibrationally excited thermal HCN transitions toward carbon-rich stars beyond IRC +10216, which paves the way towards better understanding of the innermost regions of carbon-rich CSEs with vibrationally excited HCN thermal transitions.

\subsubsection{Maser emission}\label{sec:maser}

HCN maser action is identified following the criteria that the line profiles display either higher peak intensities than expected or strong asymmetry (i.e., narrow spike features) overlaid on a broader line profile. For instance, different intensities of the $J$ = 2--1 (0, 1$^{1e}$, 0) and $J$ = 2--1 (0, 1$^{1f}$, 0) lines can be taken as evidence for maser action. Under local thermodynamic equilibrium (LTE) conditions, these two lines are expected to emit at nearly identical intensities, so the intensity difference of these two lines indicates that at least one of them deviates from LTE. In our observations, the $J$=2--1 (0, 1$^{1e}$, 0) line is significantly brighter than the $J$=2--1 (0, 1$^{1f}$, 0) line (see Figs.~\ref{fig:irc_all_spec}--\ref{fig:ii-lup_all_spec}), which is suggestive of maser actions. Symmetric profiles are generally expected from thermal emission of typical AGB CSEs and narrow asymmetric spikes are not expected in thermal line profiles by the classic models \citep[e.g.,][]{1975ApJ...197..603M,1982A&A...107..128O}. Previous observations suggested that such asymmetric spikes arise from maser action \citep[e.g.,][]{1983ApJ...267..184H,schilke..mehringer..menten2000ApJ...528L..37S,2006ApJ...646L.127F,sis..Gong_2017,2018ApJ...860..162F}, because these masers are only formed at specific regions in CSEs, which correspond to specific velocities in their line profiles. Hence, the presence of narrow asymmetric spikes are regarded as an indicator of maser action in this work.

In this study, we detect 29 HCN maser lines from 12 carbon-rich stars whereas HCN masers are not detected toward Y Pav, IRC +30374, CRL 2513, and CRL 2477 by our observations. Except for six maser lines already reported by previous observations \citep[e.g.,][]{2018AA...613A..49M}, 23 transitions are newly discovered HCN masers. Hence, our observations expand the total
number of previous known HCN masers toward carbon-rich stars by 47\% (see Table~\ref{tab:catalogue} in Appendix~\ref{app:catalog}). Our results also confirm that HCN masers are common in carbon-rich stars. Particularly, eight CSEs, which are CRL 3068, RV Aqr, R Vol, AQ Sgr, V Aql, V636 Mon, S Aur, and IRC +20370, are newly discovered to host HCN masers. The 29 HCN masers arise from 8 different transitions, which are $J=2$--$1$ (0, 1$^{1e}$, 0), $J=2$--$1$ (0, 1$^{1f}$, 0), $J=3$--$2$ (0, 1$^{1e}$, 0), $J=3$--$2$ (0, 3$^{1e}$, 0), $J=3$--$2$ (0, 2, 0), $J=4$--$3$ (0, 1$^{1e}$, 0), and $J=4$--$3$ (0, 1$^{1f}$, 0). Among them, the $J=3$--$2$ (0, 3$^{1e}$, 0), $J=3$--$2$ (0, 2, 0), $J=4$--$3$ (0, 1$^{1e}$, 0), and $J=4$--$3$ (0, 1$^{1f}$, 0) masers are first detected in the interstellar medium. 

\textbf{IRC +10216:} Five HCN masers, including $J=2$--$1$ (0, 1$^{1e}$, 0), $J=2$--$1$ (0, 1$^{1f}$, 0), $J=3$--$2$ (0, 1$^{1e}$, 0), $J=3$--$2$ (0, 1$^{1f}$, 0), and $J=4$--$3$ (0, 1$^{1f}$, 0), are detected in IRC +10216. The $J=2$--$1$ (0, 1$^{1e}$, 0) maser was reported by previous observations \citep{lucas1989cerni..discovery,2018AA...613A..49M}. Our 2018 observations show at least three peaks at $-$32.2, $-$27.9, and $-$26.3 \kms, which have peak flux densities of about 145, 295, and 351 Jy, respectively. The intensity of this line is nearly 20 times that of the $J=2$--$1$ (0, 1$^{1f}$, 0) line (see Fig.~\ref{fig:irc_all_spec}). Comparison with other measurements at different epochs are presented in Sect.~\ref{sec:variability}. Furthermore, we find that the flux density is lower for the components with LSR velocities farther from the systemic velocity of $-$26.5~\kms. The $J=2$--$1$ (0, 1$^{1f}$, 0) line was believed to show maser actions due to its intensity variation \citep{2018AA...613A..49M}.
%, and the variation is also confirmed by our study (see Sect.~\ref{sec:variability}). 
This maser peaks close to the systemic velocity, and has a flux density of about 17 Jy. The $J=3$--$2$ (0, 1$^{1e}$, 0) line shows as a narrow spike overlaid on a broad thermal profile, while such a feature is not seen in the $J=3$--$2$ (0, 1$^{1f}$, 0) line. The $J=3$--$2$ (0, 1$^{1e}$, 0) maser peaks at a redshifted LSR velocity of $-$24.0~\kms. Subtracting the broad component, we obtained a flux density of 17.5 Jy. We note that this line was observed with NRAO-12 m and IRAM-30 m  \citep{1986ApJ...300L..19Z,cerni..2011A&A...529L...3C}, but they did not detect a spike feature. On the other side, the $J=3$--$2$ (0, 1$^{1e}$, 0) maser was detected by previous Extended Submillimeter Array (eSMA) observations \citep{shinnaga2009}. We also see a presence of asymmetrical profile feature in the $J=3$--$2$ (0, 1$^{1f}$, 0) line, which is marked as a maser. \citet{2017ApJ...845...38H} monitored the line shape of this maser and reported such variability in its blue-shifted peak. %, implying that the emission might arise very close to the photosphere of the star. 
The $J=4$--$3$ (0, 1$^{1f}$, 0) line appears to show a prominent asymmetry narrow spike on the blushifted side, peaking at an LSR velocity of $-$35.9~\kms, while no asymmetric features are found in the $J=4$--$3$ (0, 1$^{1e}$, 0) line. The flux density is estimated to be 68 Jy by subtracting the broad component. The flux densities of our detected HCN masers are at least a factor of 2 lower than those ($>$800 Jy) of submillimeter HCN lasers \citep{schilke..mehringer..menten2000ApJ...528L..37S,schilke..menten2003ApJ...583..446S}.  Furthermore, we find that the intensity of the $J=3$--$2$ (0, 3$^{1e}$, 0) line may also deviate from the values expected under LTE, because its peak intensity is about 1.4 times that of the $J=3$--$2$ (0, 3$^{1f}$, 0) line. However,  the tentative maser nature of the $J=3$--$2$ (0, 3$^{1e}$, 0) line still needs to be confirmed by follow-up observations because the signal-to-noise ratios of the two lines are not good enough for the  difference to be significant. %is regarded as a tentative maser in this work.  

\textbf{RAFGL 4211:} We detected four HCN masers. Among them, the $J=2$--$1$ (0, 1$^{1e}$, 0) maser was first reported by \citet{2018AA...613A..49M}. Our $J=2$--$1$ (0, 1$^{1e}$, 0) spectrum shows at least two peaks at $-$8.1~\kms\,and $-$5~\kms\, of flux density of 24.2 Jy and 32.5 Jy, respectively. These peak flux densities are about 20 times that of the $J=2$--$1$ (0, 1$^{1f}$, 0) line (see Fig.~\ref{fig:rafgl-all}), strongly supporting its maser nature. The flux density of the $J=2$--1 (0, 1$^{1e}$, 0) maser is comparable to that of the $J=1$--0 (0, 2, 0) maser in this source \citep[see Fig.~1 in ][]{2014MNRAS.440..172S}. Although the $J=3$--2 (0, 2, 0) and $J=3$--2 (0, 3$^{1e}$, 0) masers only have flux densities of 2.4~Jy and 11.8~Jy, respectively, the detection of the two masers are exceptional, because these two masers are only discovered toward RAFGL 4211 so far. Both of the two masers peak at $-$2.1~\kms, close to the stellar systemic velocity. Different from the former three masers, the $J=4$--3 (0, 1$^{1e}$, 0) profiles exhibits a narrow spike overlaid on a broad component, while such a narrow spike does not appear in the $J=3$--2 (0, 1$^{1e}$, 0) spectrum. The $J=4$--3 (0, 1$^{1e}$, 0) maser peak at a slightly blueshifted side of $-$3.7~\kms, and its flux density is about 18.5 Jy that is fainter than the $J=2$--$1$ (0, 1$^{1e}$, 0) maser. 

\textbf{II Lup:} We detected the $J=2$--$1$ (0, 1$^{1e}$, 0) and $J=4$--$3$ (0, 1$^{1f}$, 0) masers in II Lup. The latter one is newly detected in this source. The $J=2$--$1$ (0, 1$^{1e}$, 0) maser shows at least three peaks at $-$22.7, $-$18.1, and $-$15.2~\kms that have flux densities of 21.3, 39.3, and 39.3 Jy, respectively. This maser is much brighter than the $J=2$--$1$ (0, 1$^{1f}$, 0) line (see Fig.~\ref{fig:ii-lup_all_spec}). Two peaks at $-$25.4 and $-$6.2~\kms are also identified in the $J=4$--$3$ (0, 1$^{1f}$, 0) spectrum. Their corresponding flux densities are 14.3 Jy and 8.2 Jy that are about 4 times the peak flux density of the $J=4$--$3$ (0, 1$^{1e}$, 0) spectrum.

\textbf{R Vol:} No HCN masers have been reported in this star prior to our observations. In R Vol, our observations result in the first detection of four HCN masers that are the $J=2$--$1$ (0, 1$^{1e}$, 0), $J=2$--$1$ (0, 1$^{1f}$, 0), $J=3$--$2$ (0, 1$^{1e}$, 0), and $J=4$--$3$ (0, 1$^{1e}$, 0) lines. The $J=2$--$1$, $J=3$--$2$, and $J=4$--$3$ lines in the (0, 1$^{1e}$, 0) state have peak flux densities of 8.4 Jy, 3.8~Jy, and 7.5~Jy at $-$12.9, $-$14.5, $-$11~\kms, respectively, which are much brighter than their corresponding transitions in the (0, 1$^{1f}$, 0) state (see Fig.~\ref{fig:rvol-all}). This is suggestive of their maser nature. Although the $J=2$--$1$ (0, 1$^{1f}$, 0) line only has a flux density of 1.9 Jy, this line is still regarded as a maser in this work, because its velocity coverage is much narrower than the ground state line and the LSR velocity of its peak is slightly shifted to the blueshifted side at $-$13.2~\kms. 

\textbf{AQ Sgr:} The $J=3$--$2$ (0, 1$^{1e}$, 0) line has peak flux density of 1.2 Jy at 22.6~\kms. Despite its low flux density, this line is brighter than the $J=3$--$2$ (0, 1$^{1f}$, 0) line that is not detected by the same sensitivity. Furthermore, the peak velocity is slightly redshifted with respect to the systemic velocity ($\sim$20~\kms, see Fig.~\ref{fig:aqsgr-all}). Therefore, the $J=3$--$2$ (0, 1$^{1e}$, 0) line is regarded as a maser in this study, which makes it the first HCN maser toward this source. 

\textbf{CRL 3068:} Two HCN masers are newly discovered in the source, and they are the $J=2$--1 (0, 1$^{1e}$, 0) and $J=4$--$3$ (0, 1$^{1f}$, 0) masers.  Asymmetric features indicate that these two transitions show maser actions (see Fig.~\ref{fig:crl-3068_all_spec}). The $J$ = 2--1 (0, 1$^{1e}$, 0) maser displays five distinct peaks at $-$38.5~\kms, $-$35.3~\kms, $-$33.8~\kms, $-$29.4~\kms, and $-$26.8~\kms, and their corresponding peak flux densities are 12.9 Jy, 23.8~Jy, 18.2~Jy, 9.9~Jy, and 8.6 Jy, respectively. Stronger peaks are observed in the blueshifted side. They are at least 4 times higher than the peak flux density of the $J=2$--1 (0, 1$^{1f}$, 0) line. The $J=4$--$3$ (0, 1$^{1f}$, 0) maser shows a prominent peak of flux density 70.6 Jy at $-41.7$~\kms\,that is brighter than any other transitions in the (0, 1$^{1e}$, 0) state including even the $J=2$--1 (0, 1$^{1e}$, 0) maser. The $J=4$--$3$ (0, 1$^{1f}$, 0) maser is about 10~\kms\,away from the systemic velocity of $-$31.5~\kms, but still does not reach its terminal velocity of 14.5~\kms.

%\textbf{IRC+30374:} No maser here

\textbf{V636 Mon:} The $J=2$--1 (0, 1$^{1e}$, 0) line is the only vibrationally excited HCN maser detected in V636 Mon. Its spectrum exhibits a narrow spike at 5.7~\kms, and its peak flux density is 2.2 Jy (see Fig.~\ref{fig:v636-all}). This line is at least 3 times the peak flux density of the $J=2$--1 (0, 1$^{1f}$, 0) line. Furthermore, the narrow spike is about 4~\kms\,away from the systemic velocity. These two features support that the $J=2$--1 (0, 1$^{1e}$, 0) line is a maser, which make it become the first maser detected in V636 Mon. 

\textbf{IRC +20370:} Two HCN masers from the $J=2$--1 (0, 1$^{1e}$, 0) and $J=3$--$2$ (0, 1$^{1e}$, 0) transitions are detected toward this source. Our results are the first maser detection toward this source. Both maser transitions are brighter than their corresponding lines in the (0, 1$^{1f}$, 0) state by a factor of $\sim$10 (see Fig.~\ref{fig:irc+20370-all}). The $J=2$--1 (0, 1$^{1e}$, 0) maser hosts two peaks at $-$4.6~\kms\,and $-$2~\kms, and their corresponding flux densities are 6.7 and 7.3 Jy. The $J=3$--2 (0, 1$^{1e}$, 0) maser exhibits at least one peak at $-$4~\kms\, with a flux density of 7.1 Jy, which is slightly weaker than the $J=2$--1 (0, 1$^{1e}$, 0) maser. These velocity components are slightly blueshifted from the systemic velocity of $-$0.8~\kms. The peak flux density at $-$4.6~\kms is lower than that at $-$1.7~\kms\,in the $J=2$--1 (0, 1$^{1e}$, 0) maser, which is in stark contrast to the $J=3$--2 (0, 1$^{1e}$, 0) maser. The different velocities indicates that these masers could arise from different zones or different pumping mechanisms.

\textbf{W Ori:} In this study, we report three new HCN masers from the $J=2$--1 (0, 1$^{1e}$, 0), $J=3$--$2$ (0, 1$^{1e}$, 0), and $J=4$--$3$ (0, 1$^{1e}$, 0) transitions that display a single-peak profile. Their maser nature is strongly supported by the fact that these three transitions have peak intensities about 20 times those of their corresponding (0, 1$^{1f}$, 0) transitions (see Fig.~\ref{fig:wori-all}). The respective LSR velocities of the three masers are around 0.13~\kms, $-$1.1\kms, and $-$2.5~\kms, while the corresponding flux densities are 14.5~Jy, 16~Jy, and 6.5~Jy. All of these maser peaks are within 3~\kms\,away from the systemic velocity of $-$1.0~\kms. W Ori has been found to host the HCN $J=1$--0 (0, 0, 0) maser by the spike emission at a LSR velocity of $\sim$8~\kms\,and show temporal variations in this maser \citep{olofsson1993bstudy,izumiura..1995ApJ...440..728I}. The LSR velocities indicate that our newly detected masers arise from the regions closer to the star when compared with the ground-state $J=1$--0 maser. Furthermore, the $J=2$--1 (0, 1$^{1e}$, 0) maser is more luminous than the HCN $J=1$--0 (0, 0, 0) maser.

\textbf{S Aur:} In this source, the $J=2$--1 (0, 1$^{1e}$, 0) transition exhibits a narrow component with a high flux density of 9.1~Jy at $-$17.0~\kms\,(see Fig.~\ref{fig:saur-all}). Its peak intensity is at least 30 times that of the $J=2$--1 (0, 1$^{1f}$, 0) transition, and is even twice the peak intensity of the $J=2$--1 (0, 0, 0) transitions. Furthermore, its line width is much narrower than that of the $J=3$--2 (0, 1$^{1e}$, 0) transition. These facts support that the $J$ = 2 -- 1 (0, 1$^{1e}$, 0) transition is a maser, which is the first maser reported toward this source. We also tentatively detect the $J=3$--2 (0, 0, 1) line exhibiting a peak intensity of 8.5 Jy with a very narrow line width of 0.19~\kms. The emission feature peaks close to the systematic velocity of $-$17~\kms. Since this emission feature is only detected in 2 native spectral channels above the noise level, we conservatively do not identify it as a maser.% even though the line satisfies our detection criteria. 

\textbf{V Aql:} The $J=3$--2 (0, 1$^{1e}$, 0) transition is the only vibrationally excited HCN line detected in V Aql. This transition displays a single-peaked spectrum with a peak flux density of $\sim$4.9 Jy at 55.4~\kms\,(see Fig.~\ref{fig:vaql-all}). The peak flux density is at least 3 times that of the $J=3$--2 (0, 1$^{1f}$, 0) transition. Furthermore, the LSR velocity appears to be slightly redshifted with respective to the systemic velocity of 53.5~\kms. These facts support that the $J=3$--2 (0, 1$^{1e}$, 0) transition is a maser, which is the first maser detected in V Aql.
 
\textbf{RV Aqr:} In this study, we report three new HCN masers from the $J=2$--1 (0, 1$^{1e}$, 0), (0, 1$^{1f}$, 0), and $J=3$--2 (0, 1$^{1e}$, 0) transitions. Figure~\ref{fig:rvaqr-all} shows that the $J=2$--1 (0, 1$^{1e}$, 0) and $J=3$--2 (0, 1$^{1e}$, 0) transitions are at least 10 times brighter than their corresponding (0, 1$^{1f}$, 0) transitions. Compared with the vibrationally ground-state lines, the $J=2$--1 (0, 1$^{1f}$, 0) line and also the two detected (0, 1$^{1e}$, 0) lines have much narrower velocity spreads. Furthermore, their peak velocities are about $-$1~\kms\,that are blueshifted with respect to the systemic velocity of 0.5~\kms. Therefore, all three transitions are identified as masers in this work, which are the first maser detection in RV Aqr. Their flux densities are 9.5~Jy, 1.9~Jy, and 14.7~Jy for the $J=2$--1 (0, 1$^{1e}$, 0), (0, 1$^{1f}$, 0), and $J=3$--2 (0, 1$^{1e}$, 0) masers, respectively. On the other hand, the $J=4$--3 (0, 1$^{1e}$, 0) transition is only slightly brighter than the $J=4$--3 (0, 1$^{1f}$, 0) transition; the latter has an S/N ratio of ${\sim}2$ at the peak channel and does not formally meet our detection criteria. We conservatively do not identify the $J=4$--3 (0, 1$^{1e}$, 0) transition as a maser in this work. 

\textbf{Non-detection:} HCN masers are not detected toward Y Pav, IRC +30374, CRL 2513, or CRL 2477 in our observations. Hence, we are only able to give a 3$\sigma$ upper limit of 0.9~Jy, 1.8~Jy, 0.8~Jy, and 2.6~Jy for the $J=2$--1 (0, 1$^{1e}$, 0) transitions toward Y Pav, IRC +30374, CRL 2513, and CRL 2477, respectively. Among these sources, we note that the HCN $J=1$--0 (0, 2, 0) maser was detected toward CRL 2513 by \citet{lucas1988new}, but the higher rotational transitions ($J=2$--1 and $J=3$--2) in the (0, 2, 0) state are not detected in this work (see Fig.~\ref{fig:crl2513-all}).
  
In summary, the $J=2$--$1$ (0, 1$^{1e}$, 0) maser is detected toward 12 out of 16 sources. The detection rate of 75\% is similar to that of the previous $J=2$--$1$ (0, 1$^{1e}$, 0) maser study \citep[85\%;][]{2018AA...613A..49M} and is much higher than that of the $J=1$--$0$ (0, 2, 0) maser in carbon-rich stars \citep[7\%;][]{lucas1988new}. This suggests that the $J=2$--$1$ (0, 1$^{1e}$, 0) maser is more common than the $J=1$--$0$ (0, 2, 0) maser in carbon-rich stars. 

The peak flux density of the $J=2$--$1$ (0, 1$^{1e}$, 0) maser is brighter than that of the $J=2$--$1$ (0, 0, 0) line in W Ori and S Aur, while the peak flux densities of these two lines are comparable in RAFGL 4211, R Vol, and RV Aqr. In contrast, the $J=2$--$1$ (0, 1$^{1f}$, 0) masers are much weaker and are only detected in three sources. The $J=3$--$2$ (0, 1$^{1e}$, 0) maser is detected toward 7 sources. The flux density of the $J=3$--$2$ (0, 1$^{1e}$, 0) maser is much lower than that of the $J=2$--$1$ (0, 1$^{1e}$, 0) maser toward IRC +10216 and R Vol, but is roughly comparable to that of the $J=2$--$1$ (0, 1$^{1e}$, 0) maser toward IRC +20370, W Ori, and RV Aqr. The $J=4$--$3$ (0, 1$^{1f}$, 0) masers are only detected in IRC +10216, CRL 3068, and II Lup, and all are found to show a prominent blueshifted component with a peak at $\sim$10~\kms\,from their systemic LSR velocities. 

Moreover, we find that none of the detected HCN maser peaks reach the terminal velocities, suggesting that these HCN masers arise from shells that are not yet fully accelerated. This supports the notion that these masers can be used to investigate the wind acceleration of carbon-rich stars. Furthermore, most of the bright HCN maser components appear in the blueshifted side with respect to their respective systemic LSR velocities, which may be attributed to their underlying pumping mechanisms (see~Sect.~\ref{sec:pumping}). 

\subsection{Time variability of HCN masers}
\label{sec:variability}

\begin{figure*}[!htbp]
\centering
\includegraphics[width=.45\textwidth]{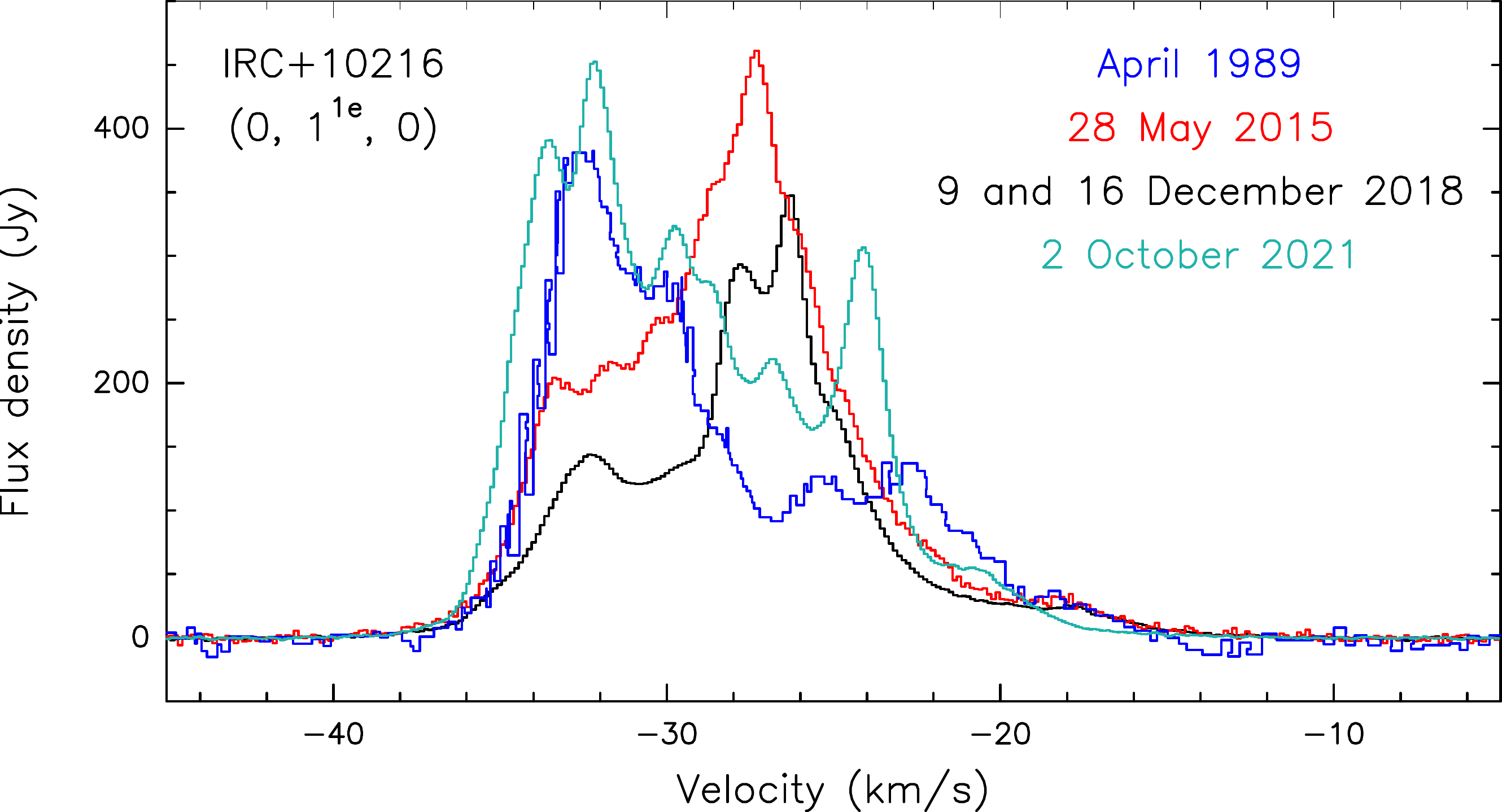}\hfill
\includegraphics[width=.45\textwidth]{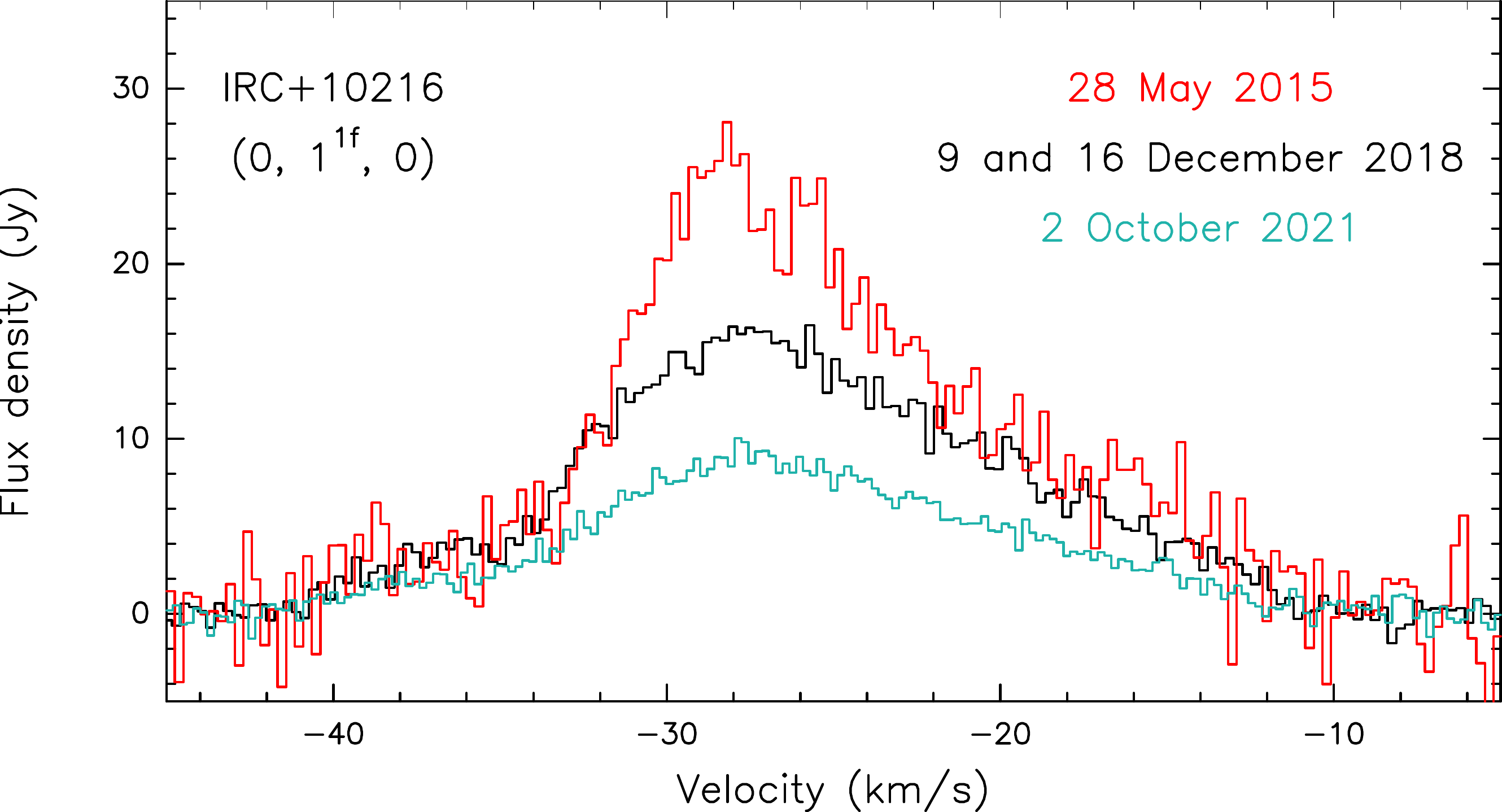}\hfill
\includegraphics[width=.45\textwidth]{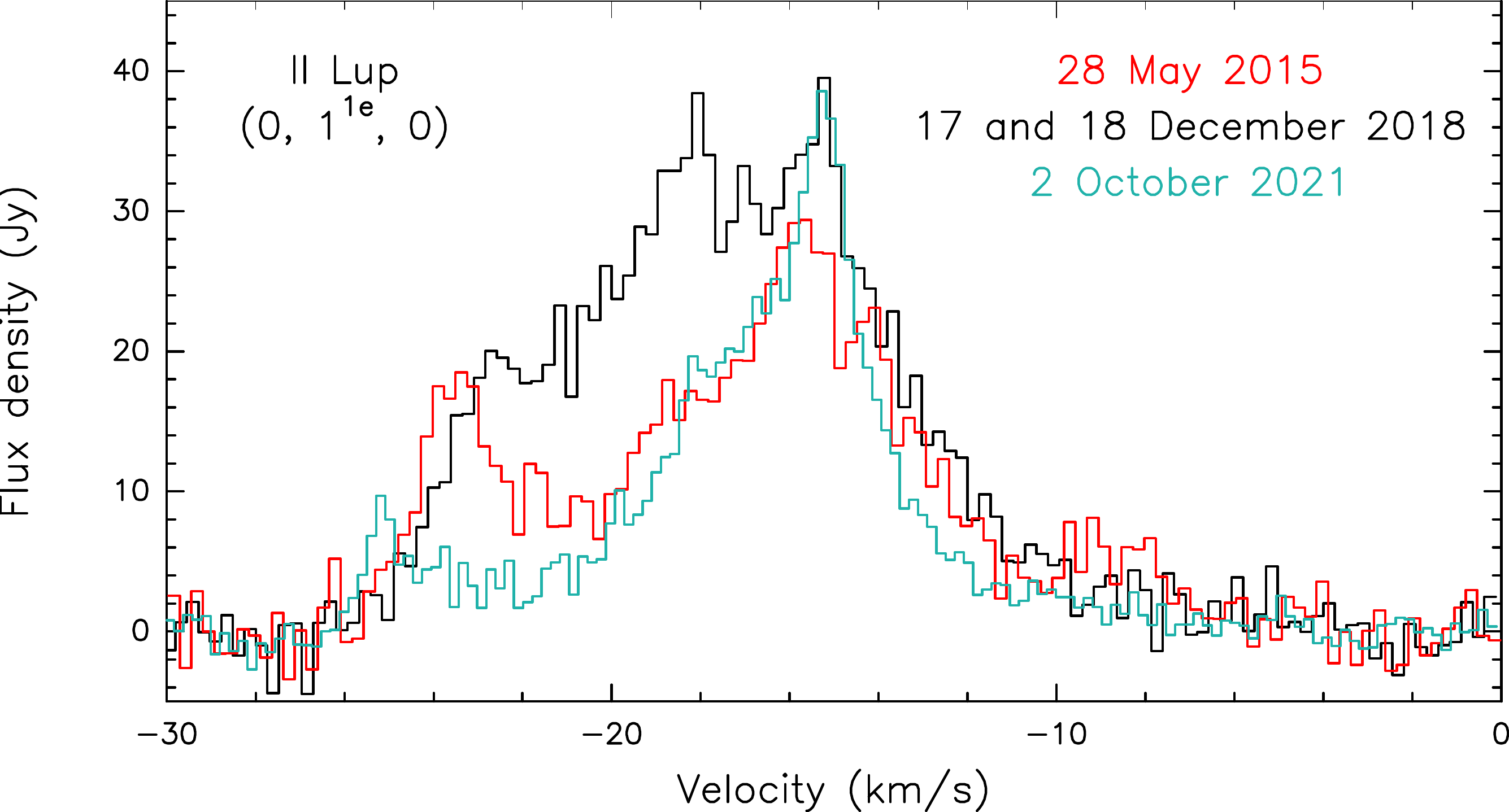}\hfill
\includegraphics[width=.45\textwidth]{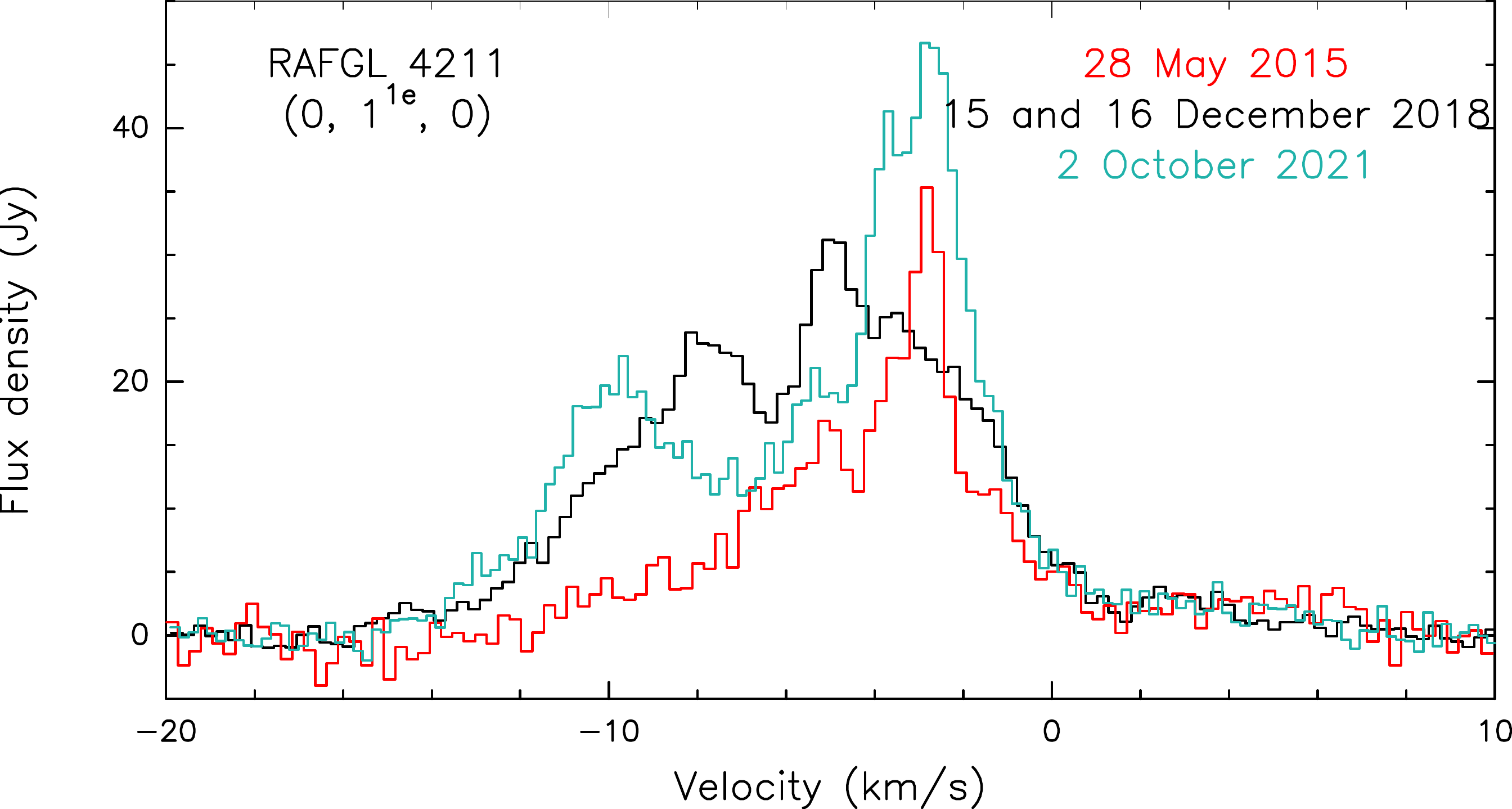}
\caption{Variability of HCN (2--1) masers seen towards IRC+10216 (top two panels) in  (0, 1$^{1e}$, 0) and (0, 1$^{1f}$, 0), II Lup (bottom left) and RAFGL 4211 (bottom right) in  (0, 1$^{1e}$, 0) vibrationally excited lines. The black spectra is from observations in 2018, cyan is from 2021, red is from \citet{2018AA...613A..49M}, and blue is from \citet{lucas1989cerni..discovery} with the date of observations indicated in the legend.}
\label{fig:variability_masers}
\end{figure*}

Three of our targets, IRC +10216, II Lup, and RAFGL 4211, have been observed in the $J=2$--$1$  (0, 1$^{1e}$, 0) and $J=2$--$1$ (0, 1$^{1f}$, 0) lines at least twice in previous works \citep{lucas1989cerni..discovery,2018AA...613A..49M}, allowing us to study the time variability of the HCN masers. Figure~\ref{fig:variability_masers} presents the HCN masers obtained at different epochs in flux density scale. %We observed these sources on 02 October 2021 and 
It can be seen that the line shape of the $J=2$--$1$ (0, 1$^{1e}$, 0) line has been changing significantly over time. For IRC +10216, the spectra of the (0, 1$^{1e}$, 0) line show four peaks at $-$32.6~\kms, $-$29.8~\kms, $-$25.1~\kms, and $-$22.6~\kms\,in 1989, two peaks at $-$33~\kms, and $-$27.2~\kms\,in 2015, three peaks at about $-$32.2~\kms, $-$26.3~\kms, and $-$27.9~\kms\,in 2018, and five peaks at $-$33.5~\kms, $-$32.1~\kms, $-$29.7~\kms, $-$26.7~\kms, $-$24.1~\kms\, in 2021 (see the top-left panel in Fig.~\ref{fig:variability_masers}). 

Toward II Lup, this line shows two peaks at $-$23.6~\kms and $-$15.9~\kms\,in 2015, three peaks at $-22.6$~\kms, $-$18.3~\kms, and $-$15.1~\kms\,in 2018, and two peaks again in 2021 at $-$25.1~\kms, and $-$15.3 ~\kms\, (see the bottom-left panel in Fig.~\ref{fig:variability_masers}). Similarly, RAFGL 4211 shows three peaks at $-$8.2~\kms, $-$5.1~\kms, $-$3.7~\kms\,in 2018, two peaks at $-$5.1~\kms\,and $-$2.9~\kms\,in 2015, and three peaks in 2021 at $-$9.7~\kms, $-$3.7~\kms, $-$2.6~\kms\, (see the bottom-right panel in Fig.~\ref{fig:variability_masers}).

Based on the measurements of the ground state HCN lines, we note that there might be about 50\% uncertainty in flux calibration. In order to minimize the effect of calibration uncertainty, we use the integrated intensity \emph{ratio} between the maser line and the ground-state $J=2$--$1$ (0, 0, 0) line to assess the temporal variation between the three epochs of APEX observations (2015--2021). For IRC +10216, the ratio changes from 0.16 $\rightarrow$ 0.15 $\rightarrow$~0.28, suggesting that the flux density has significantly increased between 2018 and 2021 ($\approx$86\%). For II Lup, the ratio changes from 0.28 $\rightarrow$ 0.35 $\rightarrow$ 0.20, indicative of a 25\% increase in the flux density of the (0, 1$^{1e}$, 0) maser between 2015 and 2018 but decrease in 2021 (42\%). For RAFGL 4211, the ratio increases from 0.16~$\rightarrow$ 0.25 $\rightarrow$ 0.36, indicating that the flux density of the (0, 1$^{1e}$, 0) maser increases by 63\% in the first two epochs and by 44\% in the last two. These suggest that the HCN $J=2$--$1$ (0, 1$^{1e}$, 0) masers are variable in the three sources over a timescale of a few years. In contrast, as shown in the top-right panel in Fig.~\ref{fig:variability_masers}, the weaker HCN $J=2$--$1$ (0, 1$^{1f}$, 0) maser is not  quite variable over the three years in IRC +10216. The spectral profiles are comparable between three epochs and the intensity ratio between the (0, 1$^{1f}$, 0) and (0, 0, 0) lines only changes by $\lesssim$10\%. However, it is worth noting that the most significant variation in the $J=2$--$1$ (0, 1$^{1e}$, 0) transition happened over a longer period in time from 1989 to 2015 \citep[see Sect. 6.3.4 in][]{2018AA...613A..49M}. 

Although the time variability of HCN masers is confirmed by our study, observations are too sparse to allow for a search for periodic variations or for association with stellar pulsation periods. Previous monitoring observations of the HCN $J=3$--$2$ (0, 1$^{1f}$, 0) maser in IRC +10216 indicates a period of $\sim$730 days \citep{2017ApJ...845...38H}, which is at least 100--200 days longer than the periods in the infrared \citep[$\sim$630$\pm$3 days;][]{1992A&AS...94..377L,menten..2012A&A...543A..73M} and centimeter bands \citep[535$\pm$50 days;][]{2006A&A...453..301M}. An anti-correlation between HCN maser luminosity and infrared luminosity was found in their single-dish observations \citep{2017ApJ...845...38H}, but was not confirmed by the follow-up monitoring observations with the Atacama Compact Array \citep{2019ApJ...883..165H}. Previous observations toward RAFGL 4211 indicate that the variability of the HCN $J=1$--$0$ (0, 2, 0) maser may be unrelated to stellar pulsation, although the observations consisted of only three epochs  \citep{2014MNRAS.440..172S}. \citet{2018AA...613A..49M} suggest that the apparent absence of correlation between the vibrationally excited HCN lines emission cycle and stellar cycle may arise from the additional contribution of collisional pumping. Further long-term monitoring observations with higher cadences and consistent flux calibration are needed to identify the exact cause(s) of the time variability of HCN masers, such as inhomogeneity in the outflowing wind or a highly anisotropic infrared field.

\section{Discussion}\label{sec:discussion}
\subsection{Population diagram}\label{sec:pm}

\begin{figure}[!htbp]
\centering
%    \resizebox{\hsize}{!}{\includegraphics{figs/IRC+10216_RT_withmasers.png}}
\includegraphics[width=0.45\textwidth]{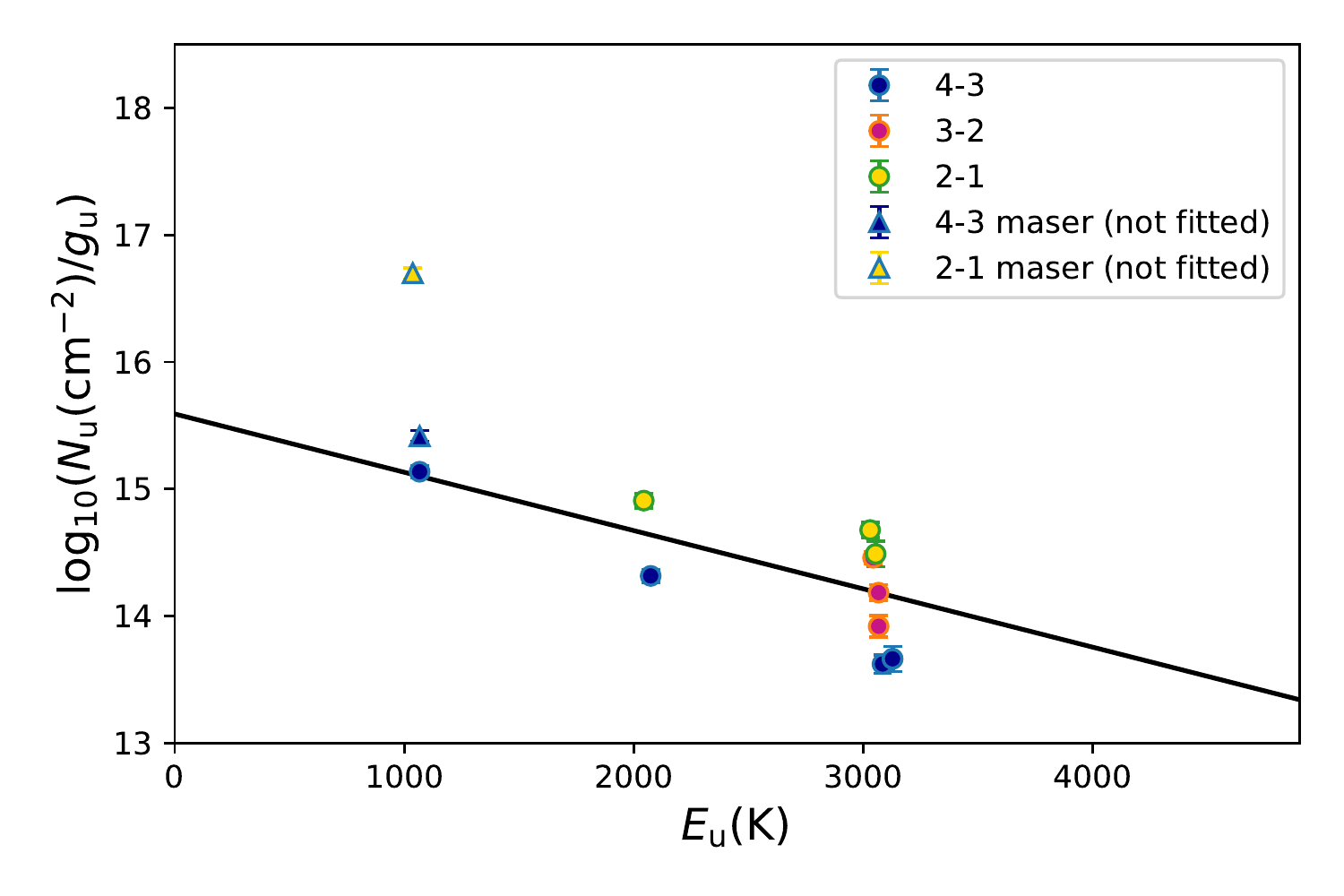}
\caption{Rotation diagram of IRC+10216 constructed from vibrationally excited HCN lines. %, indicated by circles, as in Table \ref{rotdiag_parameters}. 
Some masers are also plotted in the diagram, shown by triangles, but are excluded from the linear least-square fit.}\label{fig:irc_rotdiag}%
\end{figure}

\begin{figure}[!htbp]
\centering
%    \resizebox{\hsize}{!}{\includegraphics{CRL-3068_RT_withmasers.png}}
\includegraphics[width=0.45\textwidth]{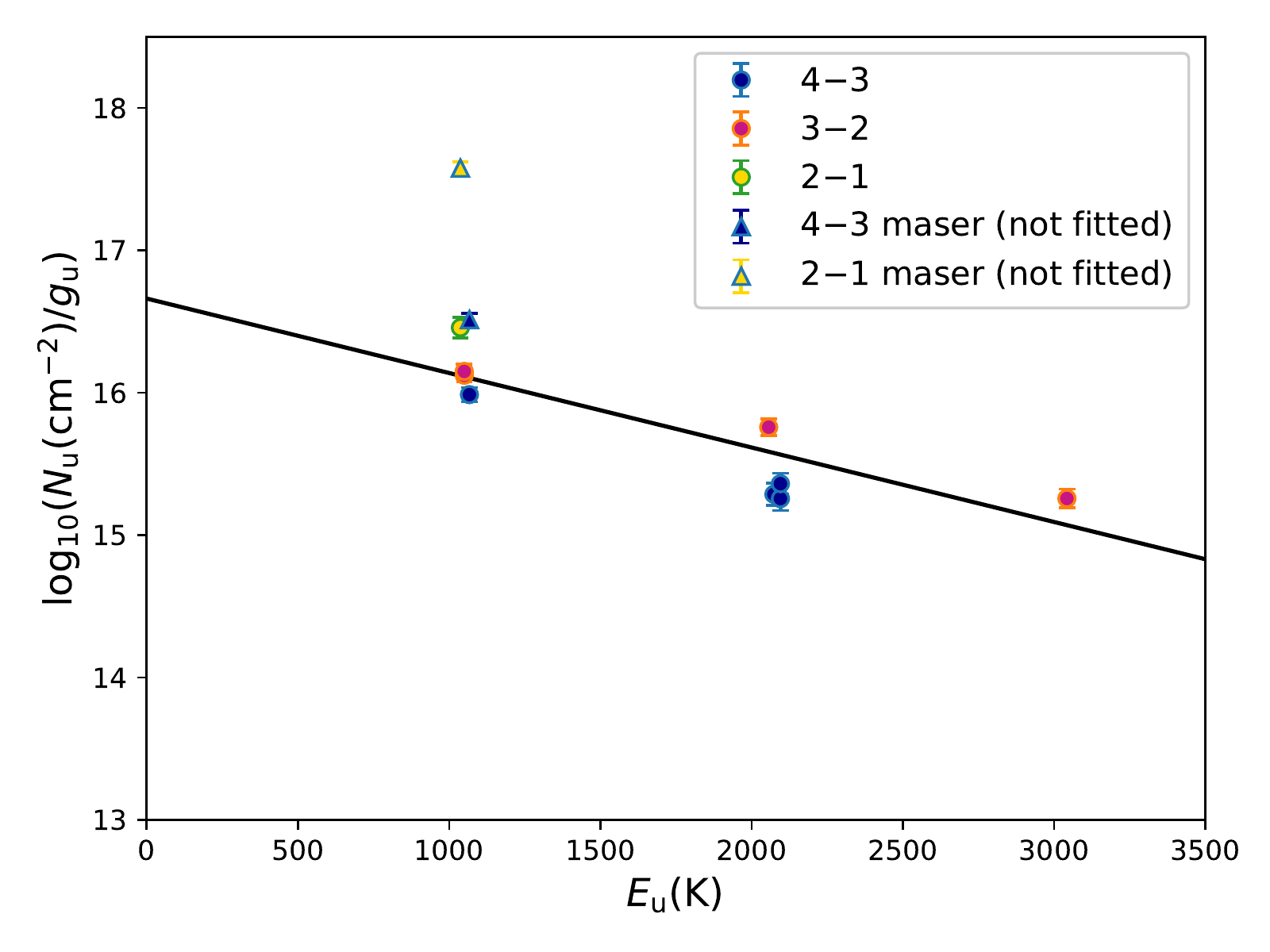}
\caption{Rotation diagram of CRL 3068. Description of the figure is same as the caption of Fig. \ref{fig:irc_rotdiag}.} \label{fig:crl-3068_rotdiag}%
\end{figure}

The vibrationally excited HCN lines detected in our observations occupy energy levels above $1000$\,K (see Table~\ref{tab:lines}). Their spectra, including those of the thermal lines, cover narrow velocity ranges (see Table \ref{tab:vib_hcn_results_updated}), suggesting that they most likely arise from a hot region where the stellar outflow has not reached its terminal velocity \citep[i.e. close to the stellar surface;][]{2018AA...613A..49M}. In order to better constrain the properties of the region traced by these transitions, we take advantage of the population diagram method to estimate the HCN excitation temperature and column density. Assuming LTE conditions and low opacities for the thermal emission of vibrationally excited HCN, we can use the standard formula to derive HCN column density and excitation temperature \citep[e.g.,][]{1999ApJ...517..209G}:
  \begin{ceqn}
     \begin{align}
         N_{\rm u}/g_{\rm u} = \frac{N_{\rm tot}}{Q(T_{\rm e})} e^{-E_{\rm u}/T_{\rm e}} = \frac{3k\int T_{\rm mb}{\rm d}V}{8\pi^3\nu\mu^2 S_{\rm ul}},
     \end{align}
  \end{ceqn}
where $N_\text{tot}$ is the HCN column density, $Q(T_{\rm e}$ is the partition function at the excitation temperature $T_{\rm e}$,  $\int T_{\rm mb} {\rm d}V$ is the velocity-integrated intensity, $k$ is the Boltzmann constant, $\nu$ is the transition's rest frequency, $\mu$ is the permanent dipole moment of HCN, and $S_{\rm ul}$ is the line strength. We obtained the values of $Q$, $\nu$, and $S_{\rm ul} \mu^2$ from CDMS \citep{2005JMoSt.742..215M} via Splatalogue\footnote{\url{https://splatalogue.online/advanced.php}} (see also Table~\ref{tab:lines}). %JPL \citep{1998JQSRT..60..883P}. 
Because the emitting size of vibrationally excited HCN lines should be small, we should correct the observed integrated intensities by dividing by the beam dilution factor, $\displaystyle \frac{\theta_\text{s}^2}{\theta_\text{s}^2 + \theta_\text{b}^2}$, where $\theta_\text{s}$ and $\theta_\text{b}$ are the source size and FWHM beam size, respectively. However, the source size is still not well constrained. The eSMA observations suggest that the vibrationally excited HCN $J=3$--$2$ (0, 1$^{1e}$, 0) emission can be as extended as 1\arcsec\,in IRC +10216 \citep[see Fig. 7 in][]{shinnaga2009} but with double peaks. Our assumption has been to adopt a Gaussian FWHM source size of $\theta_\text{s}=0.2$\arcsec\,($\sim$28 au) toward IRC +10216. Assuming the physical size to be the same as that of IRC +10216, we can estimate the source sizes for the other sources according to their distances in Table~\ref{tab:sample}. The assumed angular source sizes are given in Table~\ref{tab:rotdiag_results}.

\begin{table}[!htb]

%\centering
\caption{Column densities and excitation temperatures derived from the population diagram method.}
\label{tab:rotdiag_results}
\begin{tabular}{cccc}
\hline
\hline
Source & $\theta_{\rm s}$ & $T_\text{e}$ & $N_\text{tot}$\\  
       & (\arcsec)        & (K)       & (cm$^{-2}$)        \\
(1)    & (2)              &  (3)      &  (4)               \\
\hline
%IRC +10216 & 0.20 &  751 $\pm$ 195  & (1.0 $\pm$ 0.9) $\times$ $10^{19}$\\
%CRL 3068   & 0.02 &  735 $\pm$ 127 & (9.5 $\pm$ 3.7) $\times$ $10^{19}$\\
IRC +10216 & 0.20 &  946 $\pm$ 282  & (1.2 $\pm$ 0.9) $\times$ $10^{19}$\\
CRL 3068   & 0.02 &  830 $\pm$ 163 & (10.8 $\pm$ 4.3) $\times$ $10^{19}$\\
II Lup     & 0.04 &  700           & (1.7$\pm$0.1)$\times 10^{19}$ \\
W Ori      & 0.06 &  700           & (1.8$\pm$0.1)$\times 10^{18}$\\
S Aur      & 0.02 &  700           & (3.3$\pm$0.4)$\times 10^{19}$\\
RAFGL 4211 & 0.03 &  700           & (5.2$\pm$0.1)$\times 10^{19}$\\ 
\hline
\end{tabular}
\tablefoot{(1) Source name. (2) Assumed source size. (3) Excitation temperature. The excitation temperature is fitted for IRC +10216 and CRL 3068, while the excitation temperature assumed to be 700 K for the other sources. (4) Source-averaged HCN column density.}
\end{table}

In order to fit the population diagrams, at least two transitions covering a large enough energy range are needed to meaningfully derive excitation temperature and column density. Because a large number of vibrationally excited HCN transitions are found to show maser action for which LTE conditions do not hold (see Sect.~\ref{sec:maser}), linear least-square fits can only be applied to two sources, IRC +10216 and CRL 3068. In the case of IRC +10216, we included lines which are blended with each other, (0, 3$^{3e}$, 0) and (0, 3$^{3f}$, 0). Assuming LTE conditions, these two lines are expected to show nearly identical integrated intensities. We therefore use half the intensities integrated over the velocity range covered by the two lines for our analysis. Excluding the maser lines and other blended lines, we performed the linear least-square fits to the two sources. The results are shown in Figs.~\ref{fig:irc_rotdiag}--\ref{fig:crl-3068_rotdiag}. In each of these plots, we display, without fitting, the data points of two masers, $J=2$--$1$ (0, 1$^{1e}$, 0) and $J=4$--$3$ (0, 1$^{1f}$, 0). The $J=2$--$1$ (0, 1$^{1e}$, 0) maser lies above the fitted lines in both plots, indicating the highly non-thermal nature of its emission. The $J=4$--$3$ (0, 1$^{1f}$, 0) maser is weaker than the $J=2$--$1$ (0, 1$^{1e}$, 0) maser because its maser component only contributes a small fraction to the total integrated intensity of the transition (see Fig.~\ref{fig:irc_all_spec} and Sect.~\ref{sec:maser} for more information). The fitted excitation temperatures and HCN column densities from the population diagrams are given in Table~\ref{tab:rotdiag_results}.

In the range of $\theta_{\rm s}$=0.1\arcsec--1\arcsec, we note that the derived HCN column densities are largely dependent on the assumed source size with a relationship of $N_{\rm tot}\propto 1/\theta_{\rm s}^{2}$, that is, the derived HCN column densities can vary by at most a factor of 100. On the other hand, the excitation temperature is not sensitive to the adopted value of $\theta_{\rm s}$. %The derived excitation temperatures of the two sources are large, possibly, due to the nature of the detected vibrationally excited lines. Some thermal lines may show maser actions which cannot be clearly identified by their line profiles. Our lines, evidently, also trace different regions in the envelope, and thus constraining them in one fit would produce higher physical parameters (see Fig. 2 of \citealt{cerni..2011A&A...529L...3C}). 
The excitation temperatures of the two sources are consistent, within uncertainties, with the $T_{\rm ex}$ range of 452--753 K as derived from \emph{Herschel} observations of 8 other carbon-rich stars \citep{2018AA..Nicolaes...618A.143N} and the $T_{\rm ex}$ values of IRC +10216 ($T_{\rm ex}$=1000~K, \citealt{2000ApJ...528L..37S}; $T_{\rm ex}$=410--2465~K,  \citealt{cerni..2011A&A...529L...3C}). We also note that \citet{cerni..2011A&A...529L...3C} defined three different zones in their population diagram of IRC +10216, based on the upper energy levels of HCN lines. The $T_{\rm ex}$ value in IRC +10216 derived by our single-component fitting is higher than that of Zone III ($410{\pm}60$~K; $E_{\rm u}{\le}2000$~K), but slightly lower than that of Zone II ($1240{\pm}210$~K; $2000{<}E_{\rm u}{\le}5000$~K), probably because we fit HCN lines from both zones. Some of our identified thermal lines might in fact show (weak) maser actions because their column densities tend to deviate from other transitions of similar energies on the population diagrams (Figs.~\ref{fig:irc_rotdiag}--\ref{fig:crl-3068_rotdiag}). However, they are not easily identified from their line profiles alone. This may lead to an overestimation of the HCN excitation temperature. 
Comparing the kinetic temperature model in IRC +10216 \citep[e.g., Fig. 1 in ][]{agundez..2012A&A...543A..48A}, we infer that the vibrationally excited HCN line-emitting regions have a size of $<$20$R_{*}$ (0$\rlap{.}$\arcsec4), where the stellar radius, $R_{*}$, is 22 mas \citep{2000ApJ...543..868M}. Our assumed source size of 0$\rlap{.}$\arcsec2 for IRC +10216 is consistent with this inferred size.

For the other four stars with vibrationally excited HCN lines but without a broad coverage of line excitation energy, we derive their HCN column densities by fixing the excitation temperature to be 700 K, which is a representative value among the results from \emph{Herschel} data \citep[452--753 K;][]{2018AA..Nicolaes...618A.143N} and our fitting. The associated uncertainties in column density are derived by Monte Carlo analysis with 10,000 simulations. A Gaussian function was used to fit the resulted distribution, and the standard deviation of the Gaussian distribution was regarded as the 1$\sigma$ error. The results are shown in Table~\ref{tab:rotdiag_results}. The derived source-averaged HCN column densities range from 0.18$-$10.8 $\times 10^{19}$~cm$^{-2}$. % The derived source-averaged HCN column densities have quite a large range of (1.7$-$5.2)$\times 10^{19}$~cm$^{-2}$. 
Among these sources, W Ori has the lowest source-averaged column density, likely due to its lowest mass-loss rate of ${\sim}$3.1$\times 10^{-7}$. We note that the uncertainties in the adopted source sizes can lead to even larger uncertainties than those shown in Table~\ref{tab:rotdiag_results}.

Based on the %H$_{2}$ number density 
model of IRC +10216 \citep[e.g., Fig. 1 in ][]{agundez..2012A&A...543A..48A}, the H$_{2}$ number density is about 2$\times 10^{8}$~cm$^{-3}$ at the radius of 0.2\arcsec (i.e., 4$\times 10^{14}$~cm), corresponding to a H$_{2}$ column density of 1.6$\times$10$^{23}$~cm$^{-2}$. The HCN abundance with respect to H$_{2}$ is thus determined to be $(7.4\pm 5.5)\times 10^{-5}$. The HCN abundance is roughly consistent with previous estimate of HCN abundances in carbon-rich stars \citep{fonfria2008detailed,2013A&A...550A..78S}, but seems to be much higher than the prediction ($<3\times 10^{-6}$ at $r>$5R$_{*}$) in non-equilibrium chemical model \citep{2006A&A...456.1001C}. However, we also note that the derived HCN abundance largely relies on the assumptions, such as the emission source size, which can result in large uncertainties in the estimated HCN column density and hence its abundance.

\subsection{Pumping mechanism of HCN masers}\label{sec:pumping}
\begin{figure*}[!htbp]
\centering
\includegraphics[width=0.9\textwidth]{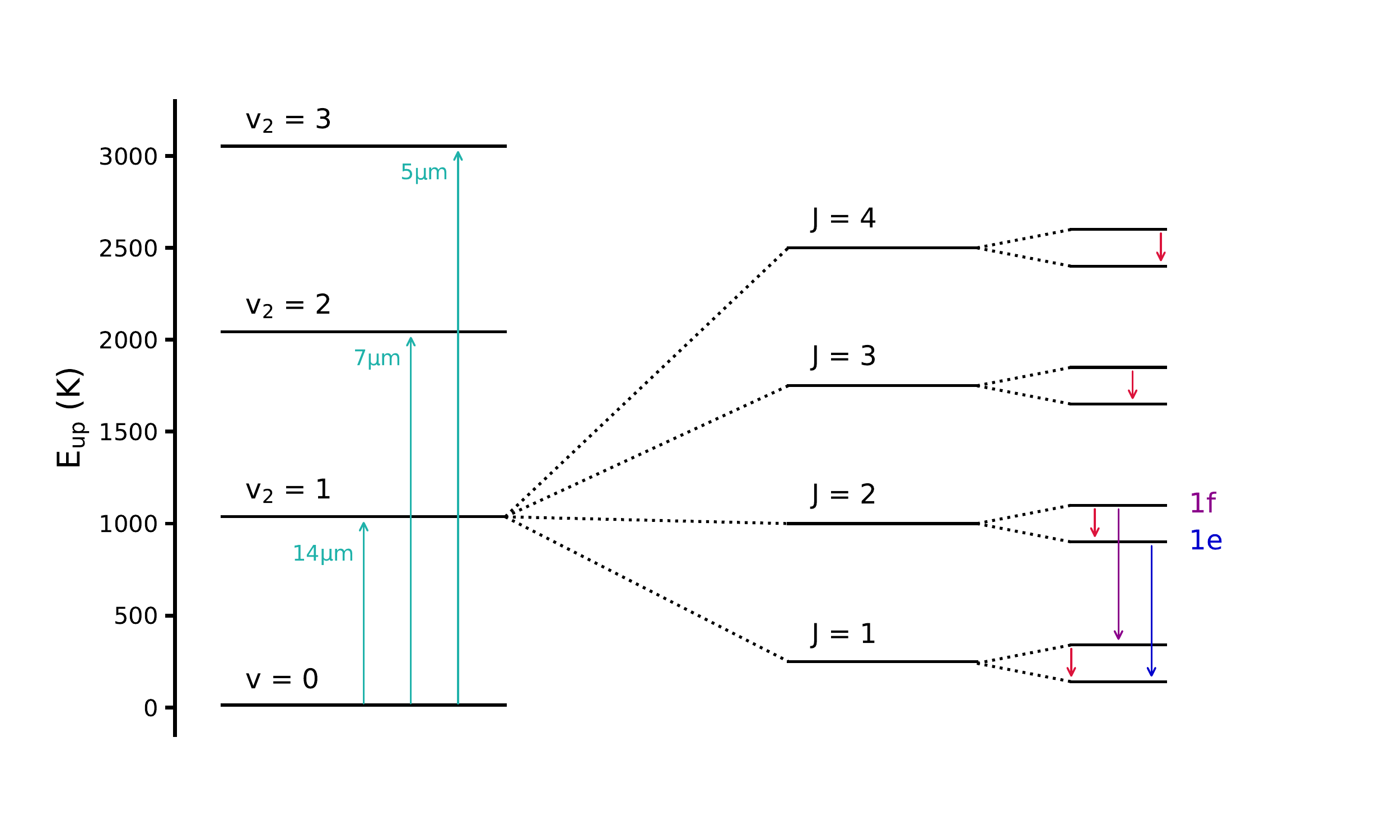}
\caption{Energy diagram of HCN. The quantum numbers, $v$ and $J$, indicate the vibrational state and rotational state, respectively. $l$-type doubling causes the splitting of every rotational level into two sub-levels, 1e and 1f, which are also denoted in this figure. Pumping photons from the ground state to highly vibrationally excited states are indicated by cyan arrows. The $J=2$--$1$ (0, 1$^{1e}$, 0) and $J=2$--$1$ (0, 1$^{1f}$, 0) transitions are indicated by the blue and purple arrows, respectively. The direct $l$-type transitions are marked in red arrows. We note that the separations of the $l$-type doublets are not drawn to scale.}\label{fig:energy_diag}%
\end{figure*}

Vibrationally excited HCN masers are generally thought to be explained by pumping through infrared radiations \citep[e.g.,][]{lucas1989cerni..discovery,2018AA...613A..49M}. Our detected HCN masers are found in the excited states of the bending mode: $\nu_{2}=1^{1e}$ or $1^{1f}$, $\nu_{2}=2$, and $\nu_{2}=3^{1e}$ (Sect.~\ref{sec:maser}). As illustrated in the energy level diagram in Fig. \ref{fig:energy_diag}, pumping HCN molecules from the ground state to these excited states requires infrared photons at 14~$\mu$m, 7~$\mu$m, and 5~$\mu$m, respectively (see Fig. \ref{fig:energy_diag}). Following \citet{2018AA...613A..49M}, we extended the comparison between the isotropic maser luminosities, $L_{\rm M}$, and the corresponding infrared luminosities, $L_{\rm IR}$, toward all detected HCN masers as shown in Table~\ref{tab:ir_maser_lumin}. Their luminosity ratios, $L_{\rm IR}/L_{\rm M}$, should be more reliable than their luminosities because the ratios are independent of the assumed distances. We find that the luminosity ratios are much greater than 50 for all the detected HCN masers. The infrared flux densities are based on the WISE and IRAS measurements at 4.6~$\mu$m and 12~$\mu$m \citep{2015A&C....10...99A}, but not at the exact wavelengths of HCN vibrational excitation of 14~$\mu$m, 7~$\mu$m, and 5~$\mu$m. While this adds to the uncertainties in the derived luminosity ratios $L_{\rm IR}/L_{\rm M}$, \citet{2018AA...613A..49M} suggest that the flux densities at 7~$\mu$m and 14~$\mu$m are within a factor of 2 of the 12~$\mu$m flux density based on the SED toward IRC +10216. Hence, our conclusion is still robust because the luminosity ratios are much higher than 2. 

The fact that the amount of IR pumping photons is much more than adequate to account for HCN masers indicates that all these masers are unsaturated. The unsaturated nature of detected HCN masers is also supported by their brightness temperature estimates. Assuming a typical maser emission size of 28 au (${\sim}0\rlap{.}$\arcsec2 for IRC +10216; Sect.~\ref{sec:pm}), the brightness temperatures of these masers are roughly 1000--300\,000~K. The map of the HCN $J=3$--$2$ (0, 1$^{1e}$, 0) emission indicates that the maser can have a smaller emission size of $<$0$\rlap{.}$\arcsec2\,toward ~IRC+10216 \citep{shinnaga2009}, which would lead to a higher estimate in brightness temperature. Accounting for the fact that the brightness temperature increases quadratically with decreasing source size and that the emission size may vary in other sources or transitions, the maser brightness temperatures are still too low to reach their saturation levels even if we assume a source size that is 10 times smaller. Unsaturated masers tend to vary on much shorter timescales than saturated masers due to their exponential amplification behavior. This may explain the observed temporal variability on a relative short timescale of a few years (Sect.~\ref{sec:variability}).

Most of the bright HCN maser components appear in the blueshifted side of their spectra with respect to their systemic LSR velocities. This may be explained by having the maser-emitting regions in the front of the stars, assuming an expanding outflow. The observed blueshifted maser may arise from the \emph{radial} amplification of background photospheric emission. This scenario has been invoked to account for the observed blueshifted SiS (1--0) maser in IRC +10216 \citep{sis..Gong_2017}. Stellar occultation may cause maser emission arising from the far side of the star to be partially blocked. Alternatively, the blueshift velocities of certain narrow maser profiles are not really significant and the maser spikes appear to peak very close to the systemic velocity. Prominent examples include RAFGL 4211, W Ori, and S Aur. These masers may be better interpreted as \emph{tangential} amplification where the emission region is on the plane of sky beside the central star. This scenario requires an outward-accelerating velocity field and has been proposed to explain the detection of strong SiO masers near the systemic velocity in oxygen-rich stars \citep{1981A&A...102...65B}.

\begin{figure*}[!htbp]
\centering
\includegraphics[width=0.34\textwidth]{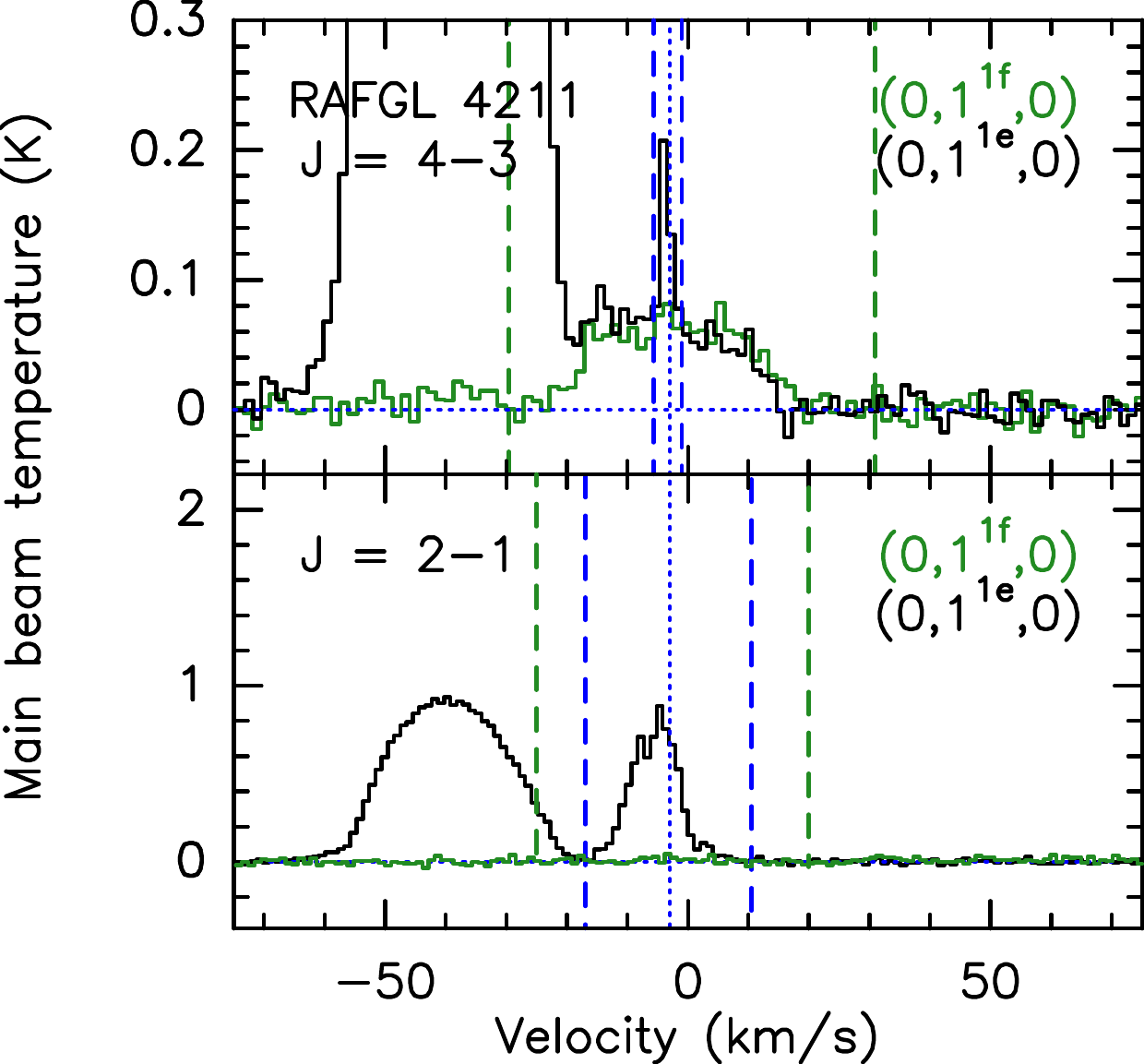}\hfill
\includegraphics[width=0.31\textwidth]{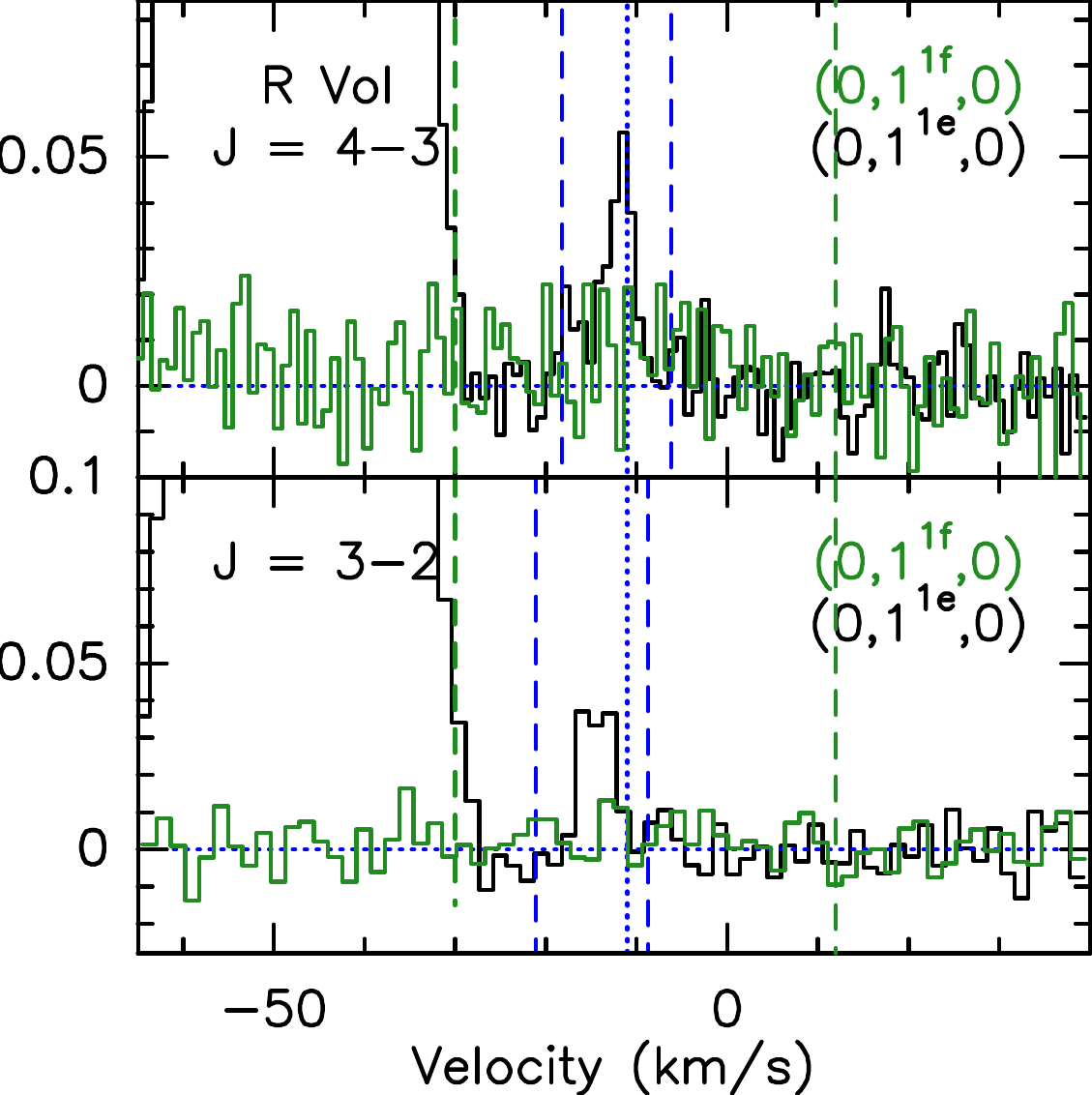}\hfill
\includegraphics[width=0.31\textwidth]{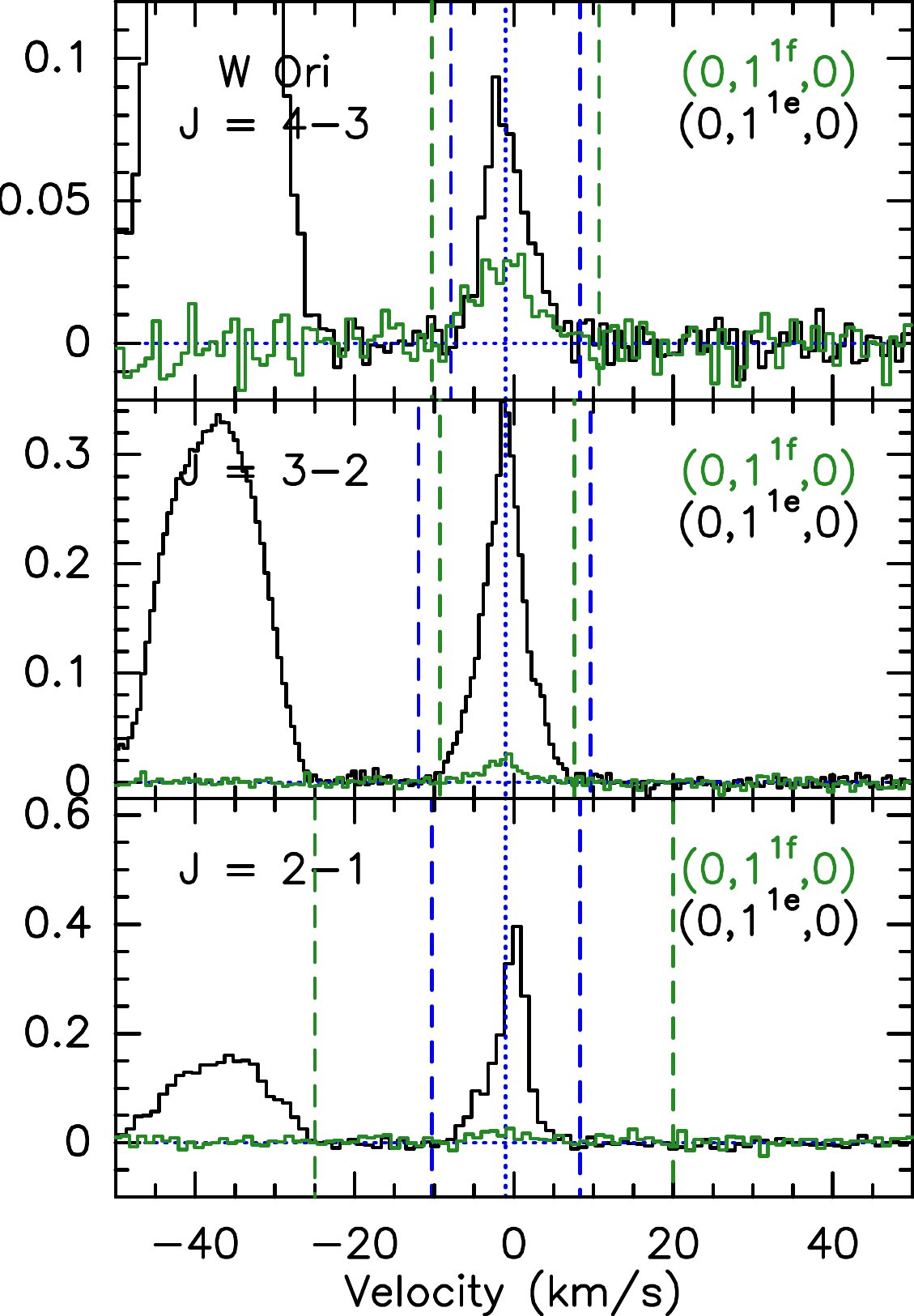}\hfill
\caption{Overlaid spectra for HCN (0, 1$^{1e}$, 0) (in black) and (0, 1$^{1f}$, 0) (in green) transitions in different rotational transitions toward RAFGL 4211 (\emph{left}), R Vol (\emph{middle}), and W Ori (\emph{right}). These sources show brighter $J=4$--$3$ (0, 1$^{1e}$, 0) emission than $J=4$--$3$ (0, 1$^{1f}$, 0). The vertical lines are labelled in the same way as Fig.~\ref{fig:irc_all_spec}.}
\label{fig:comparison_1e1f}
\end{figure*}

\begin{figure*}[t]%[!htbp]
\centering
\includegraphics[width=0.34\textwidth]{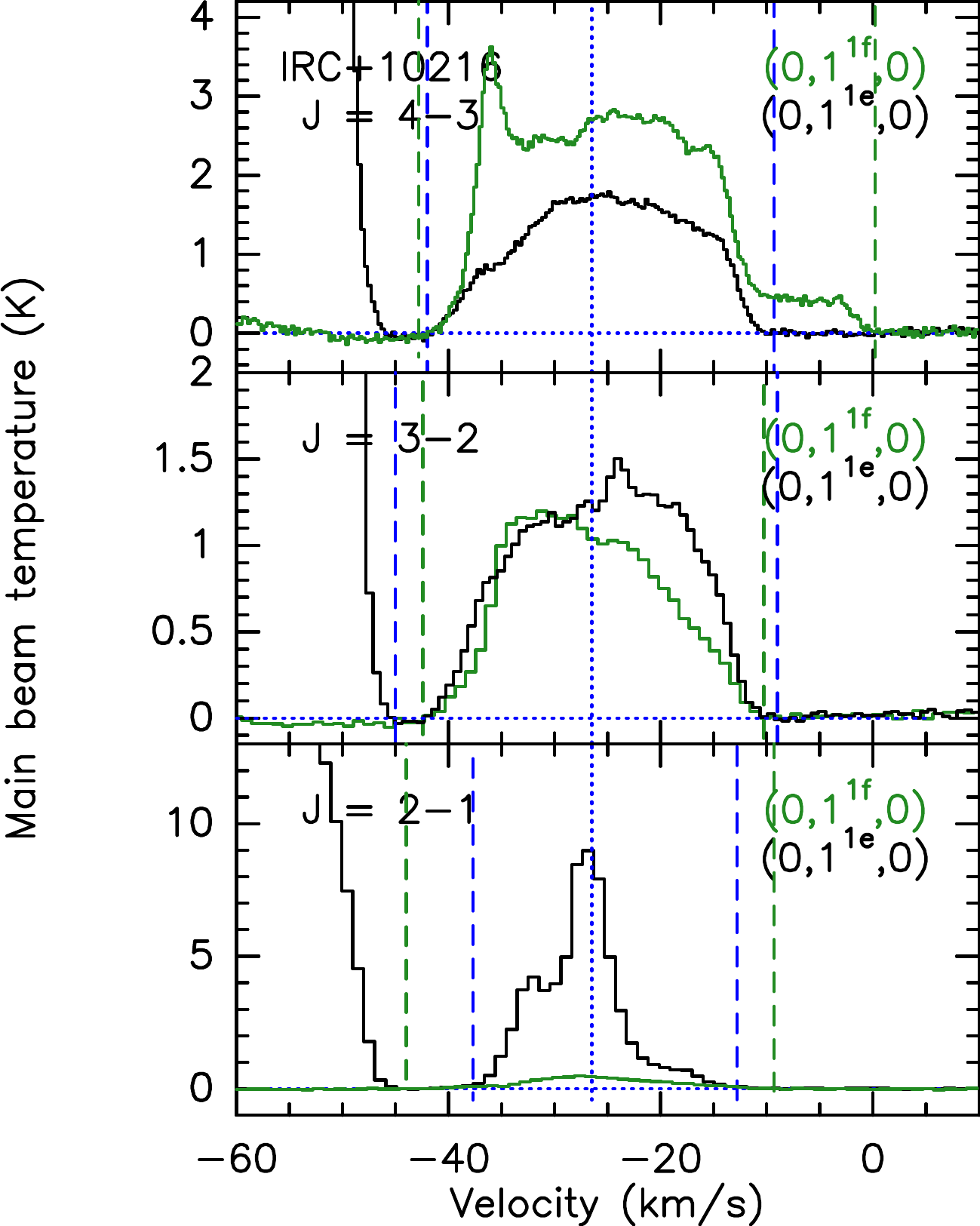}\hfill
\includegraphics[width=0.3\textwidth]{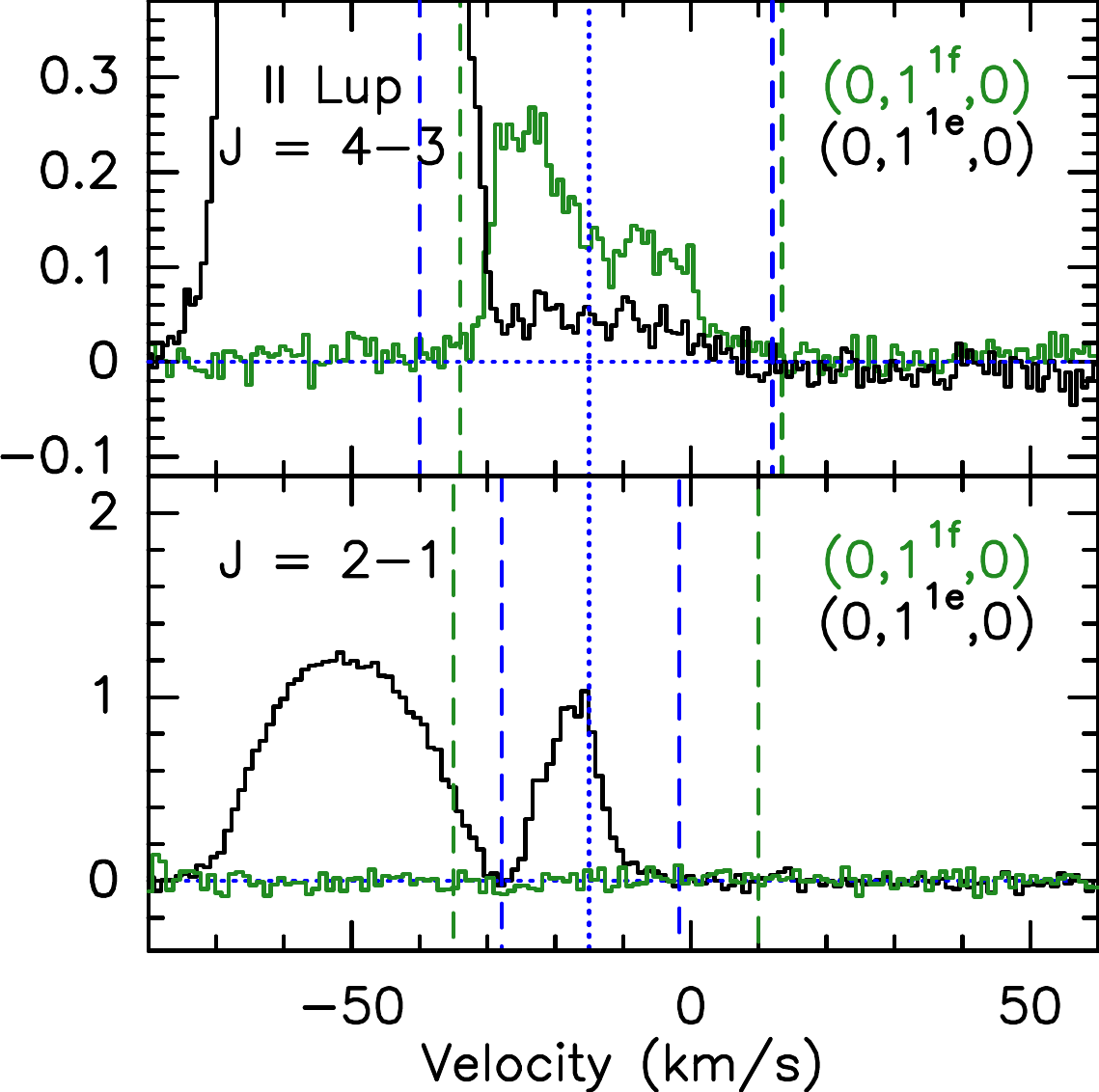}\hfill
\includegraphics[width=0.29\textwidth]{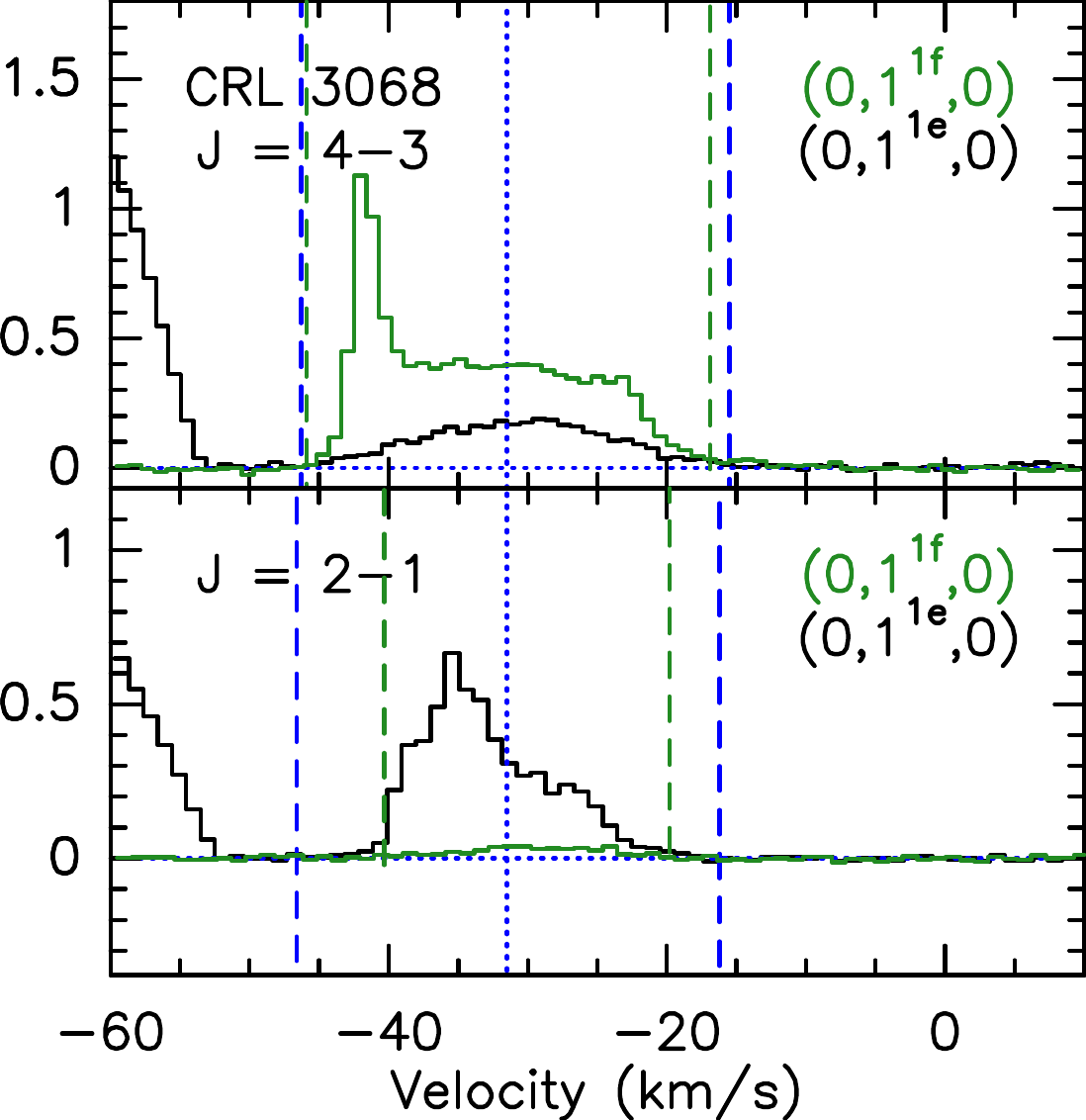}\hfill
\caption{Same as Fig.~\ref{fig:comparison_1e1f} toward IRC+10216 (\emph{left}), II Lup (\emph{middle}), and CRL 3068 (\emph{right}). These sources show brighter $J=4$--$3$ (0, 1$^{1f}$, 0) emission than $J=4$--$3$ (0, 1$^{1e}$, 0).}
\label{fig:comparison_1f1e}
\end{figure*}

The $J=2$--$1$ (0, 1$^{1e}$, 0) transition is the brightest HCN maser in our observations (see Sect.~\ref{sec:maser}). This could arise from the fact that the Einstein $A$ coefficient strongly depends on the rotational quantum number $J$. HCN molecules rapidly decay to the lower $J$ levels, and are likely to accumulate at the lowest $J$ levels. Direct $l$-type transitions have been detected in emission toward different carbon-rich stars \citep[e.g.][]{2003ApJ...586..338T,cerni..2011A&A...529L...3C}. Emission from direct $l$-type transitions can cause molecules to cascade down from the (0, $1^{1f}$, 0) level to (0, $1^{1e}$, 0) level. Furthermore, the rotational transitions in the $\nu_{2}$=1 state do not change the parity due to the selection rule. The net effects result in more HCN molecules in the (0, $1^{1e}$, 0) state than in the (0, $1^{1f}$, 0) state. The gain of masers is directly proportional to molecular column densities at their corresponding energy levels, which makes the $J=2$--$1$ (0, 1$^{1e}$, 0) transition the brightest HCN masers in the $\nu_{2}$=1 state. This may explain why HCN masers in the (0, $1^{1e}$, 0) state are typically stronger or more readily detectable than the (0, $1^{1f}$, 0) state, as shown in Figs.~\ref{fig:comparison_1e1f}, \ref{fig:comparison_1f1e}, and \ref{fig:comparison_1e1f_appendix}, which compare the (0, 1$^{1e}$, 0) and (0, 1$^{1f}$, 0) spectra in the same rotation transition at different $J$ levels and in different sources. 

The above pumping scenario is only consistent with our results for the $J=2$--$1$ and $J=3$--$2$ masers. For the $J=4$--$3$ transition, it is also consistent with the results in RAFGL 4211, R Vol, W Ori (Fig.~\ref{fig:comparison_1e1f}), and perhaps RV Aqr (Fig.~\ref{fig:comparison_1e1f_appendix}), but fails to explain the fact that (0, 1$^{1f}$, 0) maser is brighter than the (0, 1$^{1e}$, 0) maser in IRC +10216, II Lup, and CRL 3068 (Fig.~\ref{fig:comparison_1f1e}). There are two common features in these sources showing brighter (0, 1$^{1f}$, 0) maser in $J=4$--$3$. First, the peak velocities of these $J=4$--$3$ (0, 1$^{1f}$, 0) masers are all blueshifted by about 10\,{\kms} from their systemic LSR velocities, regardless of their discrepant ground-state or excited-state line widths (Tables~\ref{tab:ground_hcn_results_shell} and \ref{tab:vib_hcn_results_updated}). The peak velocities are different from those of the $J=2$--$1$ (0, 1$^{1e}$, 0) masers of the same sources or the $J=4$--$3$ (0, 1$^{1e}$, 0) masers of different sources, indicating a different pumping mechanism of this maser in these three sources. Second, all three sources show a broad emission profile in (0, 1$^{1f}$, 0), which is roughly a factor of ${\sim}2$ brighter than the corresponding, apparently thermal (0, 1$^{1e}$, 0) profile. This indicates weak maser activities in the (0, 1$^{1f}$, 0) line. The $J=4$--$3$ (0, 1$^{1f}$, 0) masers may stem from (a combination of) infrared line overlap, anisotropic infrared radiation fields, or non-negligible collisional pumping. Observations have shown a large amount of rovibrational HCN and C$_{2}$H$_{2}$ transitions at $\sim$14~$\mu$m, and a large fraction of them are overlapped \citep{2008ApJ...673..445F}. The rovibrational HCN and C$_{2}$H$_{2}$ lines at 14~$\mu$m could partially contribute to the pumping of HCN masers in the $\nu_{2}$=1 state. Such infrared line overlap has been invoked to explain the observed high-$J$ SiS masers \citep{2006ApJ...646L.127F}. Because the HCN and C$_{2}$H$_{2}$ lines are displaced in frequency, the overlap should give rise to the line asymmetries in the maser profiles. Furthermore, such a mechanism would lead to the presence of masers at nearly the same velocity shift with respect to their systemic velocities for different sources. This is in agreement with our observations where the most prominent component of the $J=4$--$3$ (0, 1$^{1f}$, 0) maser in these three sources peaks at about 10~\kms\, blueshifted from the respective systemic velocities. Alternatively, the asymmetries in the maser profiles can arise from a highly anisotropic infrared radiation field, caused by substructures in the circumstellar envelope such as spirals or broken shells. Anisotropic infrared radiation field has been indicated by the non-spherical infrared morphology toward II Lup \citep{2018MNRAS.480.1006L}. Thirdly, collisional pumping might not be negligible, although the vibrationally excited transitions are thought to be dominated by radiative pumping. As pointed out in previous studies \citep{2014MNRAS.440..172S,2018AA...613A..49M}, the time variability does not appear to correlate with stellar cycles, indicating that collisional pumping may also contribute to part of the population inversion. This may explain the second feature where the $J=4$--$3$ (0, 1$^{1f}$, 0) line exhibits weak maser over a broad velocity range in these three sources. Inhomogeneous density structures driven by stellar pulsating shocks may result in the diverse line shapes among the HCN masers.

The HCN $J=3$--$2$ (0, 2, 0) and $J=3$--$2$ (0, 3$^{1e}$, 0) masers are only detected in RAFGL 4211 among carbon stars as of today (Table~\ref{tab:catalogue}). The uniqueness of these two masers could stem from the peculiar geometry of RAFGL 4211. For instance, the relatively long coherent length of maser amplification along the line of sight may result in a uniquely high gain for this maser. Both $J=2$--$1$ and $J=4$--$3$ (0, 1$^{1e}$, 0) masers are detected, with the latter showing a narrow spike on top of a broad component. The higher-$J$ maser lines ($3-2$ and $4-3$) could partly arise from a region closer to the star, where the temperature and H$_{2}$ number density are higher than the region producing the $J=2$--$1$ (0, 1$^{1e}$, 0) maser.

\begin{table*}[!htb]
%\small
\centering
\caption{Comparison of detected HCN masers and corresponding infrared photon luminosites.}
\label{tab:ir_maser_lumin}
\begin{tabular}{ccccccc}
\hline
\hline
Source & $J$ & ($\nu_{1}$, $\nu_{2}$, $\nu_{3}$) & $S_{\rm IR}$ & $L_{\rm IR}$ & $L_{\rm M}$     & $L_{\rm IR}$/$L_{\rm M}$ \\  
       &     &     &(Jy)         & (s$^{-1}$)    & (s$^{-1}$)&  \\
(1)    & (2) & (3) & (4)          &  (5)      &  (6)              & (7) \\
\hline
\multicolumn{7}{c}{IR=12~$\mu$m v.s. 14~$\mu$m} \\
\hline
IRC +10216 & 2--1 & (0, 1$^{1e}$, 0) &4.75$\times 10^{4}$ & 1.4$\times 10^{46}$ & 2.6$\times 10^{43}$ & 531  \\
           & 2--1 & (0, 1$^{1f}$, 0) &4.75$\times 10^{4}$ & 1.9$\times 10^{46}$ & 2.9$\times 10^{42}$ & 6745 \\
           & 3--2 & (0, 1$^{1e}$, 0) &4.75$\times 10^{4}$ & 1.8$\times 10^{46}$ & 9.3$\times 10^{42}$ & 1950 \\
           & 4--3 & (0, 1$^{1f}$, 0) &4.75$\times 10^{4}$ & 2.4$\times 10^{46}$ & 2.8$\times 10^{43}$ & 874  \\
RAFGL 4211 & 2--1 & (0, 1$^{1e}$, 0) &7.93$\times 10^{2}$ & 1.2$\times 10^{46}$ & 1.4$\times 10^{44}$ & 85   \\
           & 4--3 & (0, 1$^{1e}$, 0) &7.93$\times 10^{2}$ & 2.0$\times 10^{45}$ & 1.1$\times 10^{43}$ & 181  \\
II Lup     & 2--1 & (0, 1$^{1e}$, 0) &1.32$\times 10^{3}$ & 8.5$\times 10^{45}$ & 8.0$\times 10^{43}$ & 107  \\
           & 4--3 & (0, 1$^{1f}$, 0) &1.32$\times 10^{3}$ & 1.2$\times 10^{46}$ & 4.3$\times 10^{43}$ & 282  \\
R Vol      & 2--1 & (0, 1$^{1e}$, 0) &2.03$\times 10^{2}$ & 2.0$\times 10^{45}$ & 1.3$\times 10^{43}$ & 151  \\
           & 2--1 & (0, 1$^{1f}$, 0) &2.03$\times 10^{2}$ & 3.0$\times 10^{45}$ & 2.9$\times 10^{42}$ & 1052 \\
           & 3--2 & (0, 1$^{1e}$, 0) &2.03$\times 10^{2}$ & 1.2$\times 10^{45}$ & 3.9$\times 10^{42}$ & 303  \\
           & 4--3 & (0, 1$^{1e}$, 0) &2.03$\times 10^{2}$ & 1.1$\times 10^{45}$ & 3.9$\times 10^{42}$ & 293  \\
AQ Sgr     & 3--2 & (0, 1$^{1e}$, 0) &56.6                & 9.9$\times 10^{43}$ & 2.0$\times 10^{41}$ & 506  \\
CRL 3068   & 2--1 & (0, 1$^{1e}$, 0) &7.07$\times 10^{2}$ & 2.2$\times 10^{46}$ & 2.0$\times 10^{44}$ & 111  \\
           & 4--3 & (0, 1$^{1f}$, 0) &7.07$\times 10^{2}$ & 2.1$\times 10^{46}$ & 3.5$\times 10^{44}$ & 60   \\
V636 Mon   & 2--1 & (0, 1$^{1e}$, 0) &1.25$\times 10^{2}$ & 5.8$\times 10^{44}$ & 3.1$\times 10^{42}$ & 188  \\
IRC +20370 & 2--1 & (0, 1$^{1e}$, 0) &5.34$\times 10^{2}$ & 5.6$\times 10^{45}$ & 3.3$\times 10^{43}$ & 172  \\
           & 3--2 & (0, 1$^{1e}$, 0) &5.34$\times 10^{2}$ & 7.4$\times 10^{45}$ & 2.5$\times 10^{43}$ & 299  \\
W Ori      & 2--1 & (0, 1$^{1e}$, 0) &1.84$\times 10^{2}$ & 1.0$\times 10^{44}$ & 1.8$\times 10^{42}$ & 56   \\
           & 3--2 & (0, 1$^{1e}$, 0) &1.84$\times 10^{2}$ & 1.2$\times 10^{44}$ & 2.0$\times 10^{42}$ & 59   \\
           & 4--3 & (0, 1$^{1e}$, 0) &1.84$\times 10^{2}$ & 8.7$\times 10^{43}$ & 5.1$\times 10^{41}$ & 169  \\
S Aur      & 2--1 & (0, 1$^{1e}$, 0) &1.62$\times 10^{2}$ & 1.9$\times 10^{44}$ & 2.6$\times 10^{42}$ & 73   \\
V Aql      & 3--2 & (0, 1$^{1e}$, 0) &1.50$\times 10^{2}$ & 2.0$\times 10^{44}$ & 8.7$\times 10^{41}$ & 228  \\
RV Aqr     & 2--1 & (0, 1$^{1e}$, 0) &3.08$\times 10^{2}$ & 1.1$\times 10^{45}$ & 1.0$\times 10^{43}$ & 107\\
           & 2--1 & (0, 1$^{1f}$, 0) &3.08$\times 10^{2}$ & 9.7$\times 10^{44}$ & 1.5$\times 10^{42}$ & 638  \\
           & 3--2 & (0, 1$^{1e}$, 0) &3.08$\times 10^{2}$ & 1.0$\times 10^{45}$ & 9.8$\times 10^{42}$ & 133  \\
\hline
\multicolumn{7}{c}{IR=4.6~$\mu$m v.s. 5~$\mu$m} \\
\hline
RAFGL 4211 & 3--2 & (0, 3$^{1e}$, 0) & 42.6 & 1.1$\times 10^{45}$ & 3.8$\times 10^{42}$ & 289 \\ 
\hline
\multicolumn{7}{c}{IR=12~$\mu$m v.s. 7~$\mu$m} \\
\hline
RAFGL 4211 & 3--2 & (0, 2, 0) & 7.93$\times 10^{2}$ & 4.1$\times 10^{45}$ & 7.8$\times 10^{42}$ & 533 \\ 
\hline
\end{tabular}
\tablefoot{(1) Source name. (2) Rotational levels. (3) Vibrational states. (4) Flux density at corresponding infrared wavelength. (5) Photon rates at a wavelength range according to the velocity range given by maser line (see Table~\ref{tab:vib_hcn_results_updated}). (6) Photon rates in the corresponding maser line. (7) Ratios between the infrared photon and HCN maser luminosites.}
\end{table*}
\normalsize

The physical conditions to excite the detected HCN masers can be roughly constrained. Based on the population diagram analysis (see Sect.~\ref{sec:pm}), the excitation temperature is found to be around 800$-$900~K for the vibrationally excited HCN emitting regions. These masers have to be formed in these regions that should have kinetic temperature of ${\gtrsim}700$\,K. Furthermore, the velocity widths of the maser profiles indicate that the masers form in the wind acceleration zones. Comparing with the physical model in IRC +10216 \citep[e.g.,][]{agundez..2012A&A...543A..48A}, the expected gas temperature and velocity correspond to a H$_{2}$ number density of ${>}10^{8}$~cm$^{3}$. This gives a loose constraint on the physical conditions.

\subsection{Comparison with masers in oxygen-rich stars}
OH, SiO, and H$_{2}$O masers are common in oxygen-rich stars. SiO masers are formed in the dust formation zone, OH masers are formed farther out, and H$_{2}$O masers lie between \citep[e.g.,][]{1996A&ARv...7...97H}. As discussed above, HCN masers should be formed in the innermost regions (i.e., $r<$20$R_{*}$), similar to the formation regions of SiO masers in oxygen-rich stars. Furthermore, both HCN and SiO are linear molecules that show maser actions in their ground vibrational states toward respective classes of stars (see Table~\ref{tab:catalogue} for HCN ground-state masers toward carbon-rich stars, see \citealt{2004ApJ...608..480B,2021AJ....161..111D} for SiO ground-state masers toward oxygen-rich stars). High excitation masers with lower energy levels above 4000 K are detected in both SiO and HCN, with SiO masers in vibrational states up to $v$=4 was detected in VY CMa \citep{1993ApJ...407L..33C} and up to $v$=7 was detected in Betelgeuse  \citep{2018A&A...609A..67K}; and HCN masers in the (0, 4$^{0}$, 0) state were reported toward IRC +10216, CIT 6, and Y CVn \citep{schilke..mehringer..menten2000ApJ...528L..37S,schilke..menten2003ApJ...583..446S}. These similar properties suggest that HCN masers in carbon-rich stars may be an analogue of SiO masers in oxygen-rich stars. If confirmed, HCN masers can be regarded as a powerful tool to measure the accurate proper motions and parallaxes of carbon-rich stars as performed toward oxygen-rich stars with SiO masers \citep[e.g.,][]{2012ApJ...744...23Z}, especially toward the high mass-loss carbon-rich stars, which cannot be determined by optical and infrared telescopes due to their high extinctions. VLBI measurements of HCN masers might be the only way to determine their proper motions and parallaxes. 

%\section{Stellar phase estimates}

%We determined the infrared phases of the sample of stars which are presented in the Table \ref{tab:phase}. We used the AAVSO website\footnote{\url{https://www.aavso.org/}} to determine the phases for most of the stars observed with the APEX telescope on various days in 2018. The IR maximum dates and periods, \textit{$t_{0}$}, for 7 stars were not found in the literature. 
\section{Summary and outlook}\label{sec:summary}
We performed APEX observations of multiple vibrationally excited HCN transitions toward a sample of 16 carbon-rich stars. The main results are summarised as follows:

\begin{itemize}
    \item[1.] Our observations resulted in the detection of 68 vibrationally excited HCN transitions from 13 stars, suggesting that vibrationally excited HCN lines are ubiquitous in carbon-rich stars. Among the detected transitions, 39 of them tend to be thermal, while 29 of them show maser actions. Our observations extend the spectrally resolved detection of the vibrationally excited thermal HCN transitions in carbon-rich stars beyond IRC +10216. Among the detected masers, 23 are new detection. The results expand the total number of HCN masers toward carbon-rich stars by 47\%. The masers arise from 12 different transitions, and the $J=3$--$2$ (0, 3$^{1e}$, 0), $J=3$--$2$ (0, 2, 0), $J=4$--$3$ (0, 1$^{1f}$, 0) masers are detected in the interstellar medium for the first time. 
    
    \item[2.] Temporal variations of the $J=2$--$1$ (0, 1$^{1e}$, 0) maser on a timescale of a few years have been confirmed by the dramatic change of maser line profiles. However, the cause of the time variability of HCN masers is still unclear. 
    
    \item[3.] Population diagrams of vibrationally excited thermal HCN transitions suggest an excitation temperature of about 800~K for IRC+10216 and CRL 3068, consistent with the previous studies. Assuming a typical size of 28 au for the regions showing the vibrationally excited HCN transitions, we derived the source-average HCN column densities of (0.18$-$10.8)$\times 10^{19}$~cm$^{-2}$ for six carbon-rich stars. The corresponding HCN abundance with respect to H$_{2}$ is roughly ${\sim}7.4{\pm} 5.5 \times 10^{-5}$.
    
    \item[4.] The ratios between HCN maser and corresponding infrared luminosities suggest that HCN masers are unsaturated. The unsaturated nature is supported by their temporal variation and relatively low brightness temperatures of these HCN masers. The spontaneous decay of rotational transitions and direct $l$-type transitions in the $\nu_{2}$=1 state may explain the fact that the $J=2$--$1$ (0, 1$^{1e}$, 0) transition is the brightest maser in the $\nu_{2}$=1 state. Comparing with the physical model of IRC+10216, we infer that a gas kinetic temperature of $\gtrsim$700 K and an H$_{2}$ number density of $>10^{8}$~cm$^{-3}$ are required to excite the HCN masers. A comparison between masers in carbon-rich and oxygen-rich stars suggests that HCN masers in carbon-rich stars may be an analogue of SiO masers in oxygen-rich stars.
    
\end{itemize}

Our observations suggest that both vibrationally excited HCN thermal and maser line emission are rich in the CSEs of carbon-rich stars and have quite high flux densities. This indicates that these transitions are suitable for follow-up high resolution observations to study the entire class of carbon-rich stars. With the very long baselines, ALMA observations of these vibrationally excited HCN lines can achieve an angular resolution of ${\lesssim}30$ mas \citep[e.g.,][]{2016A&A...590A.127W}, which should be able to constrain the physical conditions of the extended atmospheres and inner winds of carbon-rich stars.

\begin{acknowledgements}
      %\textcolor{red}{Python packages, matplotlib, numpy etc} 
      This publication is based on data acquired with the Atacama Pathfinder EXperiment (APEX). APEX is a collaboration between the Max-Planck-Institut f{\"u}r Radioastronomie, the European Southern Observatory, and the Onsala Space Observatory. This research has made use of NASA's Astrophysics Data System. We thank A.~V.~Lapinov and the Editor-in-chief of the \emph{Astronomicheskii Tsirkulyar} for kindly providing the scanned copies of \citet{1994msep.conf...66L} and \citet{1988ATsir1525...13Z}, respectively. We acknowledge with thanks the variable star observations from the AAVSO International Database contributed by observers worldwide and used in this research. This work has made use of Python libraries including NumPy\footnote{\url{https://www.numpy.org/}} \citep{5725236}, SciPy\footnote{\url{https://www.scipy.org/}} \citep{jones2001scipy}, Matplotlib\footnote{\url{https://matplotlib.org/}} \citep{Hunter:2007}. 
\end{acknowledgements}

\bibliography{references}
\bibliographystyle{aa}

%\appendix
\begin{appendix}

\section{Catalog of known HCN masers toward carbon-rich stars}\label{app:catalog}
Based on our new detections and literature, we have compiled the largest catalog of the HCN masers toward carbon-rich stars, and the results are listed in Table~\ref{tab:catalogue}. These masers are assigned to 36 different carbon-rich stars, and these HCN masers can be assigned to 10 different vibrational states. Our observations led to the discovery of 23 new HCN masers, increasing the total number to 75.
\begin{table*}
\caption{Catalogue of HCN maser detections in the CSEs of 36 stars till date.}\label{tab:catalogue}
\begin{minipage}[t]{0.5\textwidth}
\vspace{0pt}
\begin{tabular}{cccc}
\hline
\hline
Source name &  $\nu$ (MHz) & $J$ & $(\nu_{1}, \nu_{2}, \nu_{3})$ \\ 
(1)         & (2)          & (3) & (4)    \\
\hline
IRC +10216\tablefootmark{a} & 89087.7 & 1--0 & (0, 2, 0)  \\
IRC +10216\tablefootmark{b} & 177238.710 & 2--1 & (0,1$^{1e}$,0) \\
IRC +10216\tablefootmark{c} & 177238.655 & 2--1 & (0,1$^{1f}$,0)  \\
IRC +10216\tablefootmark{d} & 267199.283 & 3--2 & (0, 1$^{1f}$, 0)  \\
%%IRC +10216 & 264073.300 & 3--2 & (0, 0, 1) This work \\
IRC +10216\tablefootmark{e} & 356255.568 & 4--3 & (0, 1$^{1f}$, 0) \\
IRC +10216\tablefootmark{f} & 804750.9 & 9--8 & (0, 4, 0)  \\
IRC +10216\tablefootmark{g} & 890761.0 & 10--9 & (1,1$^{1}$,0)--(0, 4, 0) \\
RAFGL 4211\tablefootmark{h} & 89087.7 & 1--0 & (0, 2, 0) \\
RAFGL 4211\tablefootmark{c} & 177238.655 & 2--1 & (0,1$^{1e}$,0)  \\
RAFGL 4211\tablefootmark{e} & 266539.980 & 3--2 & (0, 3$^{1e}$, 0)\\
RAFGL 4211\tablefootmark{e} & 267243.150 & 3--2 & (0, 2, 0)   \\
RAFGL 4211\tablefootmark{e} & 354460.435 & 4--3 & (0, 1$^{1e}$, 0) \\
II Lup\tablefootmark{c} & 177238.655 & 2--1 & (0,1$^{1e}$,0)  \\
II Lup\tablefootmark{e} & 356255.568 & 4--3 & (0, 1$^{1f}$, 0)  \\
R Vol\tablefootmark{e} & 177238.656 & 2--1 & (0, 1$^{1e}$, 0) \\
R Vol\tablefootmark{e} & 178136.478 & 2--1 & (0, 1$^{1f}$, 0) \\
R Vol\tablefootmark{e} & 265852.709 & 3--2 & (0, 1$^{1e}$, 0) \\
R Vol\tablefootmark{e} & 354460.435 & 4--3 & (0, 1$^{1e}$, 0)\\
AQ Sgr\tablefootmark{e} & 177238.656 & 2--1 & (0, 1$^{1e}$, 0)\\
CRL 3068\tablefootmark{e}  & 177238.656 & 2--1 & (0,1$^{1e}$,0)  \\
CRL 3068\tablefootmark{e}  & 356255.568 & 4--3 & (0, 1$^{1f}$, 0) \\
IRC +30374\tablefootmark{i} & 89087.7 & 1--0 & (0, 2, 0)  \\
V636 Mon\tablefootmark{e} & 177238.656 & 2--1 & (0, 1$^{1e}$, 0) \\
IRC +20370\tablefootmark{e} & 177238.656 & 2--1 & (0, 1$^{1e}$, 0) \\
IRC +20370\tablefootmark{e} & 265852.709 & 3--2 & (0, 1$^{1e}$, 0) \\
W Ori\tablefootmark{j} & 88631.847 & 1--0 & (0, 0, 0)  \\
W Ori\tablefootmark{e} & 177238.656 & 2--1 & (0, 1$^{1e}$, 0) \\
W Ori\tablefootmark{e} & 265852.709 & 3--2 & (0, 1$^{1e}$, 0) \\
W Ori\tablefootmark{e} & 354460.435 & 4--3 & (0, 1$^{1e}$, 0) \\
%W Ori\tablefootmark{e} & 267199.283 & 3--2 & (0, 1$^{1f}$, 0)\\
S Aur\tablefootmark{e} & 177238.656 & 2--1 & (0, 1$^{1e}$, 0) \\
V Aql\tablefootmark{e} & 265852.709 & 3--2 & (0, 1$^{1e}$, 0) \\
RV Aqr\tablefootmark{e} & 177238.656 & 2--1 & (0, 1$^{1e}$, 0) \\
RV Aqr\tablefootmark{e} & 178136.478 & 2--1 & (0, 1$^{1f}$, 0) \\
RV Aqr\tablefootmark{e} & 265852.709 & 3--2 & (0, 1$^{1e}$, 0) \\
AFGL 2513\tablefootmark{i} & 89087.7 & 1--0 & (0, 2, 0)\\ 
Al Vol\tablefootmark{c} & 177238.655 & 2--1 & (0, 1$^{1e}$, 0)  \\
Al Vol\tablefootmark{c} & 178136.477 & 2--1 & (0, 1$^{1f}$, 0)\\
\hline
\end{tabular}

\end{minipage} \hfill
\begin{minipage}[t]{0.5\textwidth}
\vspace{0pt}
\begin{tabular}{cccc}
\hline
\hline
Source name &  $\nu$ (MHz) & $J$ & $(\nu_{1}, \nu_{2}, \nu_{3})$ \\ 
(1)         & (2)          & (3) & (4)    \\
\hline
CIT 6\tablefootmark{k} & 88006.7 & 1--0 & (1, 0, 0) \\
CIT 6\tablefootmark{k} & 88027.3 & 1--0 & (0, 0, 1) \\
CIT 6\tablefootmark{l} & 89087.7 & 1--0 & (0, 2, 0) \\
CIT 6\tablefootmark{c} & 177238.655 & 2--1 & (0, 1$^{1e}$, 0) \\
CIT 6\tablefootmark{c} & 178136.477 & 2--1 & (0, 1$^{1f}$, 0) \\
CIT 6\tablefootmark{g} & 804750.9 & 9--8 & (0, 4, 0) \\
CIT 6\tablefootmark{g} & 890761.0 & 10--9 & (1, 1$^{1}$, 0)--(0, 4, 0) \\
CRL 2688\tablefootmark{m} & 89087.7 & 1--0 & (0, 2, 0)  \\
CRL 618\tablefootmark{n} & 20181.3862 & 9--9 & (0, 1$^{1f}$, 0)--(0, 1$^{1e}$, 0) \\
CQ Pyx\tablefootmark{c} & 177238.655 & 2--1 & (0, 1$^{1e}$, 0) \\
CQ Pyx\tablefootmark{c} & 178136.477 & 2--1 & (0, 1$^{1f}$, 0) \\
FX Ser\tablefootmark{i} & 89087.7 & 1--0 & (0, 2, 0) \\
%RAFGL 2047 (IRAS 17581$-$1744)\tablefootmark{i} & 89087.7 & 1--0& (0, 2, 0)\\
RAFGL 2047\tablefootmark{i} & 89087.7 & 1--0 & (0, 2, 0) \\
IRC +50096\tablefootmark{i} & 89087.7 & 1--0 & (0, 2, 0) \\
R For\tablefootmark{c} & 177238.655 & 2--1 & (0, 1$^{1e}$, 0) \\
R Lep\tablefootmark{c} & 177238.655 & 2--1 & (0, 1$^{1e}$, 0) \\
R Lep\tablefootmark{o} & 265852.709 & 3--2 & (0, 1$^{1e}$, 0) \\
R Lep\tablefootmark{o} & 354460.435 & 4--3 & (0, 1$^{1e}$, 0) \\
R Scl\tablefootmark{o} & 265852.709 & 3--2 & (0, 1$^{1e}$, 0) \\
R Scl\tablefootmark{o} & 354460.435 & 4--3 & (0, 1$^{1e}$, 0) \\
S Cep\tablefootmark{i} & 89087.7 & 1--0 & (0, 2, 0) \\
T Dra\tablefootmark{i} & 89087.7 & 1--0 & (0, 2, 0) \\
TU Gem\tablefootmark{p} & 88631.847 & 1--0 & (0, 0, 0) \\
UU Aur\tablefootmark{p} & 88631.847 & 1--0 & (0, 0, 0) \\
V384 Per\tablefootmark{o} & 265852.709 & 3--2 & (0, 1$^{1e}$, 0) \\
V384 Per\tablefootmark{o} & 354460.435 & 4--3 & (0, 1$^{1e}$, 0) \\
V Cyg\tablefootmark{q} & 265852.709 & 3--2 & (0, 1$^{1e}$, 0) \\
V Cyg\tablefootmark{q} & 354460.435 & 4--3 & (0, 1$^{1e}$, 0) \\
V Hya\tablefootmark{c} & 177238.655 & 2--1 & (0, 1$^{1e}$, 0)\\
X Vel\tablefootmark{j} & 88631.847  & 1--0 & (0, 0, 0) \\
X Vel\tablefootmark{c} & 177238.655 & 2--1 & (0, 1$^{1e}$, 0) \\
X TrA\tablefootmark{j} & 88631.847  & 1--0 & (0, 0, 0) \\
X TrA\tablefootmark{c} & 177238.655 & 2--1 & (0, 1$^{1e}$, 0) \\
Y CVn\tablefootmark{r} & 88631.847  & 1--0 & (0, 0, 0) \\
Y CVn\tablefootmark{o} & 265852.709 & 3--2 & (0, 1$^{1e}$, 0) \\
Y CVn\tablefootmark{s} & 267199.283 & 3--2 & (0, 1$^{1f}$, 0) \\
Y CVn\tablefootmark{g} & 890761.0   & 10--9 & (1, 1$^{1}$, 0)--(0, 4, 0) \\
Y Tau\tablefootmark{p} & 88631.847  & 1--0 & (0, 0, 0) \\
\hline
\end{tabular}
\end{minipage}
\tablebib{
\tablefoottext{a}{\citet{lucas1986first}.} \tablefoottext{b}{\citet{lucas1989cerni..discovery}.} \tablefoottext{c}{\citet{2018AA...613A..49M}.} \tablefoottext{d}{\citet{2017ApJ...845...38H}.} \tablefoottext{e}{This work.} \tablefoottext{f}{\citet{schilke..mehringer..menten2000ApJ...528L..37S}.} \tablefoottext{g}{\citet{schilke..menten2003ApJ...583..446S}.} \tablefoottext{h}{\citet{2014MNRAS.440..172S}.} \tablefoottext{i}{\citet{lucas1988new}.} \tablefoottext{j}{\citet{olofsson1993bstudy}.} \tablefoottext{k}{\citet{1994msep.conf...66L}.} \tablefoottext{l}{\citet{guilloteau1987new}.} \tablefoottext{m}{\citet{1988ATsir1525...13Z}.} \tablefoottext{n}{\citet{2003ApJ...586..338T}.} \tablefoottext{o}{\citet{Bieging_2001}.} \tablefoottext{p}{\citet{izumiura..1995ApJ...440..728I}.}
\tablefoottext{q}{\citet{bieging2000vcygsubmillimeter}.} \tablefoottext{r}{\citet{izumiura1987}.} \tablefoottext{s}{\citet{2021AA...651A...8F}.}
}
\end{table*}

\section{Estimates of the stellar variability phase}\label{app:phasetext}
The phases of stellar variability on the dates of our 2018 observations were calculated for a number of sources in Table\,\ref{tab:phase}, primarily using the information compiled by AAVSO. In the used convention, phase zero is at the maximum light. While pulsation periods are available for the majority of sources (cf. Table\,\ref{tab:sample}), the epochs of maxima are scarce in the literature, limiting the number of stars with known phases at the epochs of our observations. In addition to AAVSO records, we used data from \citet[][for IRC+10216]{menten..2012A&A...543A..73M}, \citet[][for II\,Lup]{Feast2003}, \citet[][for IRC+30374]{Whitelock2006}, and \citet[][for CRL-2513]{2006MNRAS...menzies..369..783M}.

\begin{table*}
\caption{Pulsation phase at the time of the observation.}\label{tab:phase}
\begin{minipage}[t]{0.5\textwidth}
\vspace{0pt}
\begin{tabular}{ccccc}
\hline
\hline
 Source & $\Delta J$ & JD & $\phi$ & MM/DD \\
\hline
IRC+10216 & 4$-$3 & 2458302 & 0.95 & 07/02\\
          & 2$-$1 & 2458462 & 0.20 & 12/09\\
          & 3$-$2 & 2458467 & 0.21 & 12/14\\
          & 2$-$1 & 2458469 & 0.21 & 12/16\\
          & 4$-$3 & 2458473 & 0.22 & 12/20\\
\hline
RAFGL4211 & 4$-$3 & 2458278 & $-$ & 06/08 \\
          & 4$-$3 & 2458303 & $-$ & 07/03\\
          & 3$-$2 & 2458308 & $-$ & 07/08\\
          & 3$-$2 & 2458315 & $-$ & 07/15\\
          & 3$-$2 & 2458316 & $-$ & 07/16\\
          & 3$-$2 & 2458397 & $-$ & 10/05\\
          & 2$-$1 & 2458468 & $-$ & 12/15 \\
          & 2$-$1 & 2458469 & $-$ & 12/16\\
\hline          
II Lup & 2$-$1 & 2458470 & 0.70 & 12/17 \\
       & 2$-$1 & 2458471 & 0.70 & 12/18 \\
       & 3$-$2 & 2458308 & 0.41 & 07/08\\
 & 3$-$2 & 2458315 & 0.43 & 07/15\\
 & 3$-$2 & 2458398 & 0.57 & 10/06\\
 & 4$-$3 & 2458302 & 0.40 & 07/02\\
 & 4$-$3 & 2458303 & 0.41 & 07/03\\
\hline
R Vol & 4$-$3 & 2458365 & 0.8 & 09/03 \\
 & 4$-$3 & 2458367 & 0.8 & 09/05 \\
 & 4$-$3 & 2458448 & 0.9 & 11/25 \\
 & 4$-$3 & 2458451 & 0.9 & 11/28 \\
 & 3$-$2 & 2458452 & 0.9 & 11/29 \\
 & 2$-$1 & 2458462 & 1.0 & 12/09 \\
 & 2$-$1 & 2458464 & 1.0 & 12/11 \\
 & 2$-$1 & 2458469 & 1.0 & 12/16 \\
 & 4$-$3 & 2458473 & 1.0 & 12/20 \\
\hline
Y Pav & 3$-$2 & 2458308 & 0.03 & 07/08\\
 & 3$-$2 & 2458312 & 0.04 & 07/12 \\
 & 3$-$2 & 2458314 & 0.04 & 07/14 \\
 & 3$-$2 & 2458398 & 0.23 & 10/06 \\
 & 4$-$3 & 2458447 & 0.34 & 11/24 \\ 
 & 4$-$3 & 2458448 & 0.34 & 11/25 \\
 & 4$-$3 & 2458449 & 0.34 & 11/26\\
 & 3$-$2 & 2458450 & 0.34 & 11/27\\
 & 2$-$1 & 2458451 & 0.35 & 11/28\\
\hline
AQ Sgr & 3$-$2 & 2458308 && 07/08\\
 & 3$-$2 & 2458311 &$-$ & 07/11\\
 & 4$-$3 & 2458367 &$-$ & 09/05\\
 & 4$-$3 & 2458369 &$-$ & 09/07\\
 & 3$-$2 & 2458370 &$-$ & 09/08\\
 & 3$-$2 & 2458375 &$-$ & 09/13\\
 & 3$-$2 & 2458398 &$-$ & 10/06\\ 
\hline
CRL 3068 & 2$-$1 & 2458451 &$-$ & 11/28\\
 & 2$-$1 & 2458462 &$-$ & 12/09\\
 & 2$-$1 & 2458463 &$-$ & 12/10\\
 & 2$-$1 & 2458464 &$-$ & 12/11\\
 & 3$-$2 & 2458304 &$-$ & 07/04 \\
 & 3$-$2 & 2458313 &$-$ & 07/13\\
 & 3$-$2 & 2458370 &$-$ & 09/08\\
 & 3$-$2 & 2458398 &$-$ & 10/06\\
 & 4$-$3 & 2458367 &$-$ & 09/05\\
 & 4$-$3 & 2458368 &$-$ & 09/06\\
 & 4$-$3 & 2458369 &$-$ & 09/07\\
 & 4$-$3 & 2458459 &$-$ & 12/06\\
\hline

\end{tabular}

\end{minipage} \hfill
\begin{minipage}[t]{0.5\textwidth}
\vspace{0pt}
\begin{tabular}{ccccc}
\hline
\hline
 Source & $\Delta J$ & JD & $\phi$ & MM/DD \\
\hline
IRC+30374 & 2$-$1 & 2458467  &$-$ & 12/14\\
 & 3$-$2 & 2458308  &$-$ & 07/08\\
 & 4$-$3 & 2458446  &$-$ & 11/23\\
 \hline
V636 Mon & 2$-$1 & 2458455  &$-$ & 12/02\\
 & 2$-$1 & 2458463  &$-$ & 12/10\\
 & 3$-$2 & 2458449  &$-$ & 11/26\\
 & 4$-$3 & 2458365  &$-$ & 09/03\\
 & 4$-$3 & 2458367  &$-$ & 09/05\\
 & 4$-$3 & 2458374  &$-$ & 09/12\\
 & 4$-$3 & 2458451  &$-$ & 11/28\\
\hline
IRC+20370 & 3$-$2 & 2458308 & 0.45 & 07/08\\
 & 3$-$2 & 2458315 & 0.46 & 07/15\\
 & 3$-$2 & 2458397 & 0.62 & 10/05\\
 & 3$-$2 & 2458398 & 0.62 & 10/06\\
 & 2$-$1 & 2458452 & 0.73 & 11/29\\
 & 4$-$3 & 2458459 & 0.74 & 12/06\\
\hline
W Ori & 3$-$2 & 2458304 & 0.92 & 07/04\\
 & 3$-$2 & 2458313 & 0.96 & 07/13\\
 & 3$-$2 & 2458314 & 0.96 & 07/14\\
 & 3$-$2 & 2458315 & 0.97 & 07/15\\
 & 4$-$3 & 2458365 & 0.20 & 09/03\\ 
 & 4$-$3 & 2458367 & 0.21 & 09/05\\ 
 & 4$-$3 & 2458373 & 0.24 & 09/11\\ 
 & 3$-$2 & 2458375 & 0.25 & 09/13\\ 
 & 4$-$3 & 2458448 & 0.59 & 11/25\\
 & 3$-$2 & 2458452 & 0.61 & 11/29\\
 & 2$-$1 & 2458453 & 0.62 & 11/30\\
 & 4$-$3 & 2458460 & 0.65 & 12/07\\
 & 2$-$1 & 2458463 & 0.67 & 12/10\\
\hline
S Aur & 4$-$3 &2458366 & 0.73 & 09/04\\
 & 3$-$2 & 2458398 & 0.79 & 10/06 \\
 & 4$-$3 &2458460 & 0.89 & 12/07\\
 & 2$-$1 & 2458463 & 0.90 & 12/10 \\
\hline
V Aql & 3$-$2 & 2458308 & 0.91 & 07/08\\
 & 3$-$2 & 2458313 & 0.92 & 07/13\\
 & 4$-$3 & 2458366 & 0.05 & 09/04\\
 & 4$-$3 & 2458367 & 0.05 & 09/05\\
 & 3$-$2 & 2458377 & 0.08 & 09/15\\
\hline
CRL 2513 & 2$-$1 & 2458452 &$-$  & 11/29\\
 & 2$-$1 & 2458463 &$-$  & 12/10\\
 & 3$-$2 & 2458308 &$-$  & 07/08\\
 & 3$-$2 & 2458313 & $-$ & 07/13\\
\hline
CRL 2477 & 2$-$1 & 2458468 &$-$  & 12/15\\
 & 3$-$2 & 2458370 & $-$ & 09/08 \\
\hline
RV Aqr & 3$-$2 & 2458313 & 0.74	& 07/13\\
 & 3$-$2 & 2458314 & 0.74	& 07/14\\ 
 & 4$-$3 & 2458368 & 0.86	& 09/06\\
 & 3$-$2 & 2458371 & 0.87	& 09/09\\
 & 4$-$3 & 2458446 & 0.04	& 11/23\\
 & 4$-$3 & 2458448 & 0.04	& 11/25\\
 & 4$-$3 & 2458449 & 0.04	& 11/26\\ 
 & 2$-$1 & 2458452 & 0.05	& 11/29\\
 & 2$-$1 & 2458453 & 0.05 & 11/30\\
\hline
\end{tabular}

\end{minipage}
\end{table*}

%\clearpage

\section{Vibrationally excited HCN lines in other sources}
\label{app:spectra}
\begin{figure*}[t]%[!htb]
\centering
\includegraphics[width=0.335\textwidth]{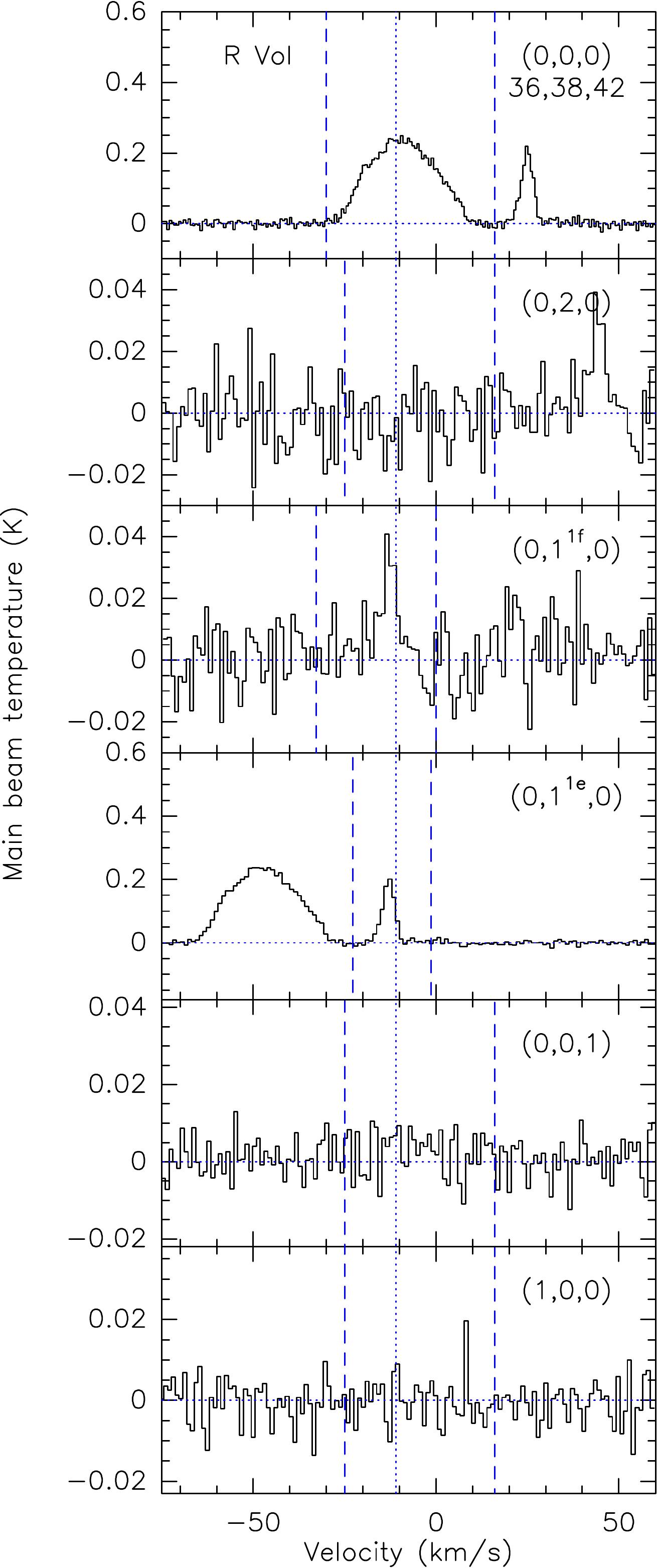}\quad
\includegraphics[width=0.30\textwidth]{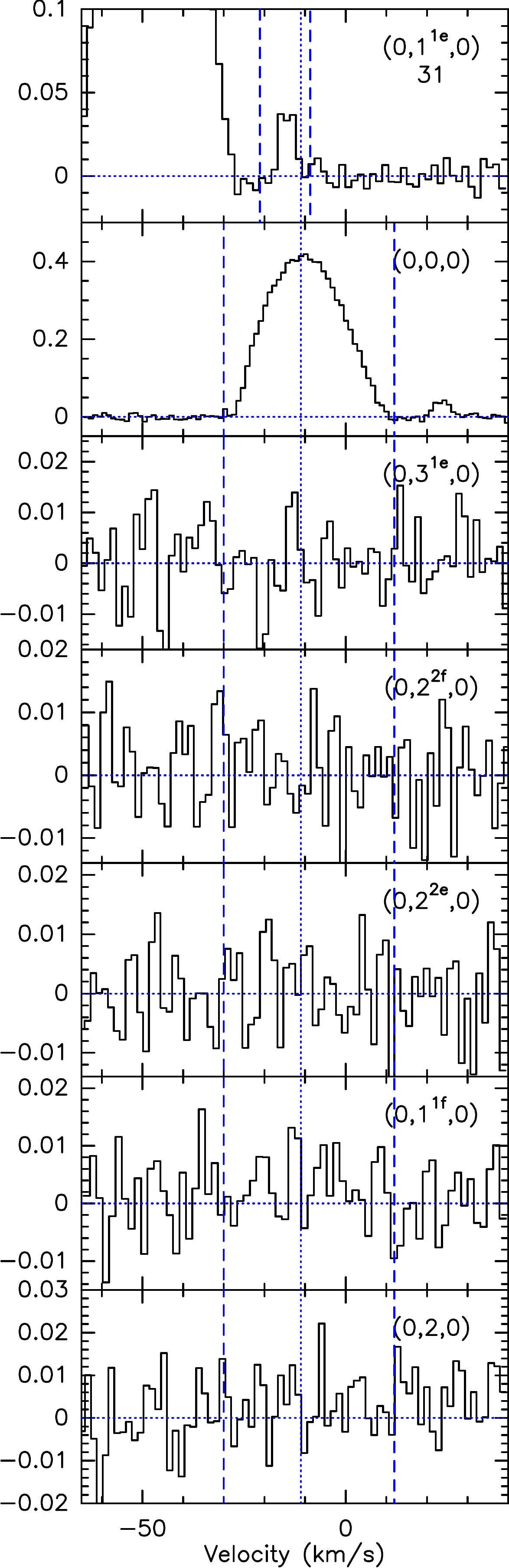}\quad
\includegraphics[width=0.30\textwidth]{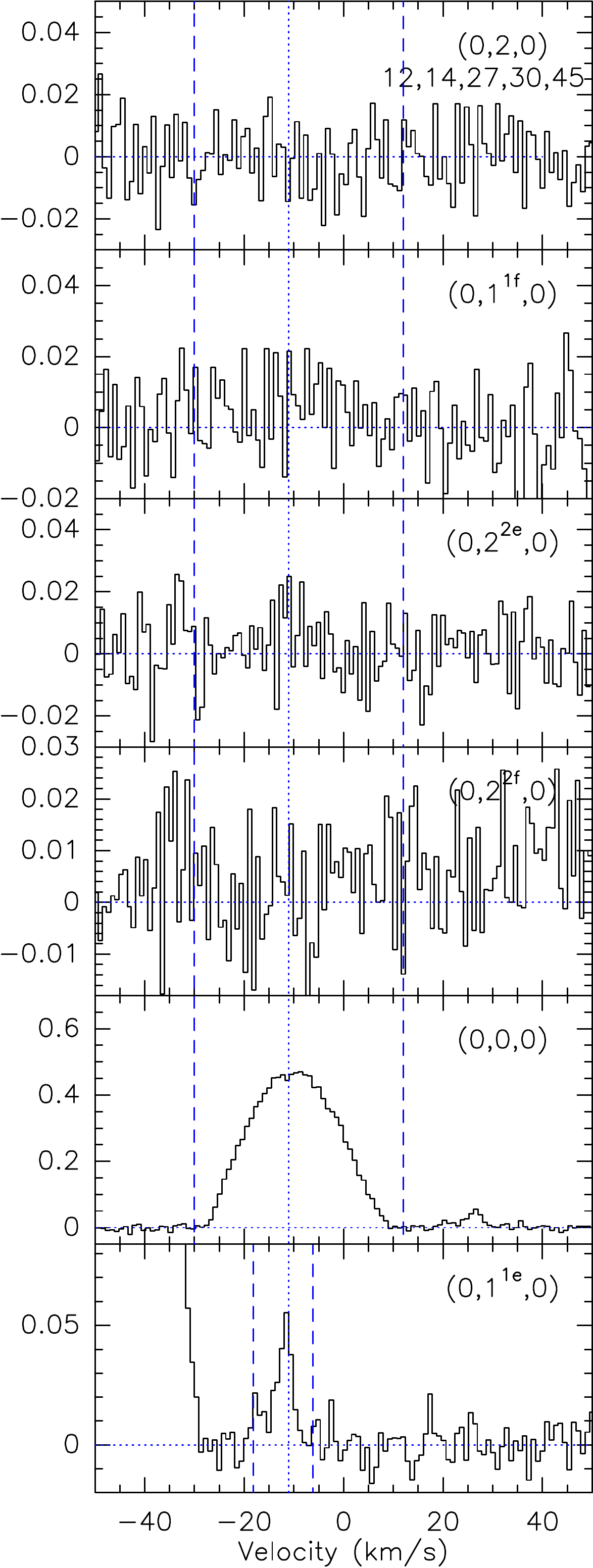}
\caption{Spectra for $J=$2--1 (left), 3--2 (center) and 4--3 (right) transitions of HCN towards R Vol. Description of the figure is the same as in the Fig. \ref{fig:irc_all_spec} caption.}
\label{fig:rvol-all}
\end{figure*}

\begin{figure*}%[t]%[!htb]
\centering
\includegraphics[width=0.335\textwidth]{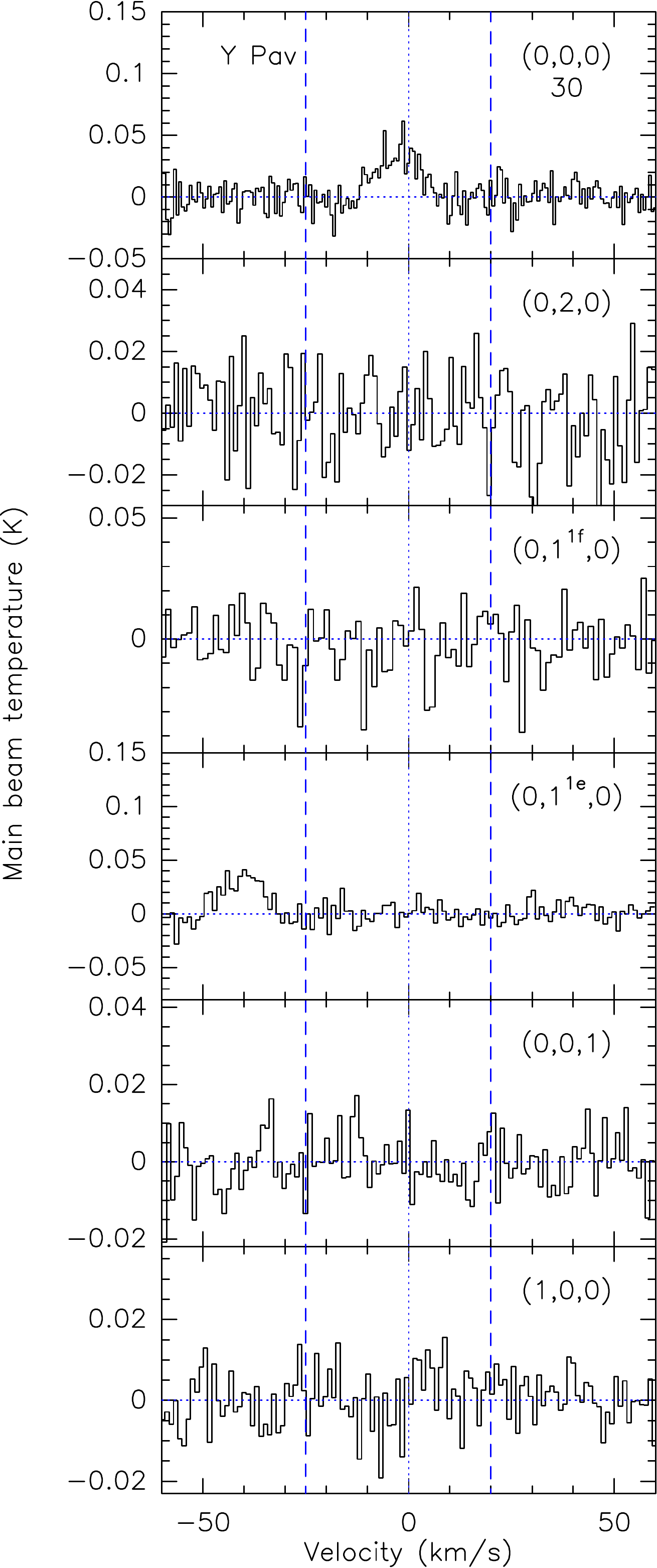}\quad
\includegraphics[width=0.30\textwidth]{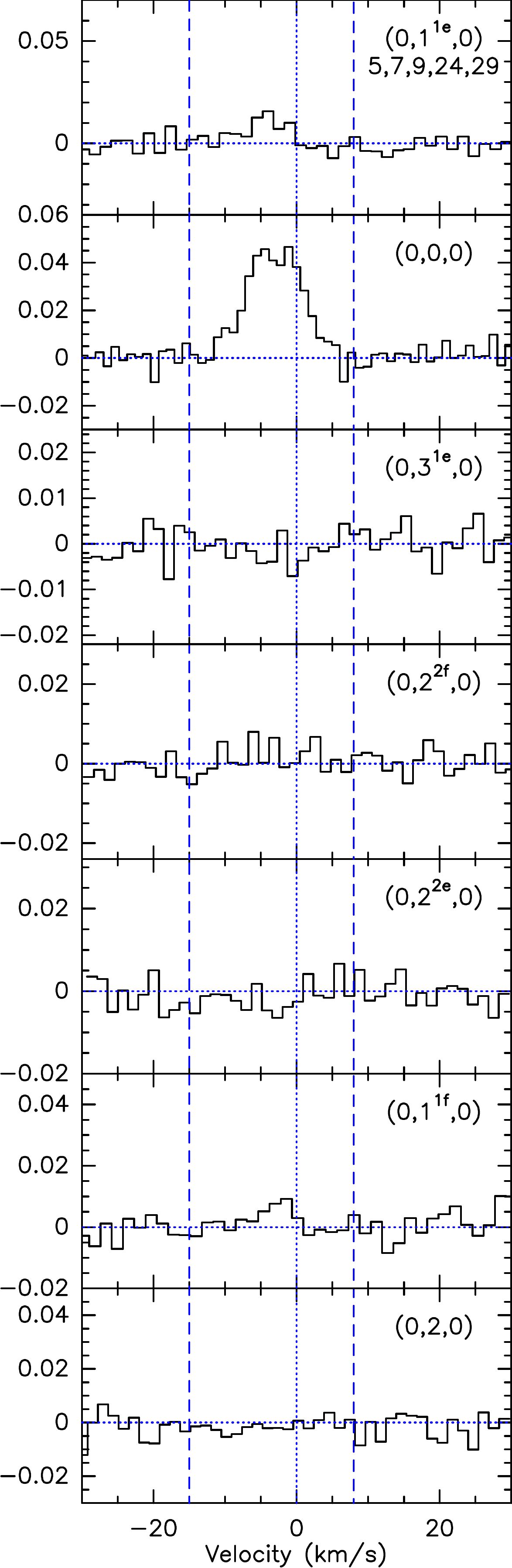}\quad
\includegraphics[width=0.30\textwidth]{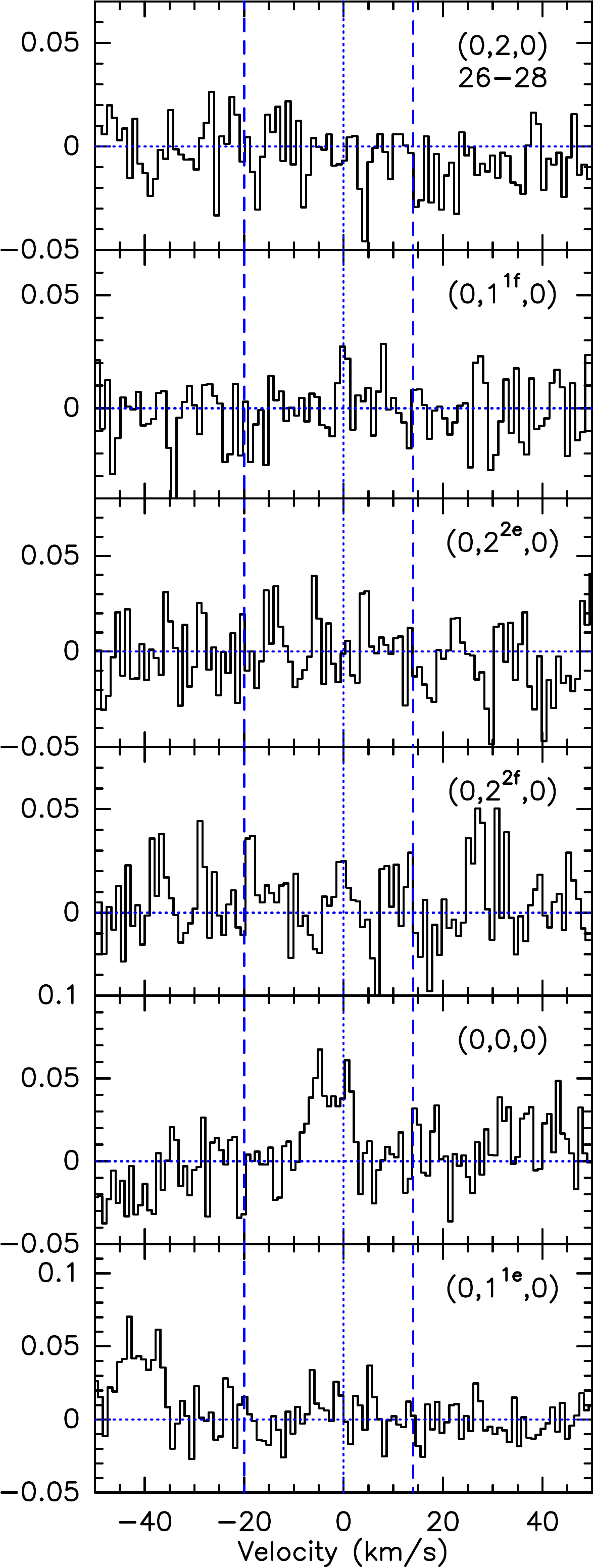}
\caption{Spectra for $J=2$--1 (left), 3--2 (center) and 4--3 (right) transitions of HCN towards Y Pav. Description of the figure is the same as in the Fig. \ref{fig:irc_all_spec} caption.}
\label{fig:ypav-all}
\end{figure*}

\begin{figure*}%[t]%[!htb]
\centering
\includegraphics[width=0.33\textwidth]{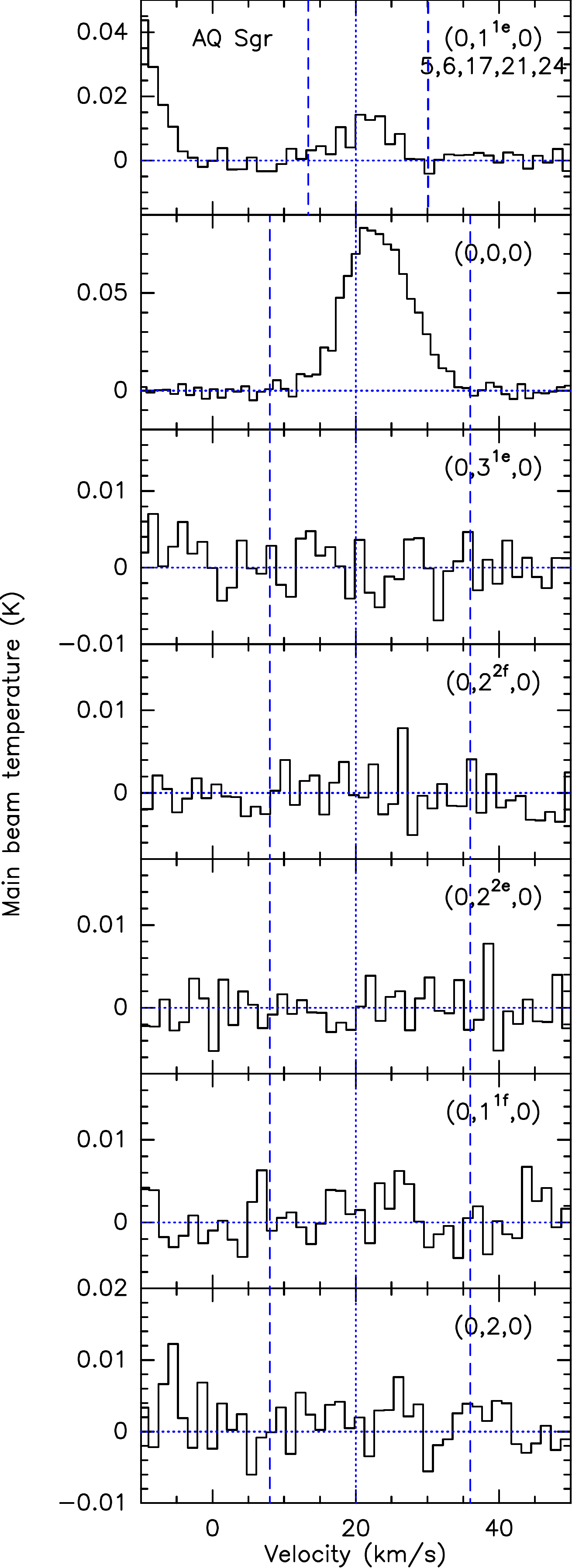}\quad
\includegraphics[width=0.30\textwidth]{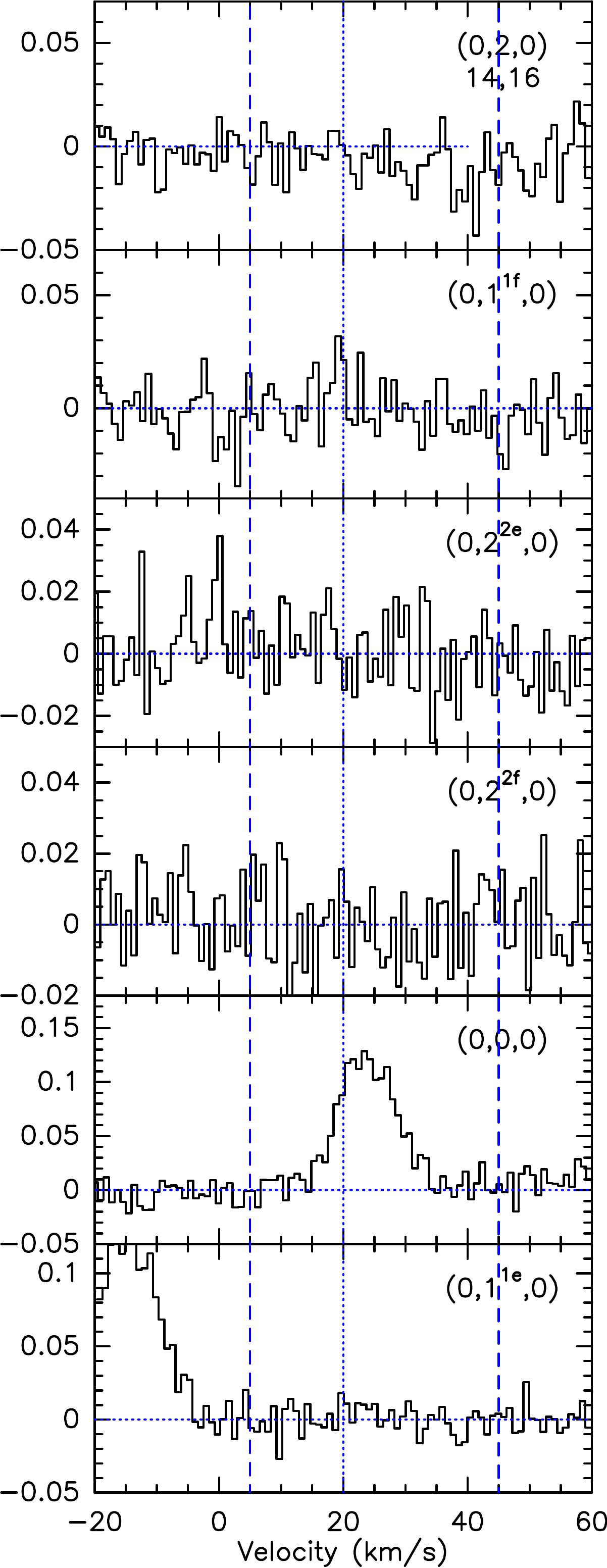}
\caption{Spectra for $J=3$--2 (left) and 4--3 (right) transitions of HCN towards AQ Sgr. Description of the figure is the same as in the Fig. \ref{fig:irc_all_spec} caption.}
\label{fig:aqsgr-all}
\end{figure*}

\begin{figure*}%[t]%[!htb]
\centering
\includegraphics[width=0.335\textwidth]{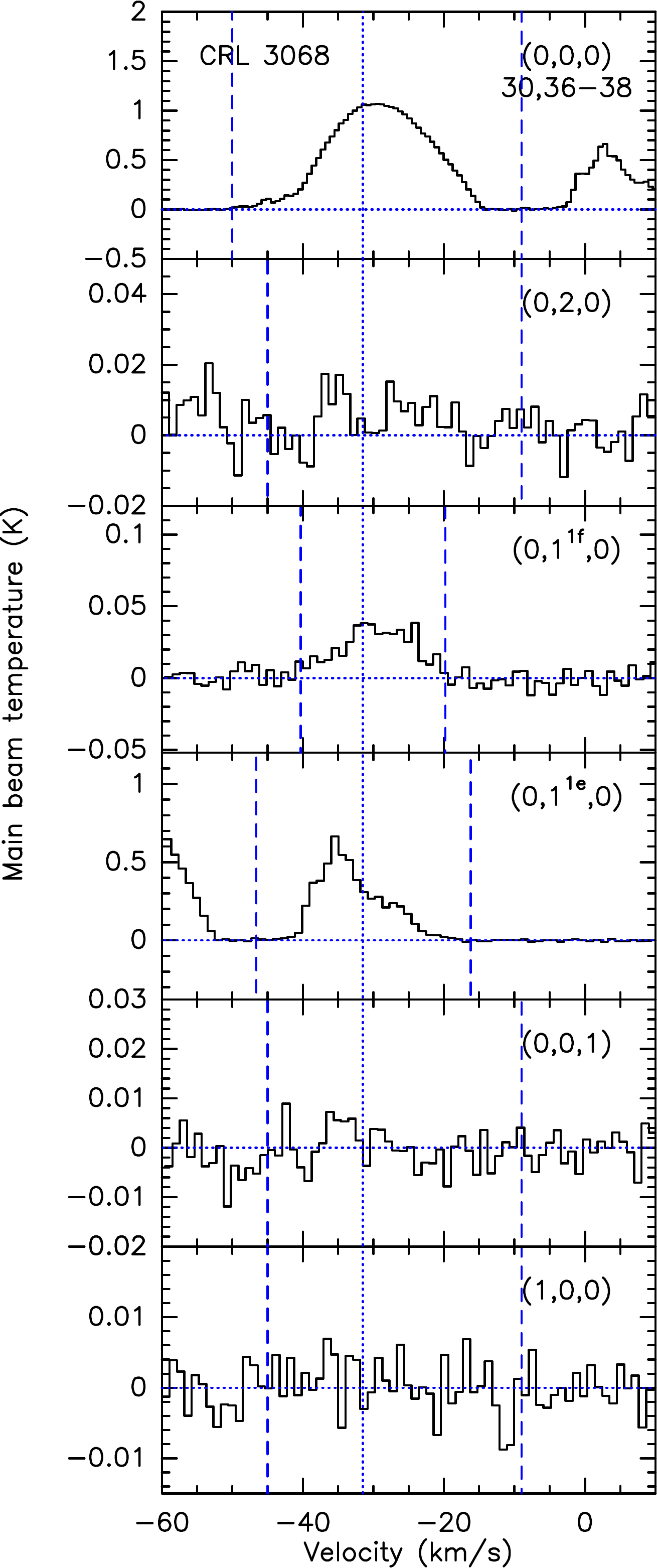}\quad
\includegraphics[width=0.30\textwidth]{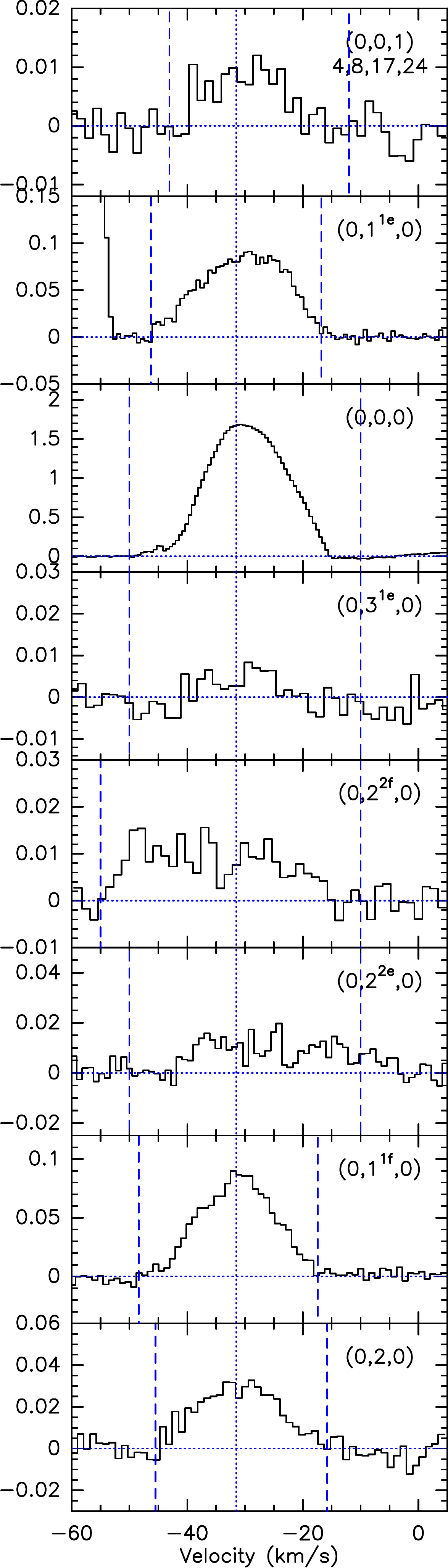}\quad
\includegraphics[width=0.30\textwidth]{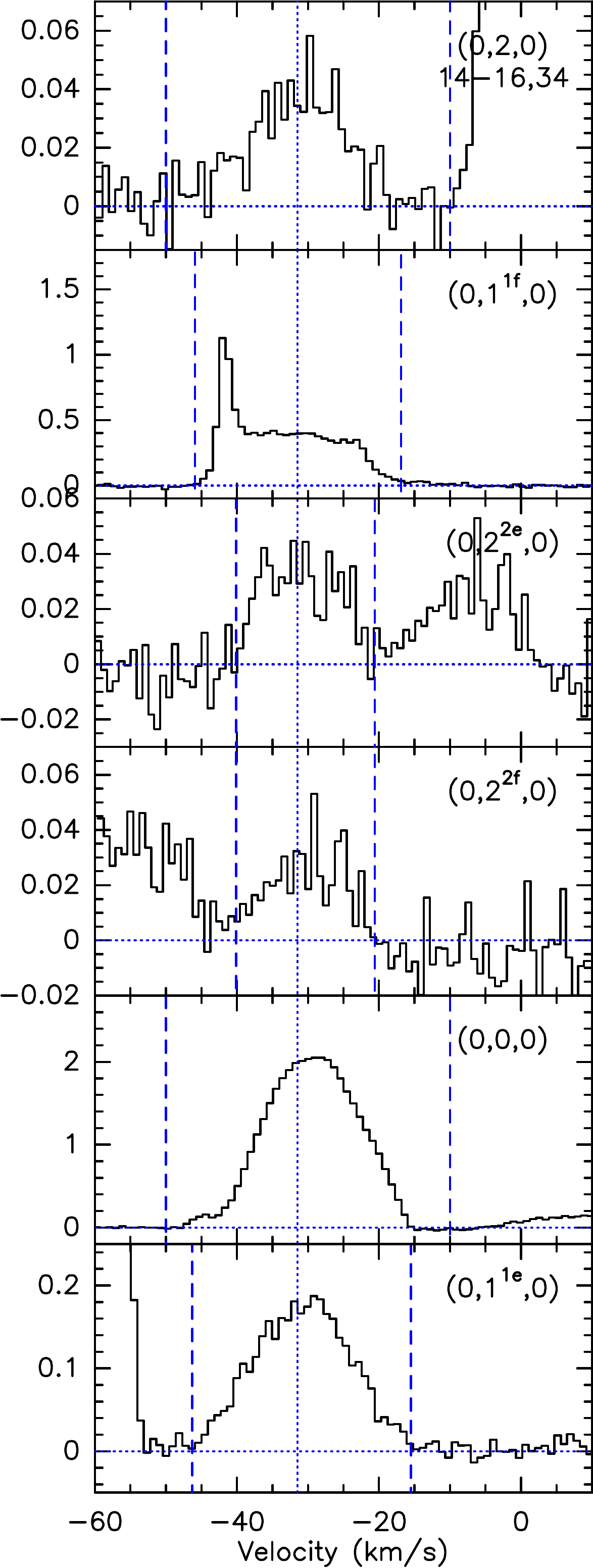}
\caption{Spectra for $J=2$--1 (left), 3--2 (center) and 4--3 (right) transitions of HCN towards CRL 3068. Description of the figure is the same as in the Fig. \ref{fig:irc_all_spec} caption.}
\label{fig:crl-3068_all_spec}
\end{figure*}

\begin{figure*}%[t]%[!htb]
\centering
\includegraphics[width=0.335\textwidth]{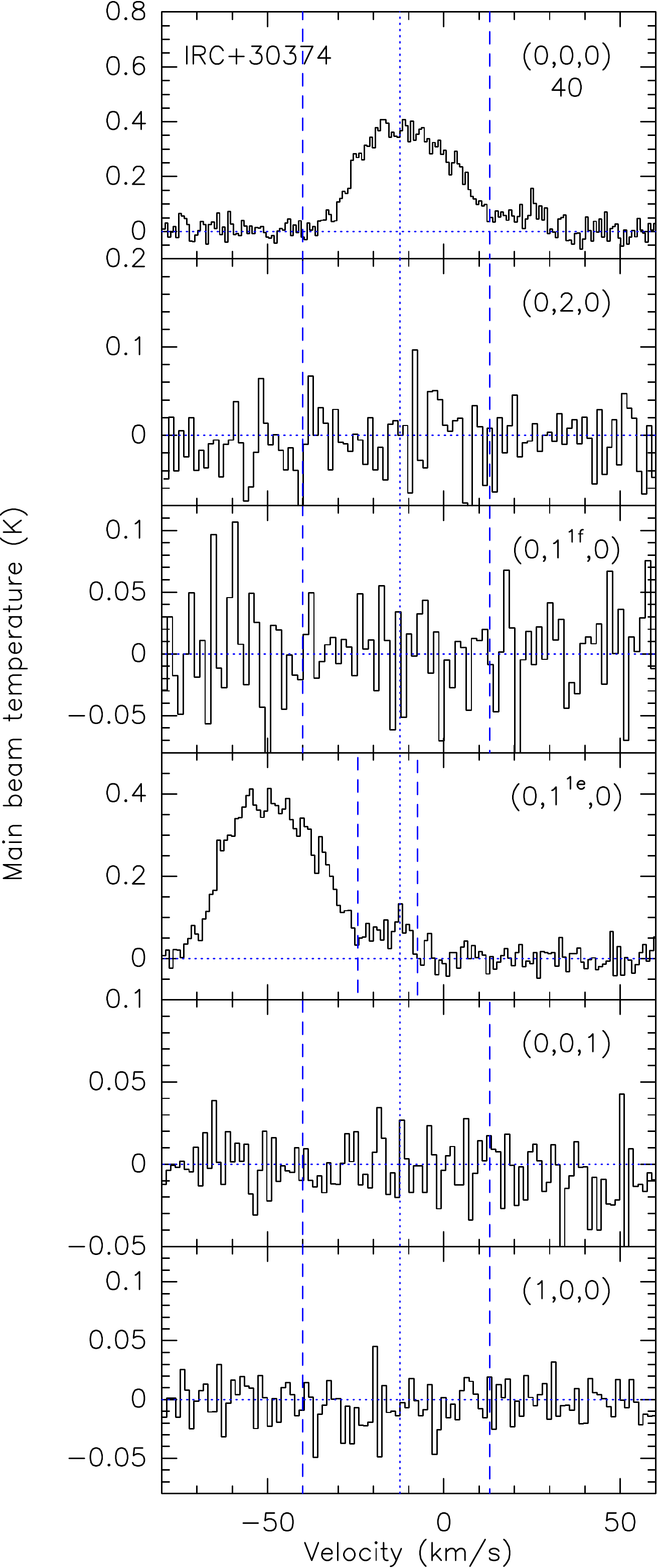}\quad
\includegraphics[width=0.31\textwidth]{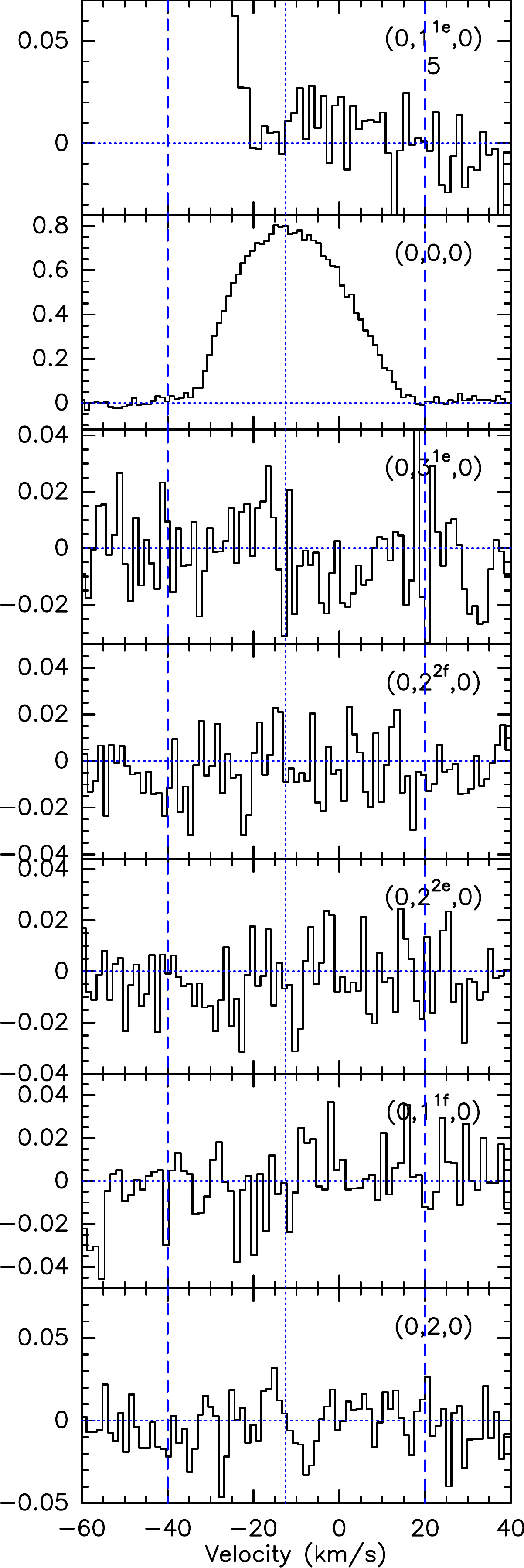}\quad
\includegraphics[width=0.31\textwidth]{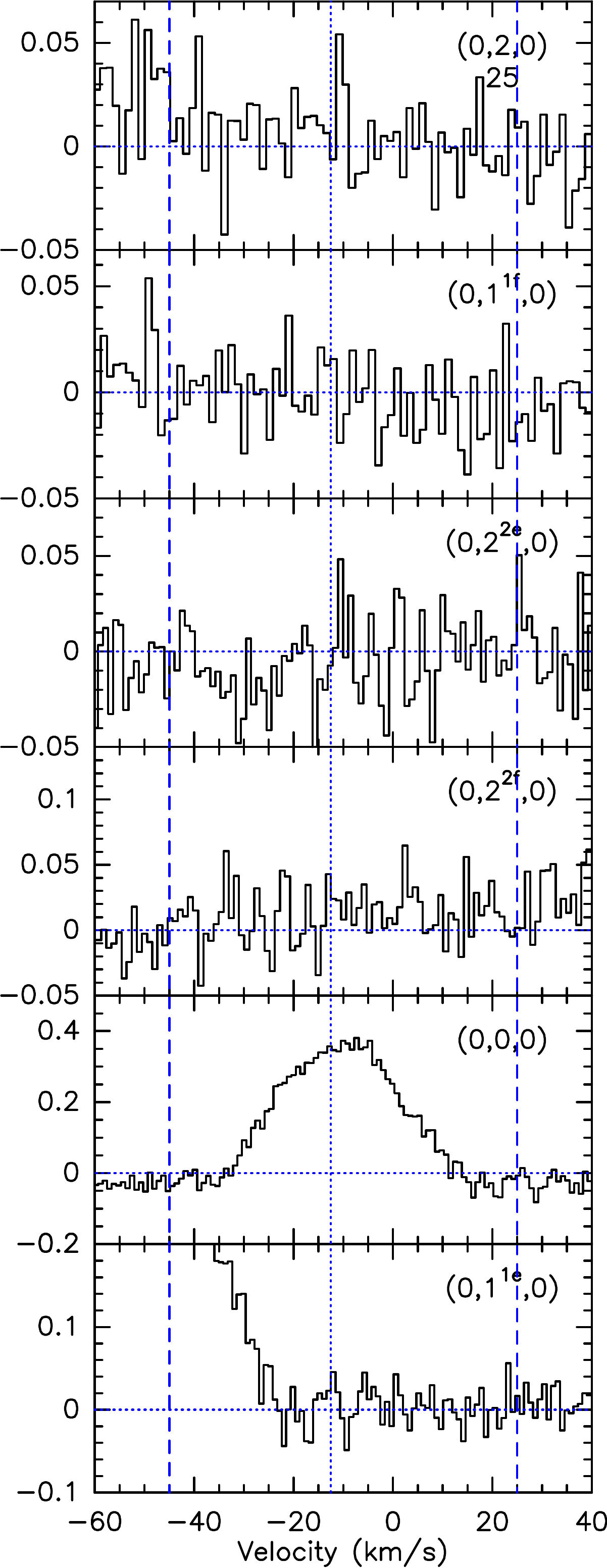}
\caption{Spectra for $J=2$--1 (left), 3--2 (center) and 4--3 (right) transitions of HCN towards IRC +30374. Description of the figure is the same as in the Fig. \ref{fig:irc_all_spec} caption. }
\label{fig:irc+30374-all}
\end{figure*}

\begin{figure*}%[t]%[!htb]
\centering
\includegraphics[width=0.335\textwidth]{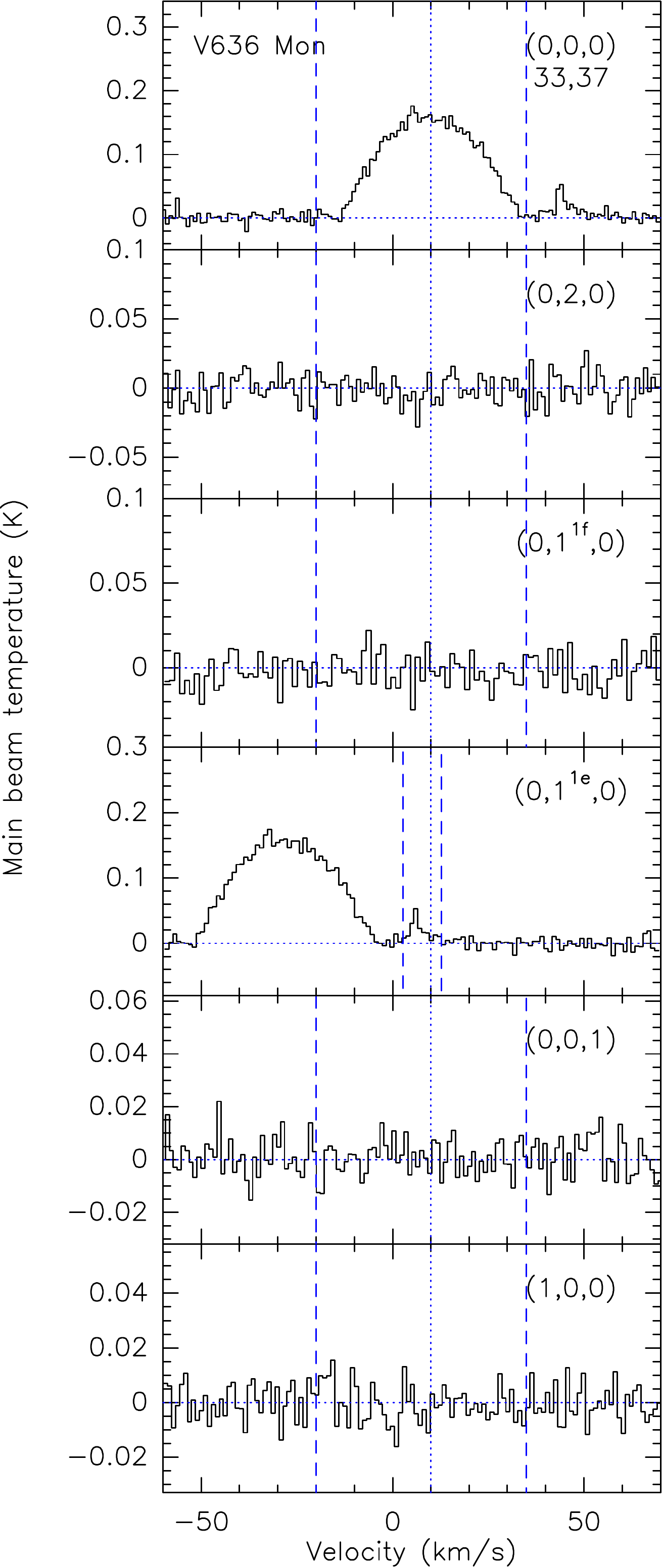}\quad
\includegraphics[width=0.31\textwidth]{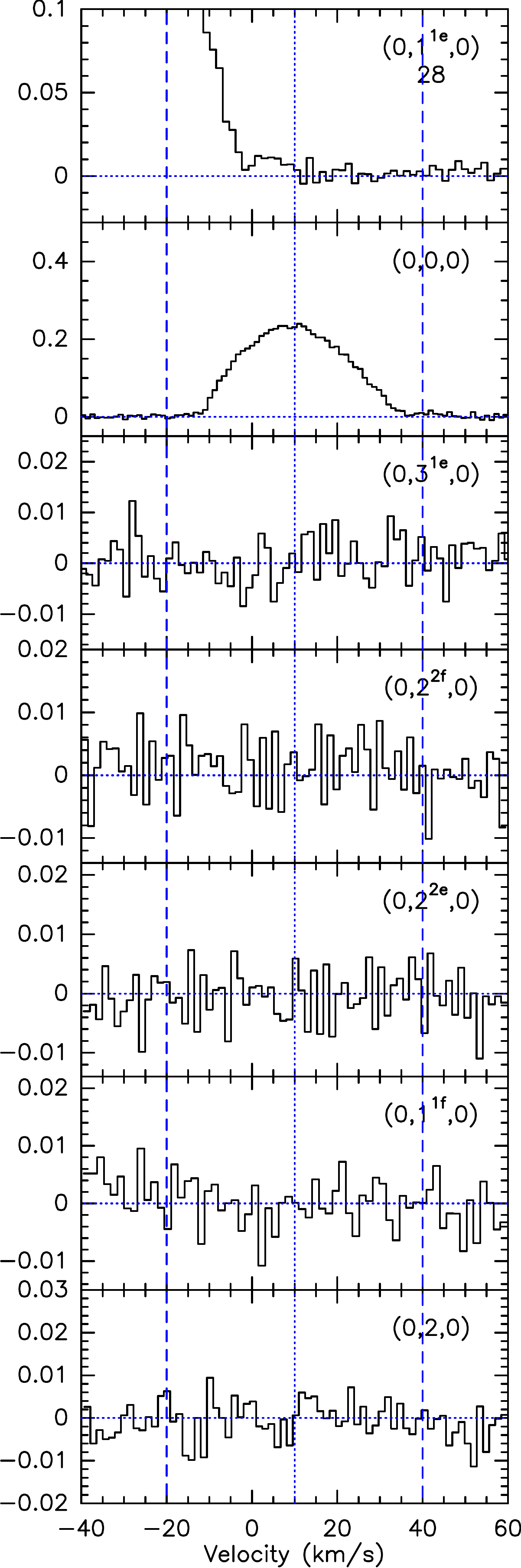}
\includegraphics[width=0.30\textwidth]{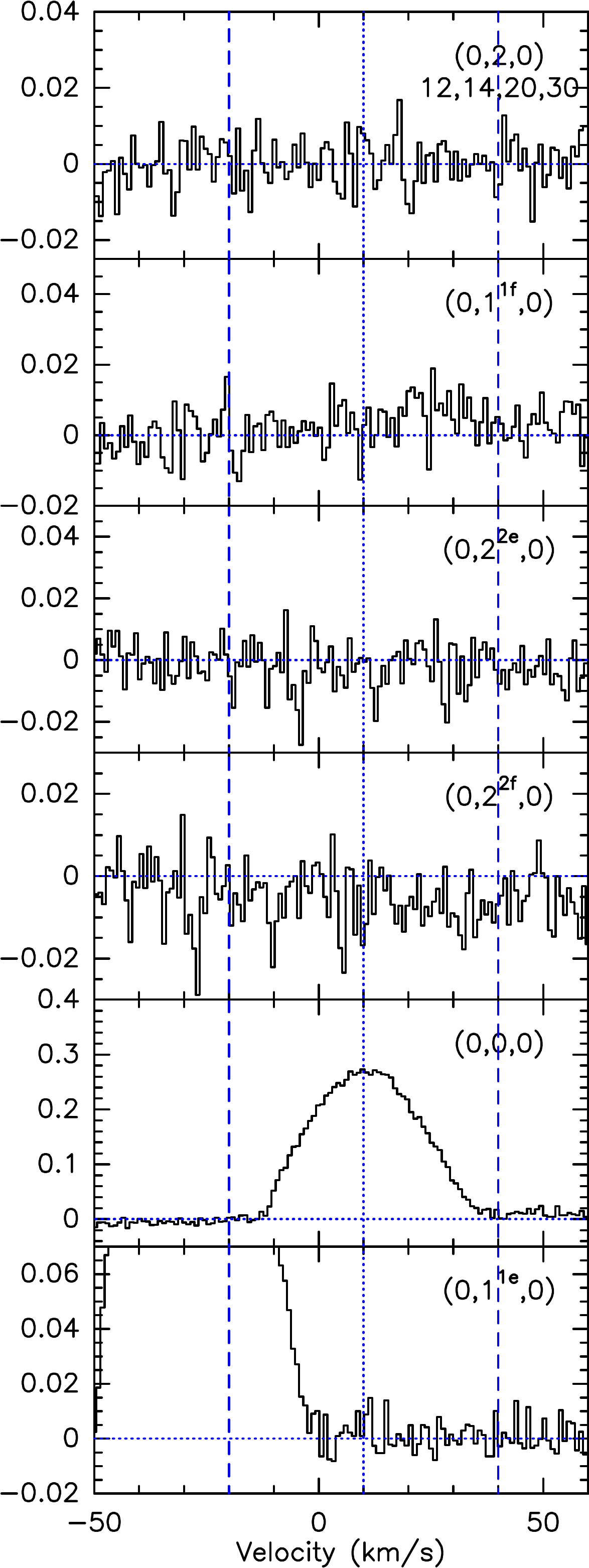}\quad
\caption{Spectra for $J=2$--1 (left), 3--2 (center) and 4--3 (right) transitions of HCN towards V636 Mon. Description of the figure is the same as in the Fig. \ref{fig:irc_all_spec} caption.}
\label{fig:v636-all}

\end{figure*}

\begin{figure*}%[t]%[!htb]
\centering
\includegraphics[width=0.343\textwidth]{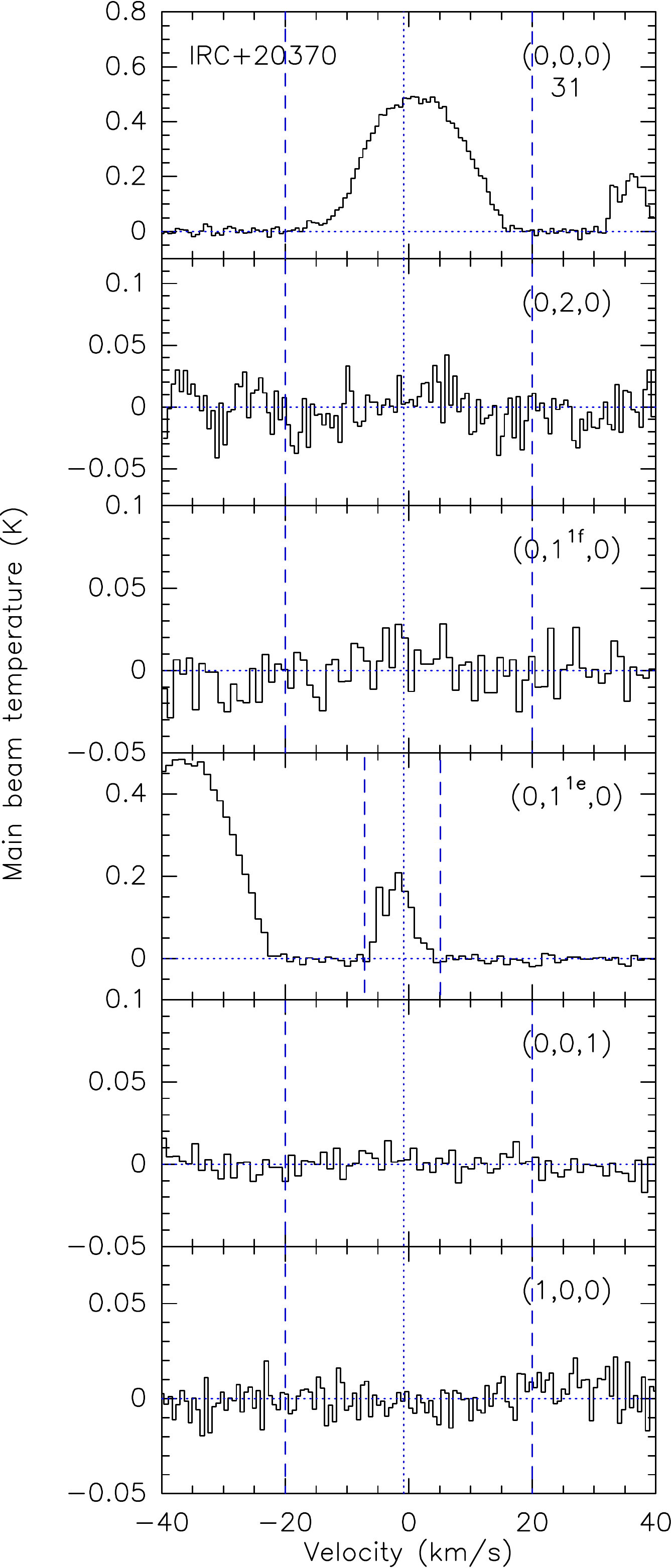}\quad
\includegraphics[width=0.31\textwidth]{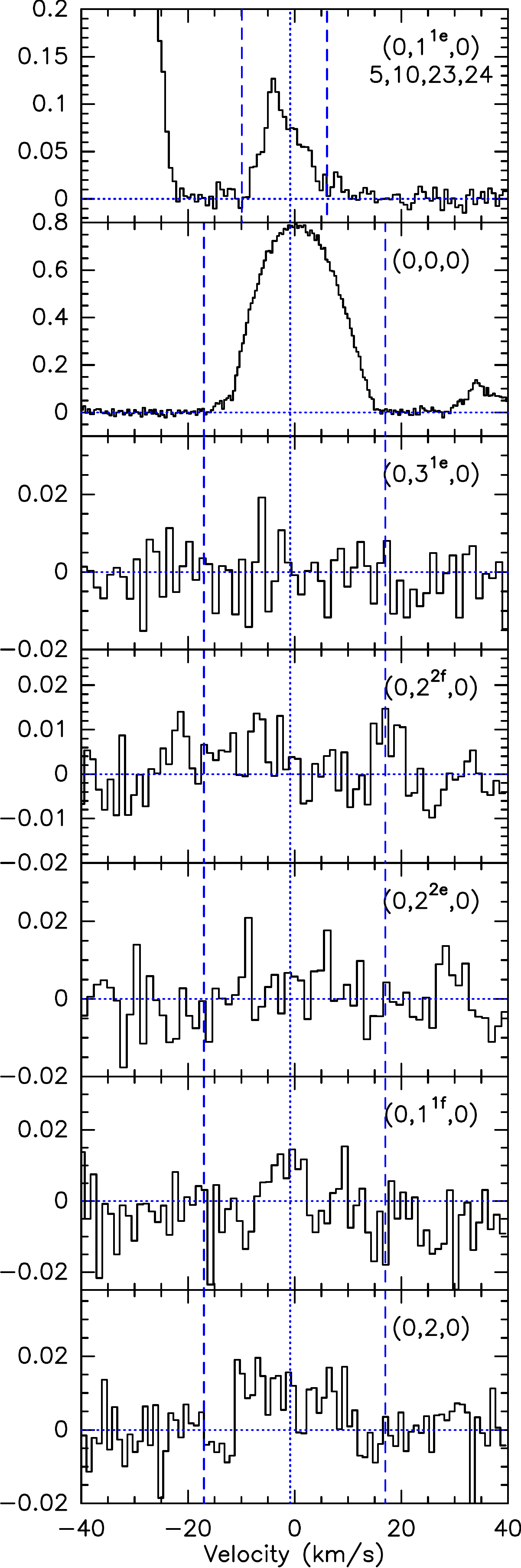}\quad
\includegraphics[width=0.30\textwidth]{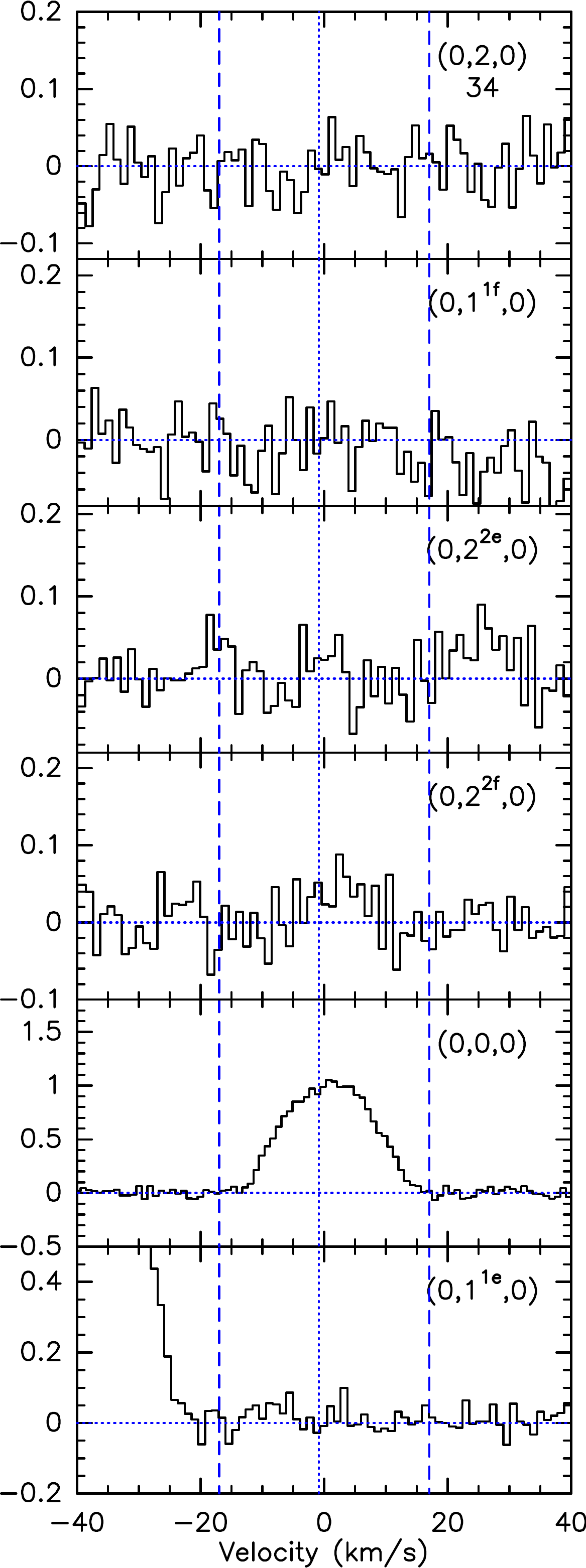}
\caption{Spectra for $J=2$--1 (left), 3--2 (center) and 4--3 (right) transitions of HCN towards IRC +20370. Description of the figure is the same as in the Fig. \ref{fig:irc_all_spec} caption.}
\label{fig:irc+20370-all}
\end{figure*}

\begin{figure*}%[t]%[!htb]
\centering
\includegraphics[width=0.335\textwidth]{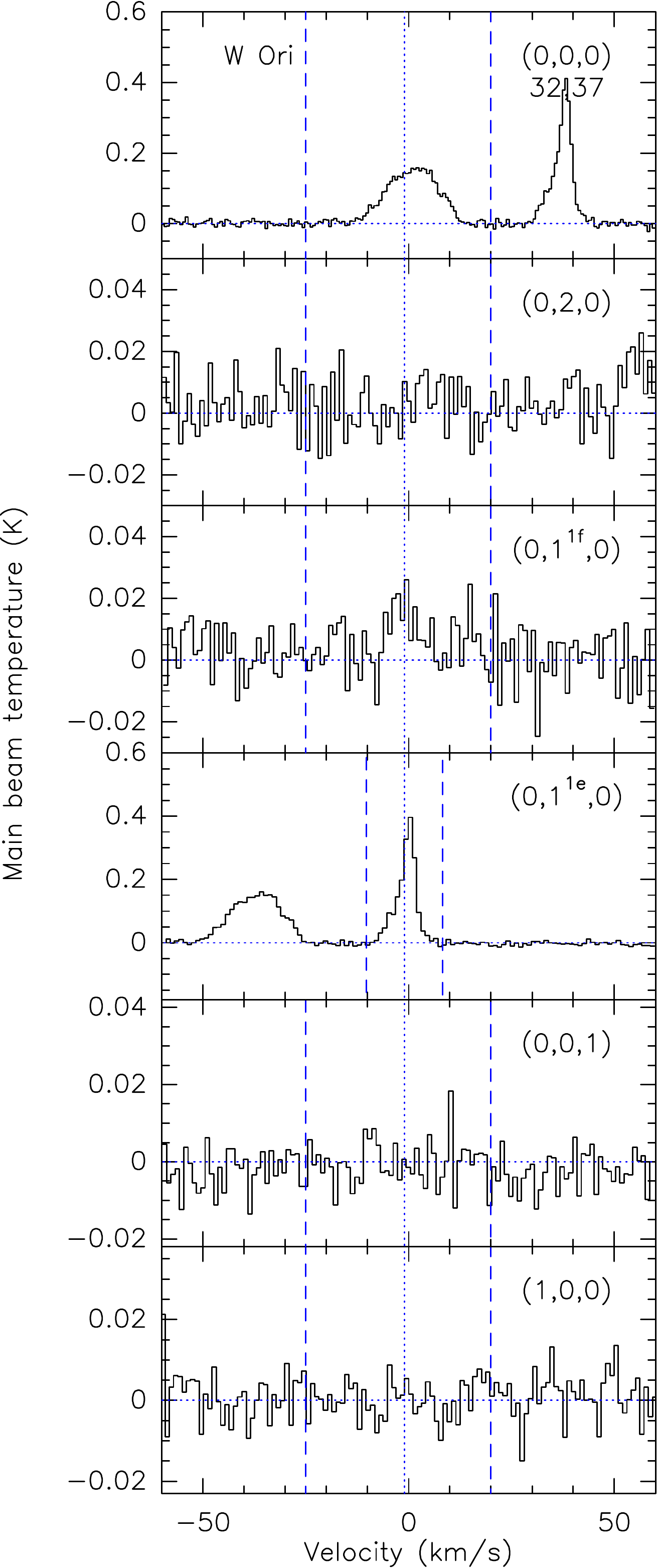}\quad
\includegraphics[width=0.31\textwidth]{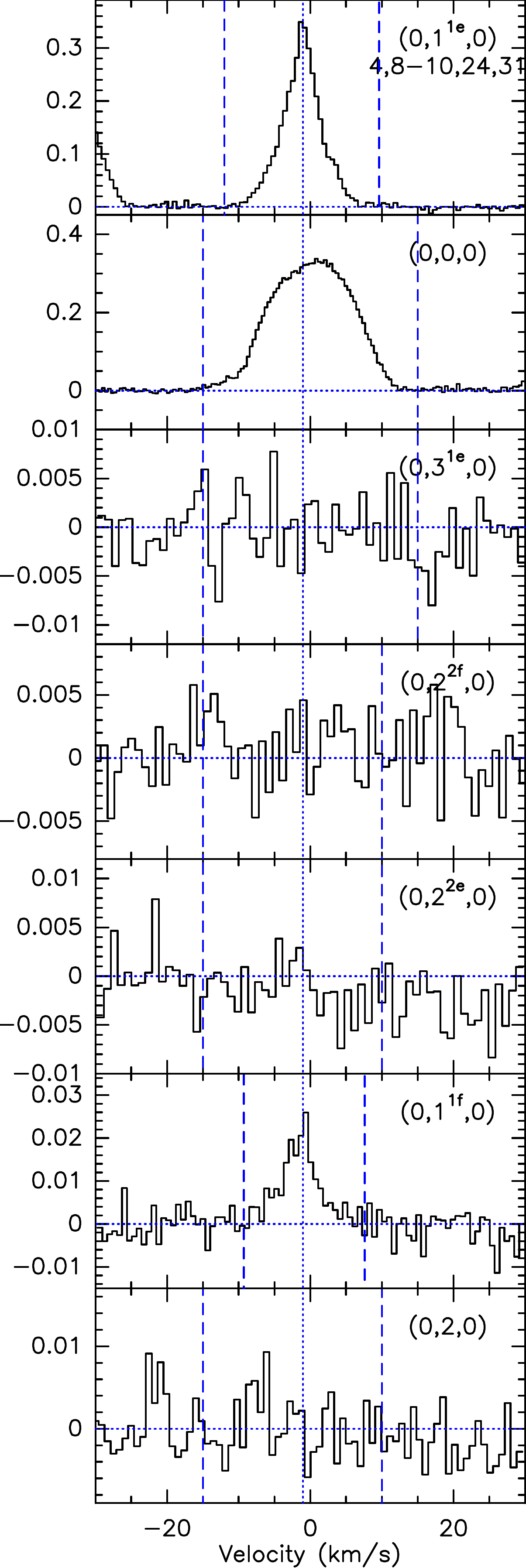}\quad
\includegraphics[width=0.3\textwidth]{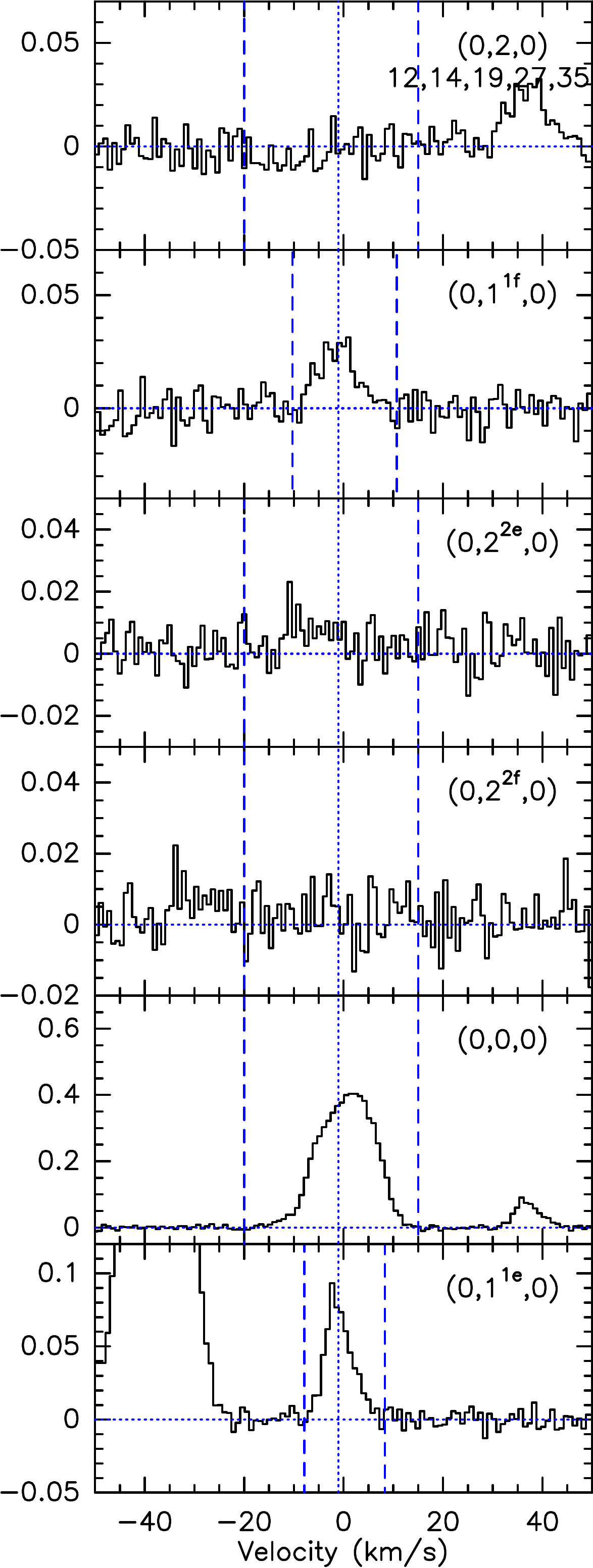}
\caption{Spectra for $J=2$--1 (left), 3--2 (center) and 4--3 (right) transitions of HCN towards W Ori. Description of the figure is the same as in the Fig. \ref{fig:irc_all_spec} caption.}
\label{fig:wori-all}
\end{figure*}

\begin{figure*}%[t]%[!htb]
\centering
\includegraphics[width=0.335\textwidth]{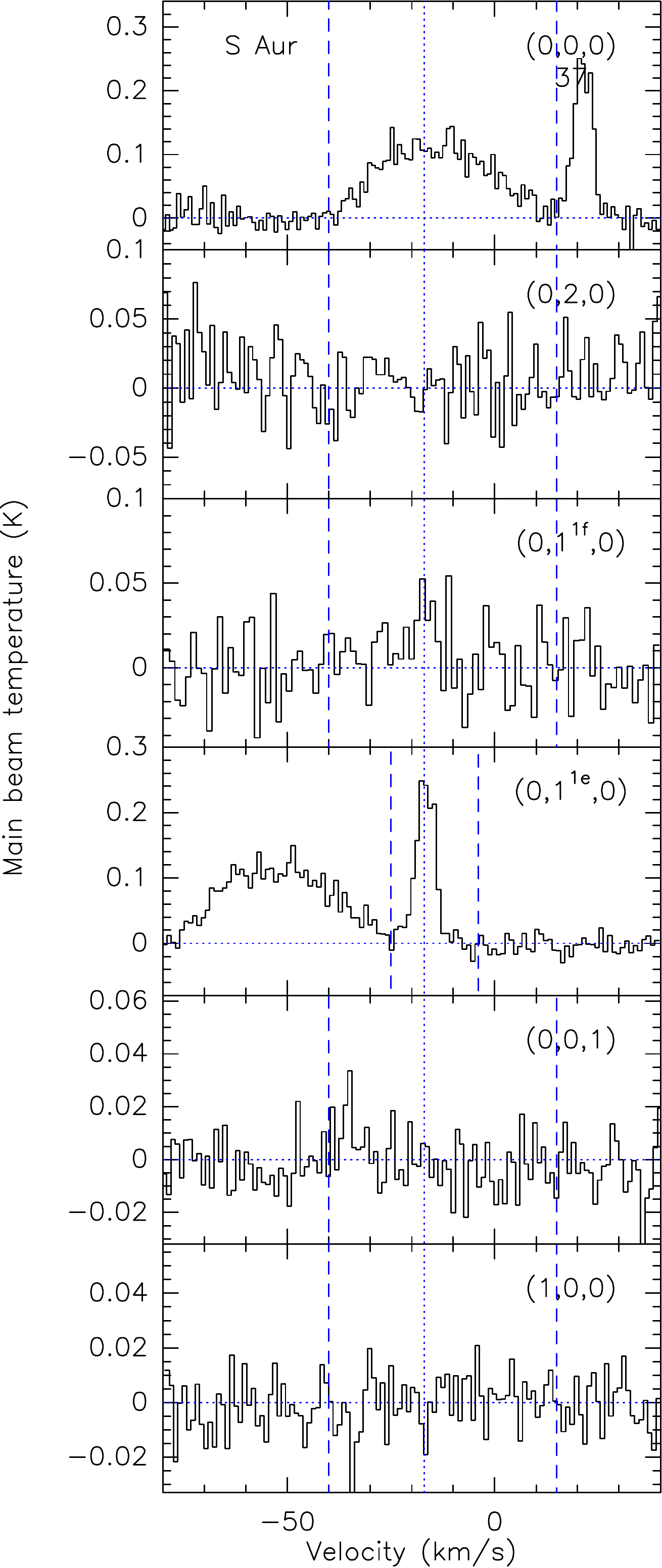}\quad
\includegraphics[width=0.31\textwidth]{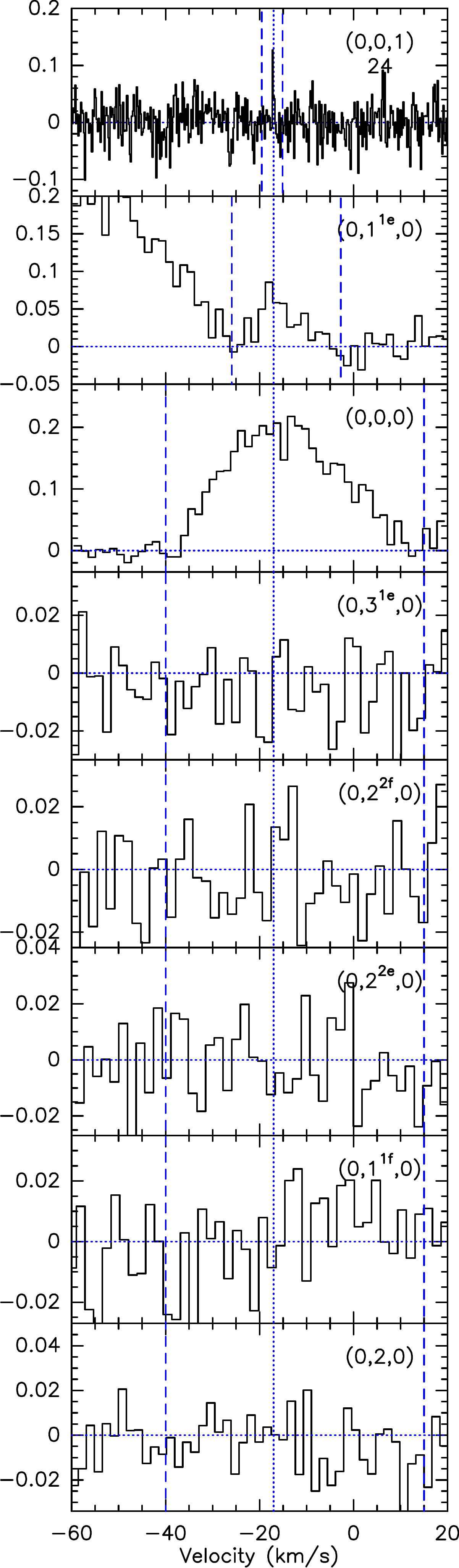}\quad
\includegraphics[width=0.31\textwidth]{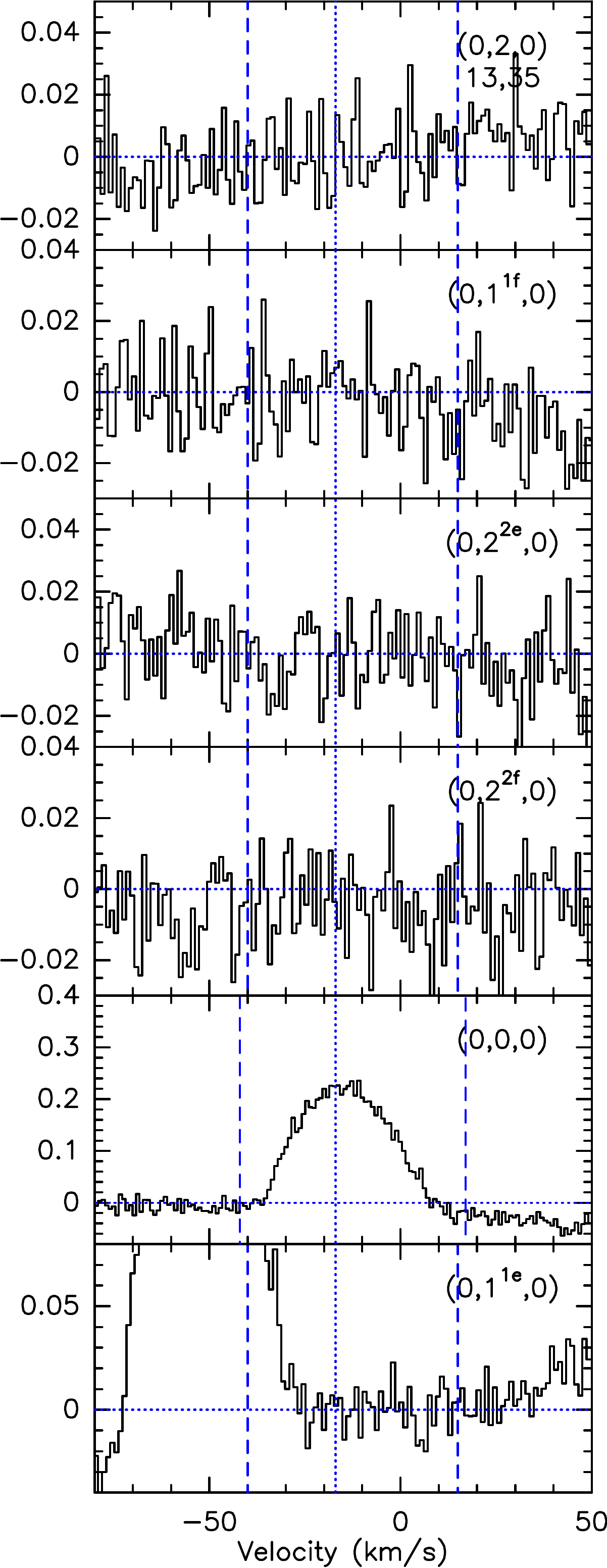}
\caption{Spectra for $J=2$--1 (left), 3--2 (center) and 4--3 (right) transitions of HCN towards S Aur. Description of the figure is the same as in the Fig. \ref{fig:irc_all_spec} caption.}
\label{fig:saur-all}
\end{figure*}

\begin{figure*}%[t]%[!htb]
\centering
\includegraphics[width=0.33\textwidth]{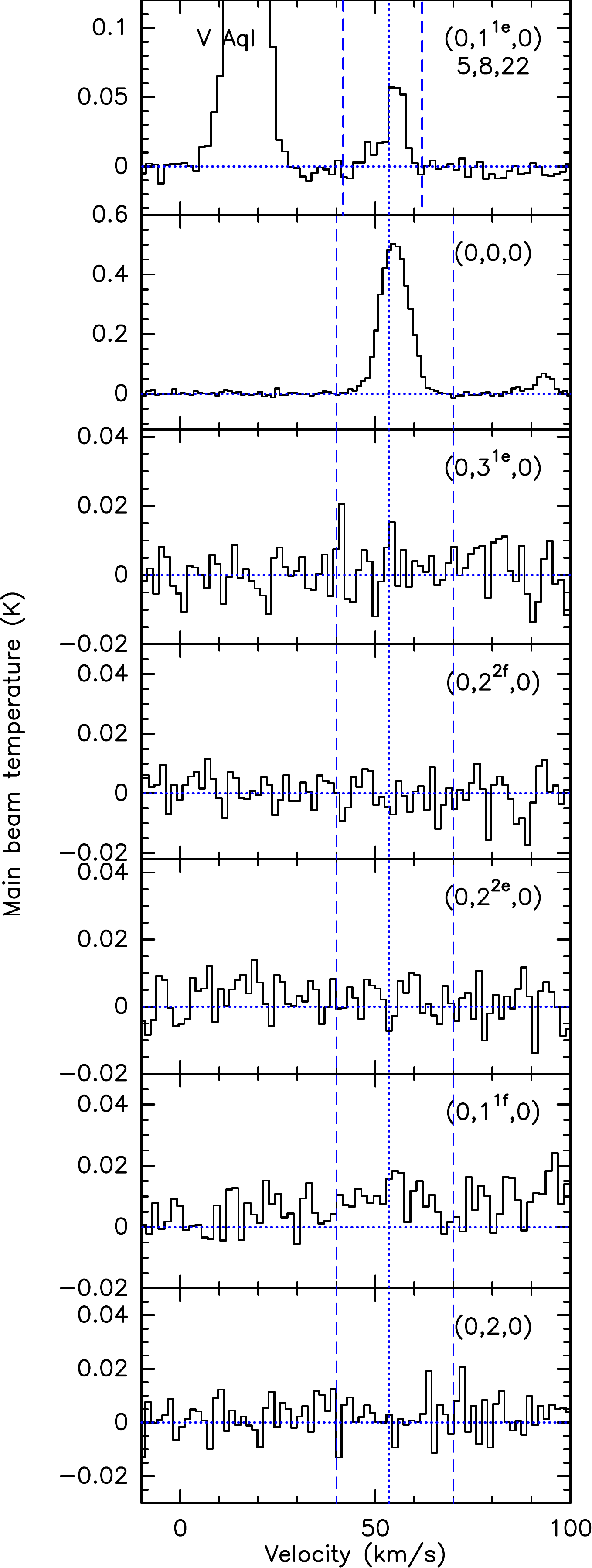}\quad
\includegraphics[width=0.30\textwidth]{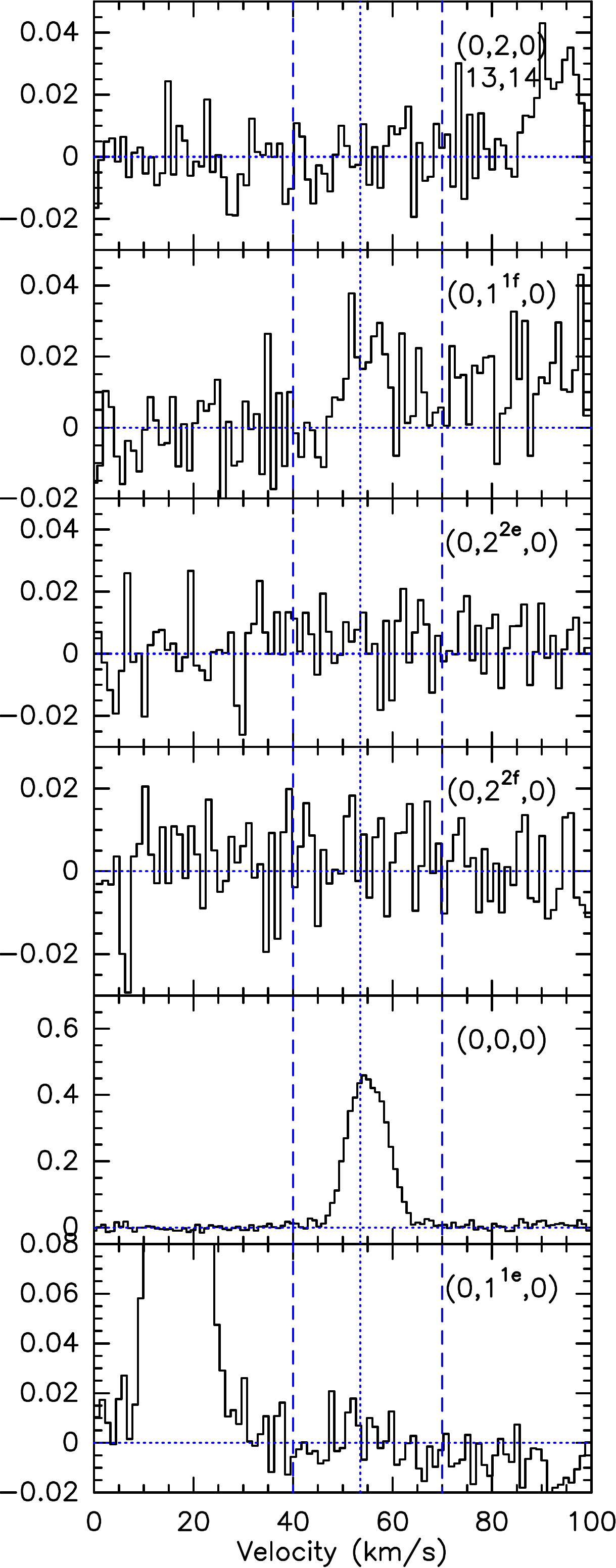}
\caption{Spectra for $J=3$--2 (left) and 4--3 (right) transitions of HCN towards V Aql. Description of the figure is the same as in the Fig. \ref{fig:irc_all_spec} caption.}
\label{fig:vaql-all}
\end{figure*}

\begin{figure*}%[t]%[!htb]
\centering
\includegraphics[width=0.33\textwidth]{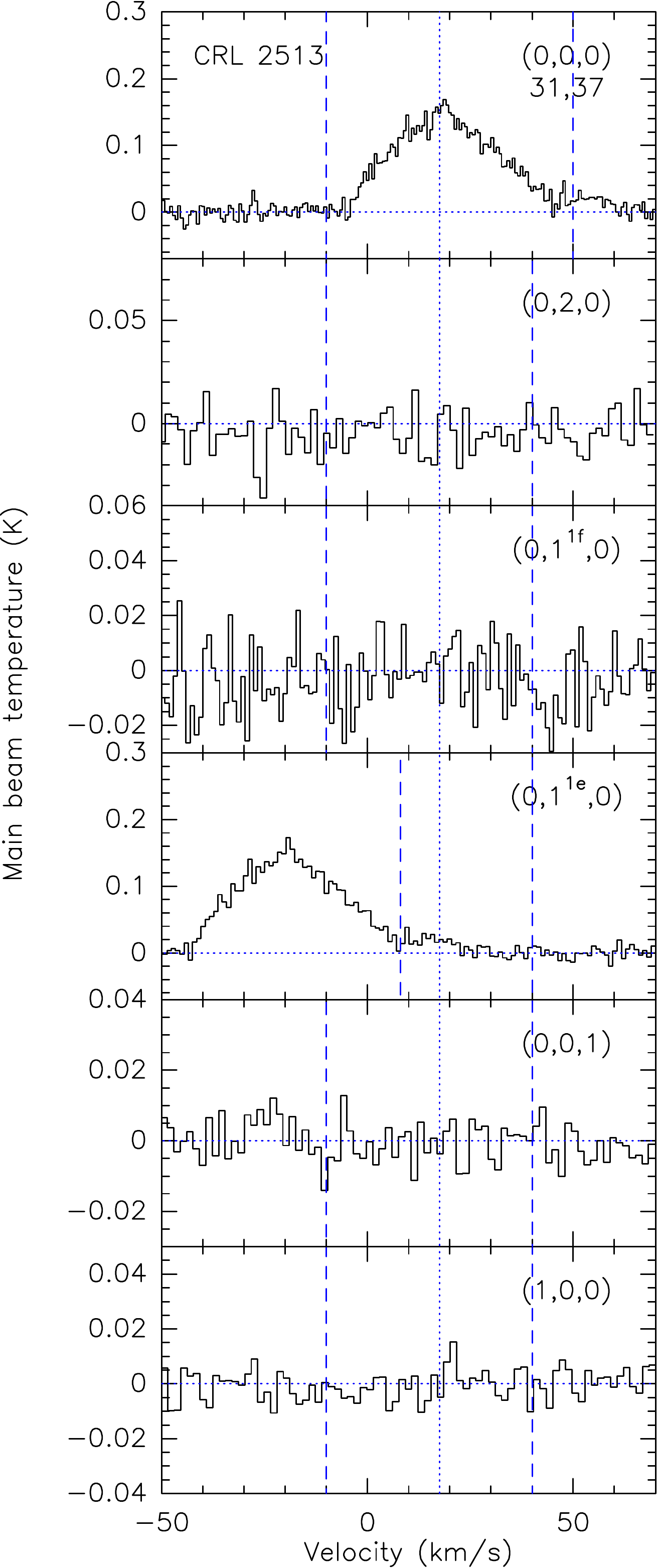}\quad
\includegraphics[width=0.3\textwidth]{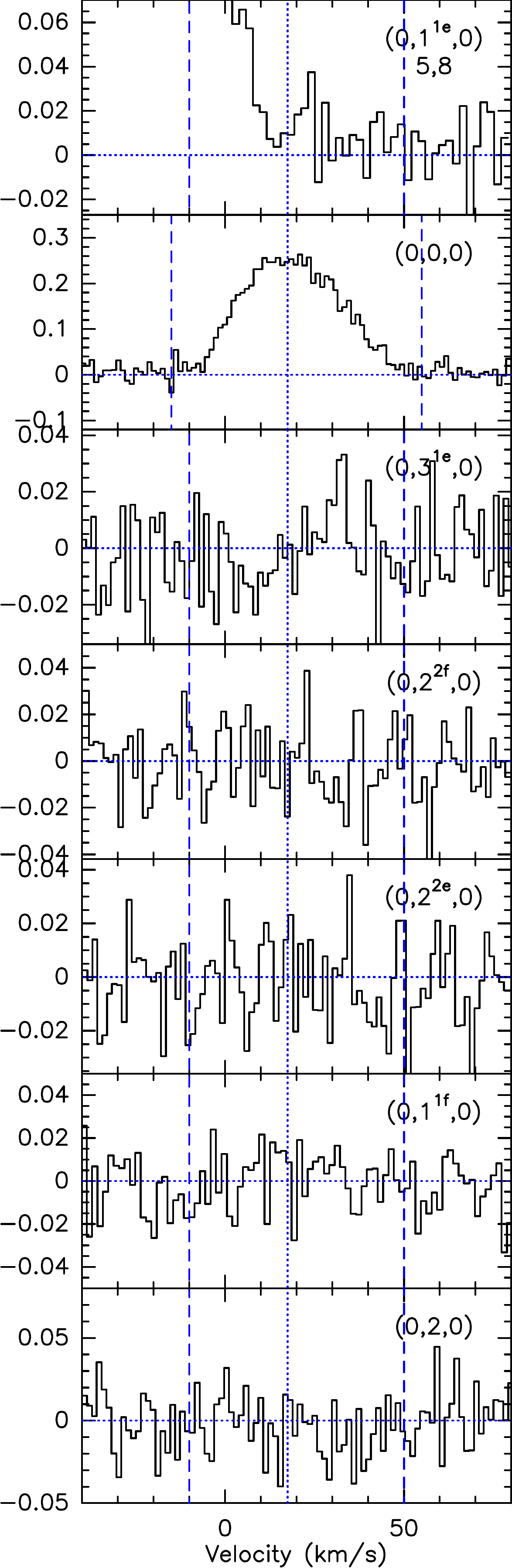}
\caption{Spectra for $J=2$--1 (left) and 3--2 (right) transitions of HCN towards CRL 2513. Description of the figure is the same as in the Fig. \ref{fig:irc_all_spec} caption.}
\label{fig:crl2513-all}
\end{figure*}

\begin{figure*}%[t][!htb]
\centering
\includegraphics[width=0.33\textwidth]{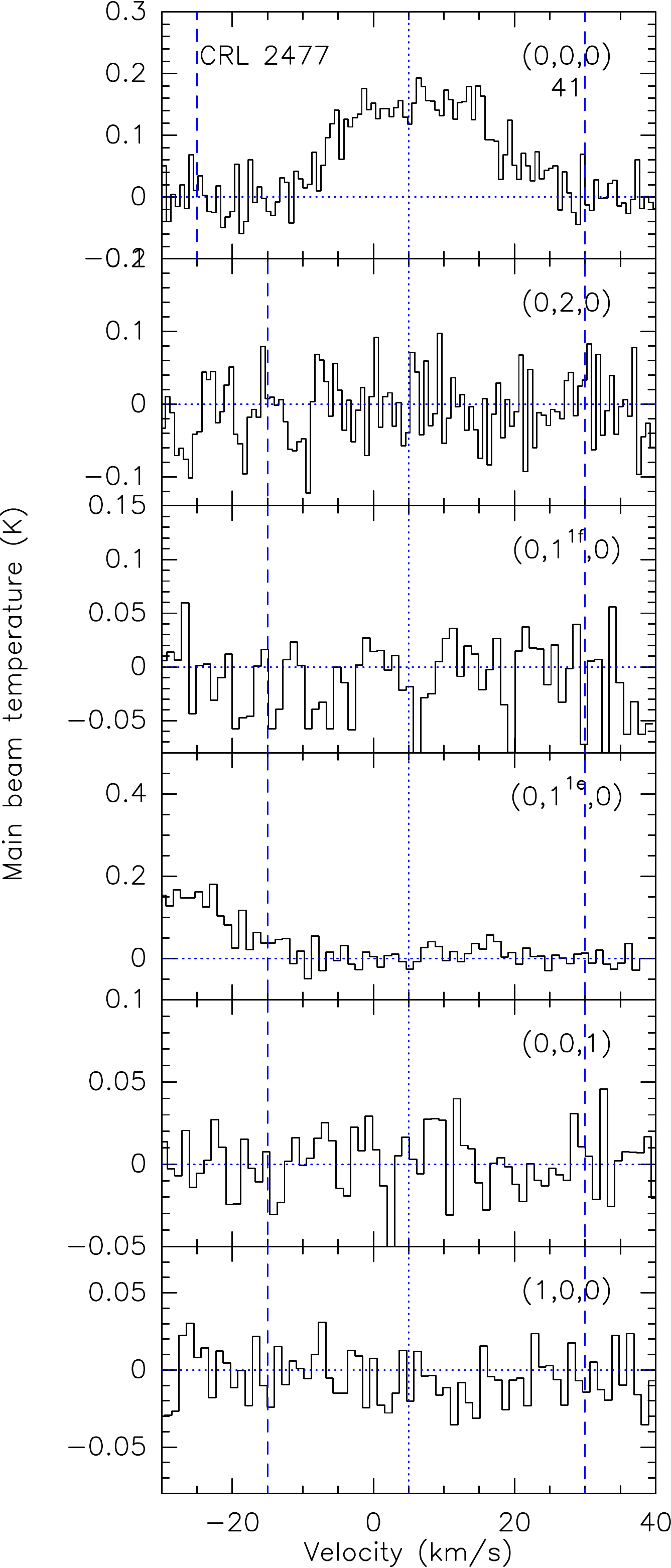}\quad
\includegraphics[width=0.3\textwidth]{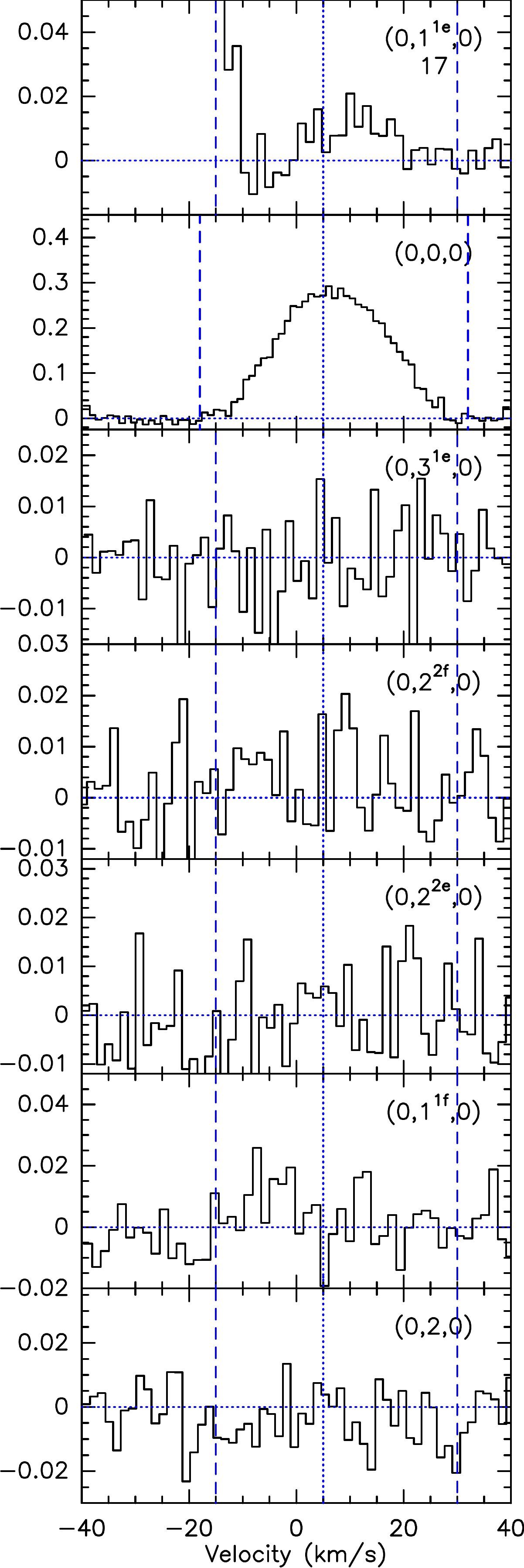}
\caption{Spectra for $J=2$--1 (left) and 3--2 (right) transitions of HCN towards CRL 2477. Description of the figure is the same as in the Fig. \ref{fig:irc_all_spec} caption.}
\label{fig:crl2477-all}
\end{figure*}

\begin{figure*}%[t]%[!htb]
\centering
\includegraphics[width=0.335\textwidth]{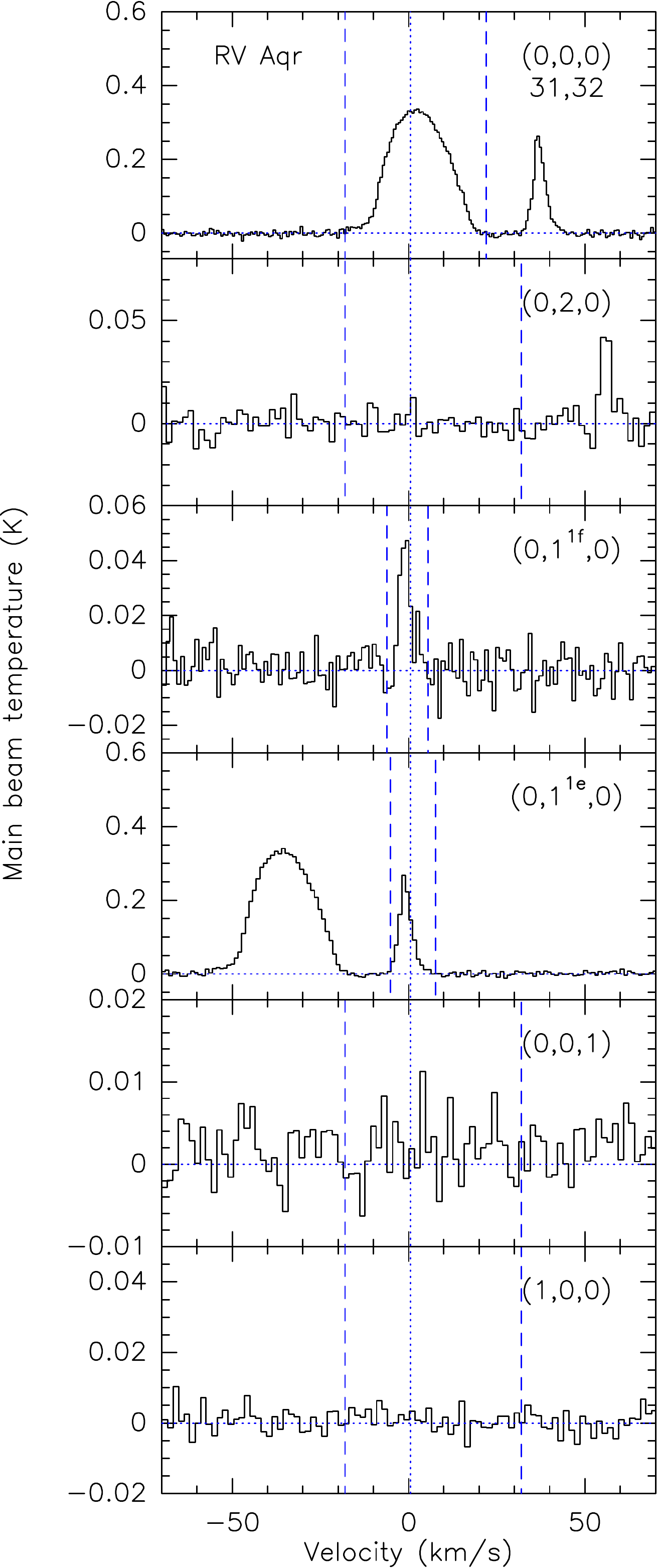}\quad
\includegraphics[width=0.30\textwidth]{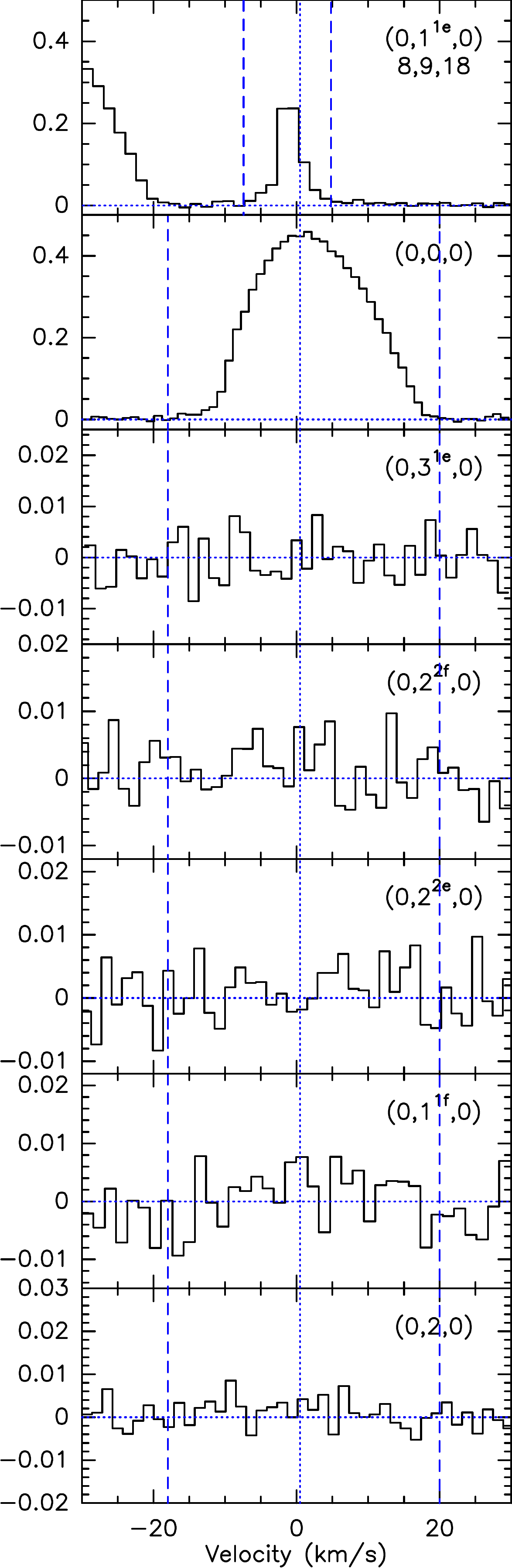}\quad
\includegraphics[width=0.30\textwidth]{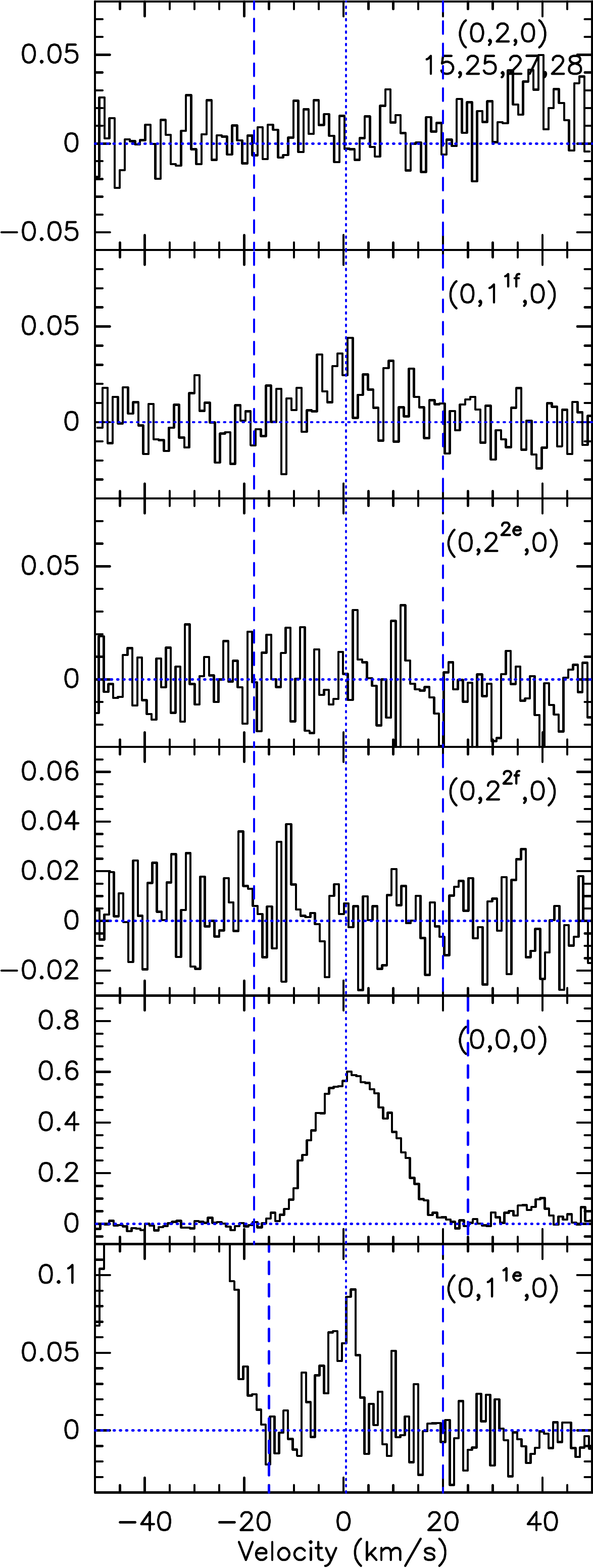}
\caption{Spectra for $J=2$--1 (left), 3--2 (center) and 4--3 (right) transitions of HCN towards RV Aqr. Description of the figure is the same as in the Fig. \ref{fig:irc_all_spec} caption.}
\label{fig:rvaqr-all}
\end{figure*}

In addition to the HCN spectra of IRC +10216, CRL 3068, and II Lup (see Figs.~\ref{fig:irc_all_spec}--\ref{fig:ii-lup_all_spec}), the observed APEX HCN spectra in the other 13 sources are presented in Figs.~\ref{fig:rvol-all}--\ref{fig:rvaqr-all}.


\section{Comparison of HCN emission in $(0, 1^{1e}, 0)$ and $(0, 1^{1f}, 0)$ vibrational states}
\label{app:1e1f}

\begin{figure*}%[t]%[!htbp]
\centering
\includegraphics[width=.34\textwidth]{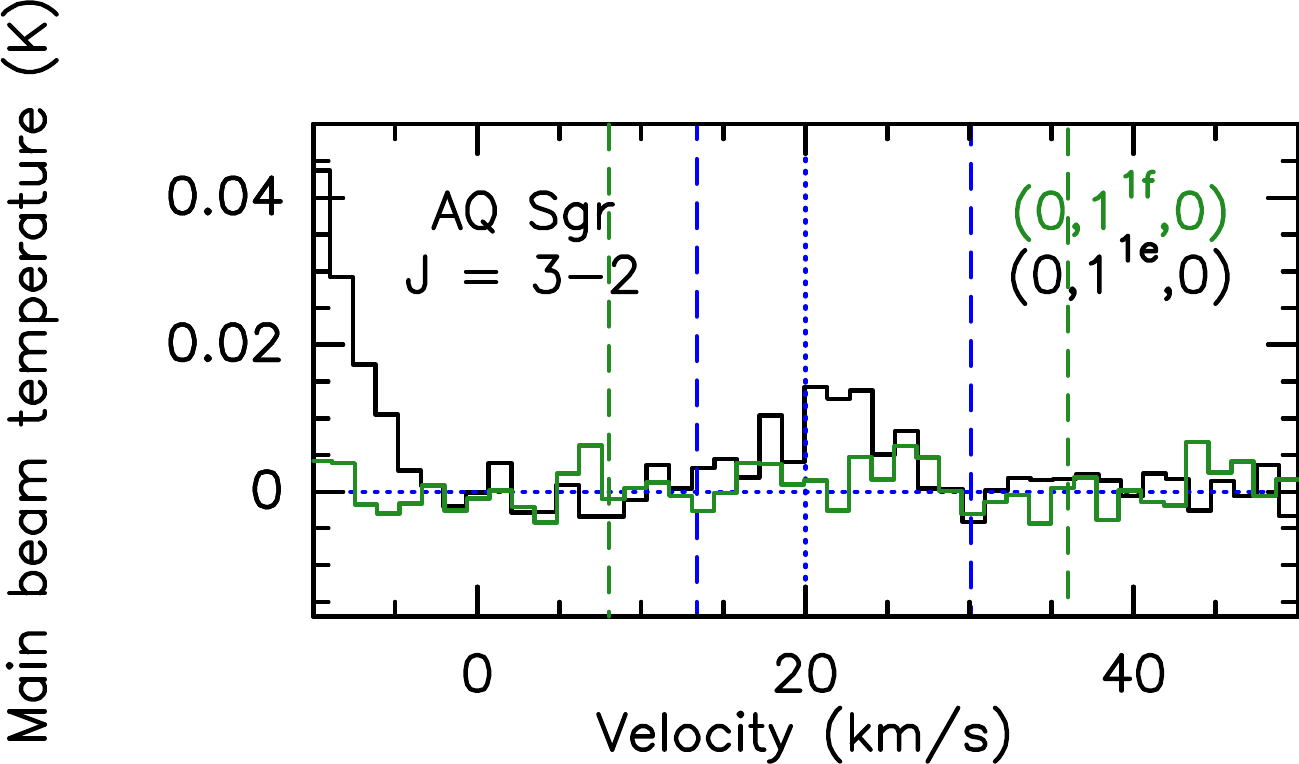}\hfill
\includegraphics[width=.31\textwidth]{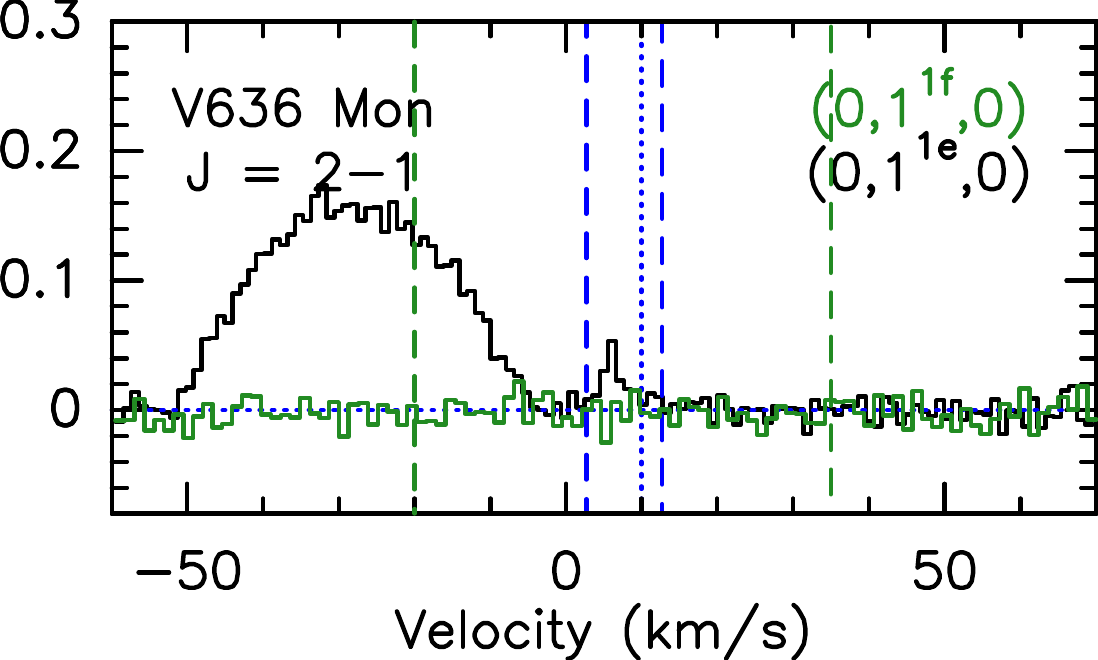}\hfill
\includegraphics[width=.31\textwidth]{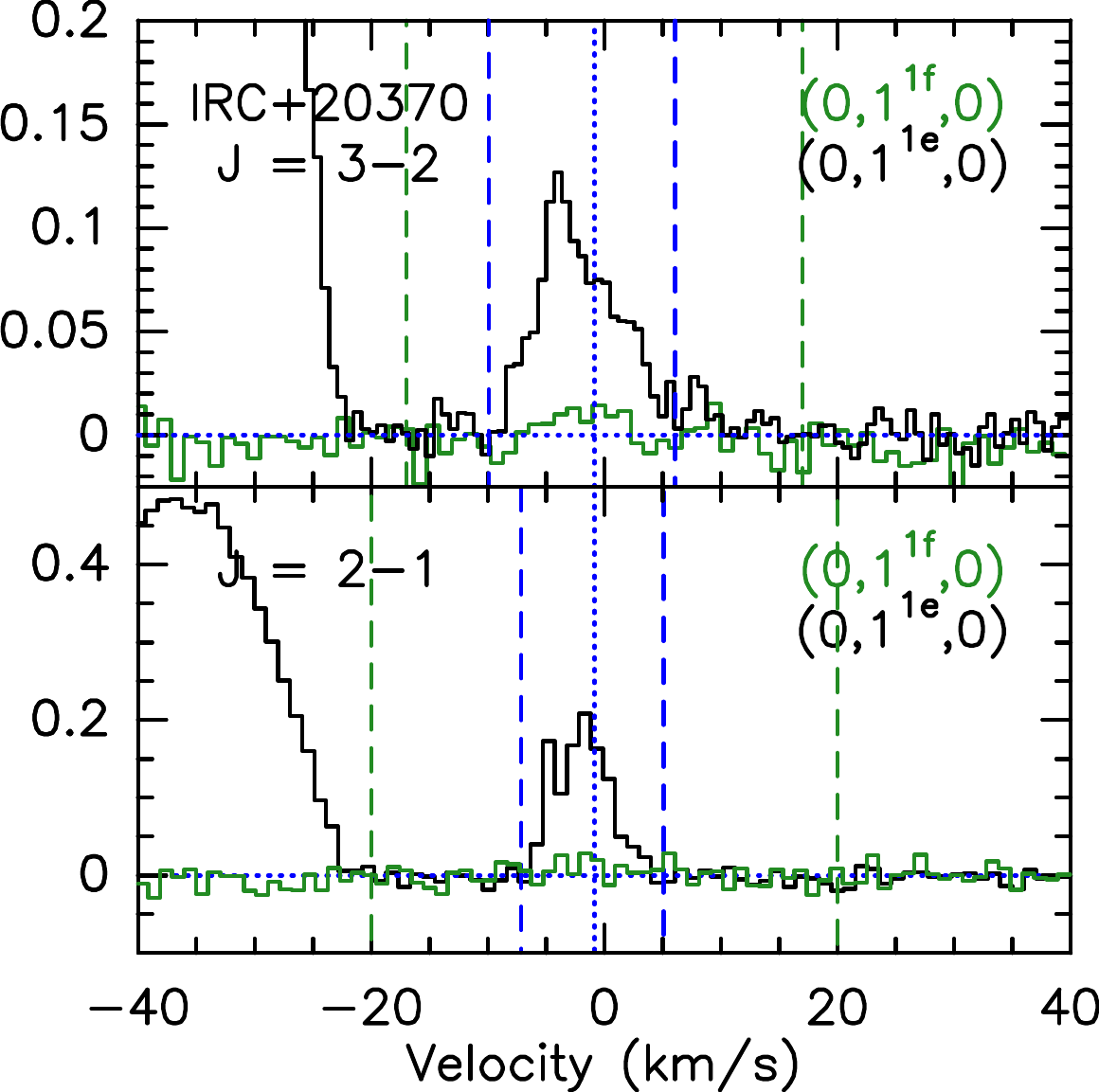}\hfill \\
\includegraphics[width=.34\textwidth]{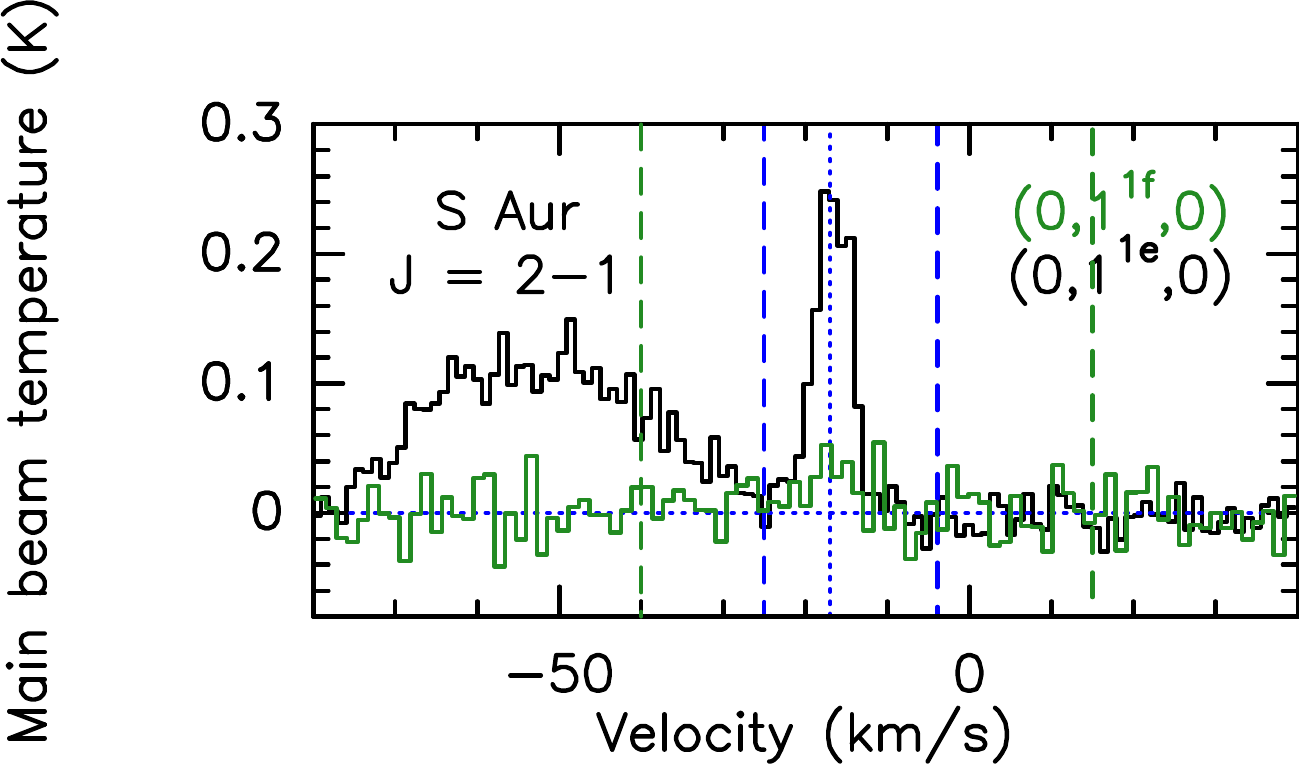}\hfill
\includegraphics[width=.31\textwidth]{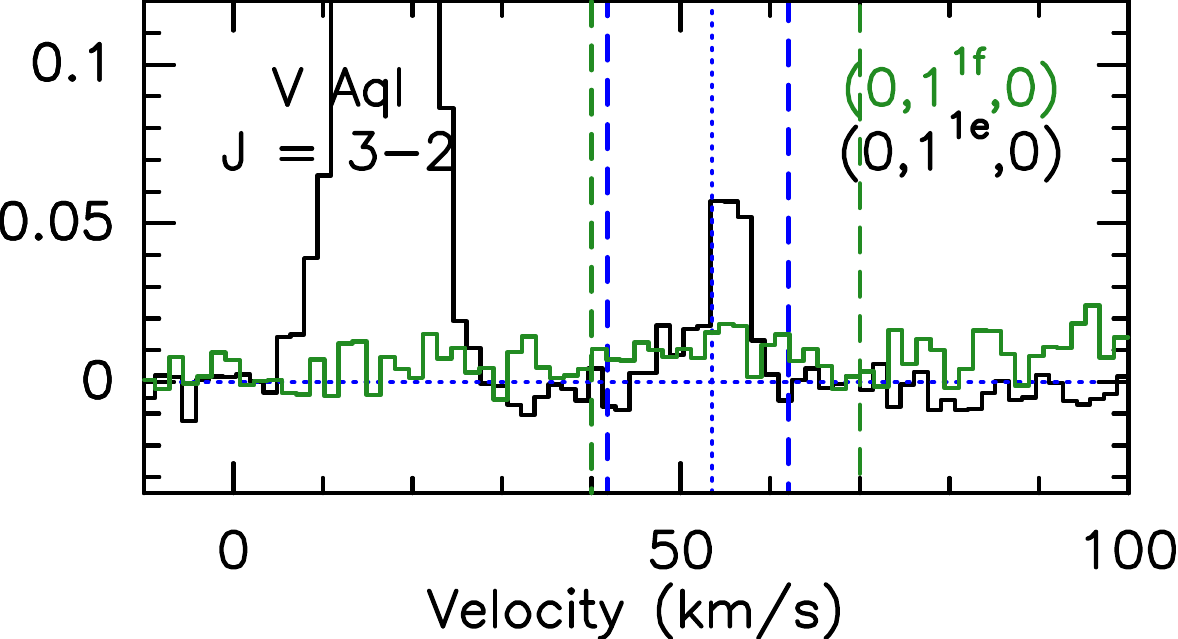}\hfill
\includegraphics[width=.31\textwidth]{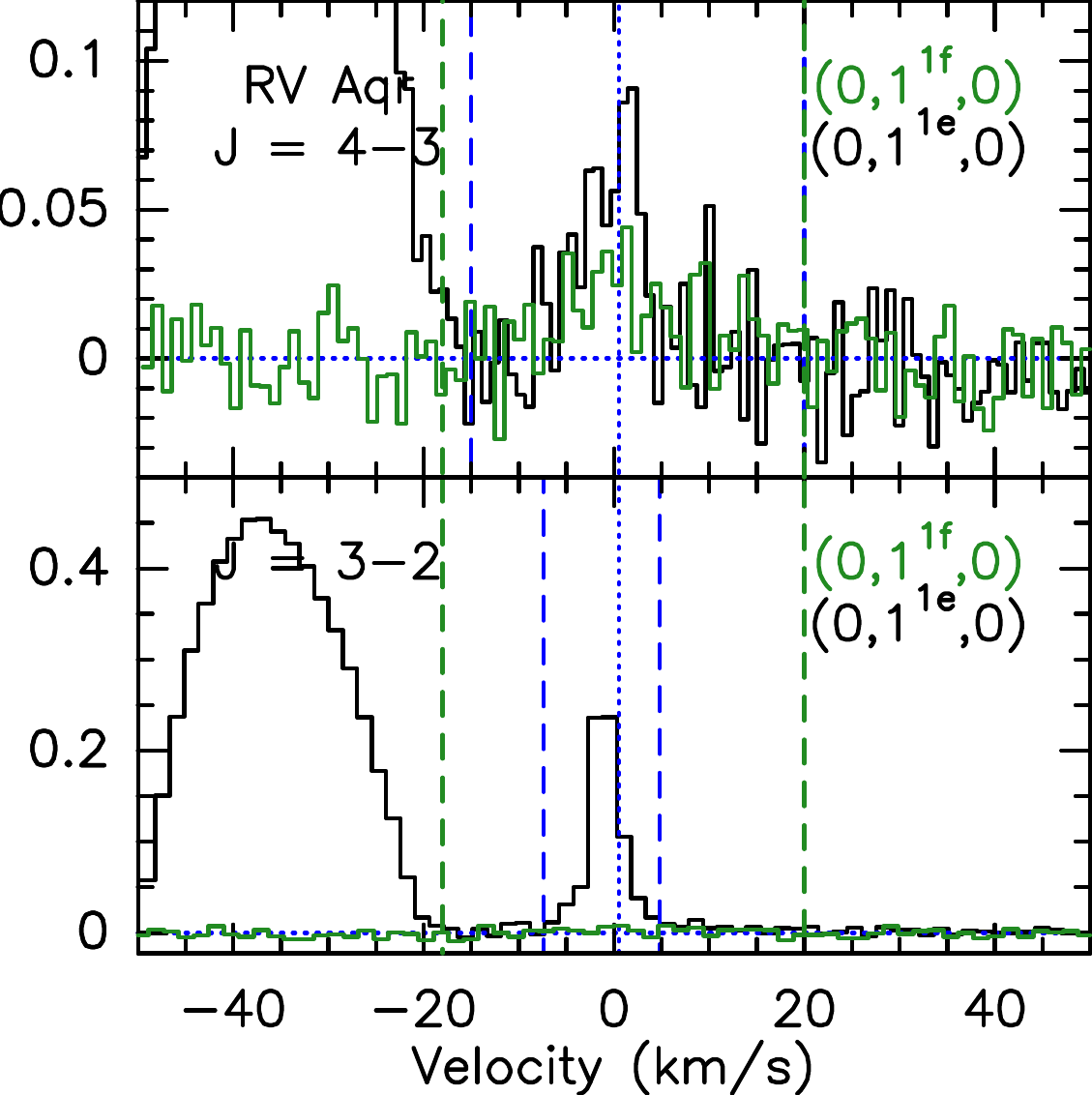}\hfill
\caption{Same as Figs.~\ref{fig:comparison_1e1f} and \ref{fig:comparison_1f1e} toward AQ Sgr, V636 Mon, IRC+20370, S Aur, V Aol, and RV Aqr.}
\label{fig:comparison_1e1f_appendix}
\end{figure*}

\end{appendix}
\end{document}